\documentclass[11pt]{article} 

\usepackage[top=1in, left=1.in, right=1.in, bottom=1in]{geometry}

\usepackage{setspace}
\usepackage[T1]{fontenc}       % Proper font encoding for better output
\usepackage[utf8]{inputenc}    % UTF-8 encoding for special characters
\usepackage{lmodern}           % Improved font rendering (Latin Modern)
\usepackage{microtype}         % Enhances typography for better spacing

\usepackage{amsmath, amssymb, amsthm}  % Core math packages
\usepackage{mathtools}                 % Enhancements for amsmath
\usepackage{bbm}                        % Indicator function and blackboard bold symbols
\usepackage{bm}                         % Bold math symbols
\usepackage{commath}                    % Provides \dif for derivatives and other useful macros
\usepackage{nicefrac}                   % Compact fractions
\usepackage{xfrac}                      % Alternative fraction styles

\usepackage{mathrsfs}        % Ralph Smith’s Formal Script, provides \mathscr
\DeclareMathAlphabet{\mathscr}{OMS}{rsfs}{m}{n}  % Ensure \mathscr uses rsfs font family

\theoremstyle{plain}
\newtheorem{theorem}{Theorem}[section]
\newtheorem{lemma}[theorem]{Lemma}

\newtheorem{remark}[theorem]{Remark}

\theoremstyle{definition}

\usepackage{graphicx}         % Include images
\usepackage{caption}          % Better caption handling
\usepackage{subcaption}       % Subfigures (grouped figures)
\usepackage{booktabs}         % Professional-looking tables
\usepackage{multirow}         % Merging rows in tables
\usepackage{array}            % Improved table layout
\usepackage{float}            % Better figure placement
\usepackage{wrapfig}          % Text wrapping around figures

\usepackage[authoryear]{natbib}  % Citation style (numeric)
\usepackage{hyperref}         % Clickable links in PDF
\hypersetup{
    colorlinks=true,          % Enable colored links
    linkcolor=blue,           % Color for internal links
    citecolor=blue,           % Color for citations
    urlcolor=blue             % Color for URLs
}

\usepackage{tikz}             % Core TikZ package for graphics
\usetikzlibrary{
    decorations.pathmorphing, 
    positioning,              % Easier node placement
    arrows,                   % Arrows for diagrams
    decorations.pathreplacing,% Curly braces for braces
    calc                     % Mathematical calculations
}

\usepackage{pgfplots}
\usepgfplotslibrary{fillbetween} % For filling areas under curves
\pgfplotsset{compat=1.17} % Compatibility setting

\usepackage{algorithm}        % Floating environment for algorithms
\usepackage{algpseudocode}    % Pseudocode syntax

\usepackage{colortbl}      % Color tables: cell shading and colored rows
\usepackage{hhline}        % Enhanced horizontal lines in tables
\usepackage{calc}          % Perform arithmetic calculations in LaTeX lengths
\usepackage{tabularx}      % Tables with adjustable-width columns
\usepackage{comment}       % Environment for commenting out blocks
\usepackage{longtable}
\usepackage{xcolor}           % Colors for text and highlighting
\usepackage{enumitem}         % Custom bullet points and numbering
\usepackage{xspace}           % Smart spacing in custom macros
\usepackage{fancyvrb}         % Improved verbatim environment
\usepackage{tcolorbox}        % Colored boxes for notes and highlights
\usepackage{pifont}
\usepackage{changepage}       % Add indent paragraph

\usepackage{todonotes}        % Inline and margin to-do notes, with \listoftodos

\newcommand{\E}{\mathbb{E}}           % Expectation
\newcommand{\Var}{\mathrm{Var}}       % Variance
\newcommand{\Cov}{\mathrm{Cov}}       % Covariance
\newcommand{\indep}{\perp\!\!\!\perp} % Independence symbol

\allowdisplaybreaks        % Allow page breaks in multi-line display math

\title{\bf Coarsening Bias from Variable Discretization in Causal Functionals}

\author{
     Xiaxian Ou and Razieh Nabi \\[1em]
    {\small Department of Biostatistics and Bioinformatics, Emory University, Atlanta, GA, U.S.A.} \\ [0.25em]
}

\date{}

\begin{document}

\AddToHookNext{shipout/foreground}{%
  \begin{tikzpicture}[remember picture,overlay]
    \node[anchor=north west, font=\small\itshape, align=left]
      at ([xshift=1in,yshift=-0.55in]current page.north west)
      {Accepted to the Forty-Second Annual Conference on Uncertainty in Artificial Intelligence (UAI 2026).};
  \end{tikzpicture}%
}

\maketitle

\begin{abstract}
Causal identification functionals often require integration over conditional densities of continuous variables, such as those arising in nonparametric identification theory of total and mediated causal effects in DAGs with hidden variables. Estimating these densities and evaluating the resulting integrals can be statistically and computationally demanding. A common workaround is to discretize the continuous variable and replace integrals with finite sums. Although convenient, discretization alters the population-level functional and can induce non-negligible approximation bias, even when identification is correct. Under smoothness conditions, we show that the resulting coarsening error is first order in the bin width and arises at the level of the target functional, distinct from statistical estimation error. We propose a simple debiased coarsened functional that evaluates the outcome regression at within-bin conditional means, eliminating the leading coarsening error term and yielding a second-order approximation error. We derive plug-in and one-step estimators for this debiased coarsened functional. Simulations demonstrate substantial bias reduction and near-nominal confidence interval coverage, even under coarse binning. Our results provide a simple framework for controlling the impact of variable discretization on both parameter approximation and statistical estimation. 
\end{abstract}

\noindent%

%%%%%%%%%%%%%%%%%%%%%%%%%%%%%%%%%%%%%%%%%%%%%%%%%%%%%%%%%%%%
\section{Introduction}
\label{sec:intro} 

Several causal parameters admit identification formulas that require averaging an outcome regression over the distribution of mediating variables. This structure arises in mediation analysis \citep{pearl01direct, vanderweele2014mediation}, path-specific effects \citep{avin2005identifiability, shpitser2013counterfactual}, and nonparametric identification of causal effects in hidden-variable DAGs \citep{tian02general, bhattacharya2022semiparametric, richardson2023nested, guo2023flexible, guo2024average, guo2025causal}. When the mediator is continuous or high-dimensional, as is common in biomedical applications involving biomarkers such as BMI, blood pressure, or gene expression profiles and in social science applications involving cognitive test scores, psychometric measurements, or socioeconomic indices \citep{salinas2021discovery, huang2022causal, hemade2025revisiting}, evaluating these identification formulas requires estimating conditional mediator densities and computing integrals with respect to them. Both tasks can be statistically challenging, and Monte Carlo approximations may be computationally intensive and unstable in moderate samples. 

A range of strategies have been proposed to address these challenges. 
One class of approaches directly models the conditional mediator density and evaluates the identifying integral numerically, using (semi/non-)parametric estimators \citep{ma2014multiple, tingley2014mediation, devick2022role, linero2022simulation, li2023effect}. These methods can be computationally demanding, and typically rely on sufficiently accurate estimation of the conditional mediator densities and, in parametric implementations, require correct density specification for consistency. Sequential regression approaches avoid explicit density estimation by rewriting the identifying functional as iterated conditional expectations, thereby reducing the problem to estimation of a collection of regression functions \citep{liu2025bayesian, wang2025targeted}. However, consistency of such estimators often requires correct specification of all the nuisance regression components. Alternatively, the identifying functional can often be expressed using inverse probability weights involving treatment probabilities and mediator density ratios, which may be rewritten via Bayes’ rule. Such weighting approaches can suffer from instability when treatment probabilities are small or poorly estimated \citep{hong2025ratio, zhou2022semiparametric, liu2024two}. Influence-function based estimators, including one-step and targeted minimum loss estimators \citep{van2000asymptotic, van2006targeted}, provide a principled approach to estimation bias reduction and inference with flexible nuisance estimation, and can offer robustness to some forms of model misspecification. However, such estimators are often avoided in practice due to methodological complexity. 

In practice, a simpler workaround is often adopted by discretizing the mediator and replacing integrals with finite sums \citep{chung2022association, morera2023discretizing, piccininni2023effect}. This avoids estimation of a continuous conditional density and simplifies computation. However, discretization changes the population-level parameter. Even when the causal identification assumptions hold, the discretized functional is, in general, a different parameter from the original one. In many applied analyses, the bias induced by mediator discretization is either overlooked or treated as secondary relative to computational considerations, and its impact on the target parameter is rarely quantified.  

Here, we study the difference between the original causal functional and its discretized analogue, which we refer to as the coarsening error (or, when viewed relative to the original target parameter, the coarsening bias). Our primary focus is not on first-order estimation bias arising from nuisance estimation, but rather on the population-level approximation error induced by mediator discretization itself. We show that, under smoothness conditions, the naive discretized functional incurs a first-order error proportional to the maximum bin width, driven by within-bin shifts in mediator means across treatment levels. We then construct a debiased coarsened functional that evaluates the outcome regression at within-bin conditional means. A Taylor expansion argument shows that this modification eliminates the leading first-order coarsening bias term, yielding a second-order remainder in the bin width. For statistical estimation, we derive plug-in and one-step estimators and establish their asymptotic properties. Simulation studies demonstrate substantial bias reduction and improved coverage relative to naive discretization, even under coarse binning. More broadly, our results provide a formal foundation for controlling the impact of discretization on both parameter approximation and statistical estimation. 
%%%%%%%%%%%%%%%%%%%%%%%%%%%%%%%%%%%%%%%%%%%%%%%%%%%%%%%%%%%%
\section{Target estimands}
\label{sec:estimands}
Let $A$ denote a binary exposure taking values in $\{a_0, a_1\}$, $C$ observed covariates with support $\mathcal{C}$ and marginal distribution $P_C$, $M$ a mediator, and $Y$ the outcome. Define the outcome regression $\mu(m,a,c) \!=\! \E(Y \,|\, M \!\!=\!\! m, A \!\!=\!\! a, C \!\!=\!\! c)$, the conditional mediator density $f_{M\,|\, A,C}(m \,|\, a,c) \!=\! p(M \!\!=\!\! m \,|\, A \!\!=\!\! a, C \!\!=\!\! c)$, and the propensity score $\pi(a \,|\, c) \!=\! p(A \!\!=\!\! a \,|\, C \!\!=\!\! c)$. Let $Q = \{\mu, f_{M\,|\, A,C}, \pi\}$ collect the nuisance functions. The specific causal functionals considered below may depend only on subsets of $Q$. For each $c \in \mathcal{C}$, define 
\begin{align}
    \theta(Q)(c) = \int \mu(m, a_1, c)\, f_{M|A,C}(m \,|\, a_0, c)\, dm \, . 
    \label{eq:estimand}
\end{align}%
The conditional functional $\theta(Q)(c)$ serves as a building block for several causal parameters. In particular, define
\begin{equation}\label{eq:causal_targets}
\begin{aligned}
\psi(Q) 
&= \int \theta(Q)(c) \, p(c) \, dc \, , 
\\
\gamma(Q)
&= p(a_0) \, \E(Y \,|\, a_0) +  \int \theta(Q)(c) \, \pi(a_1 \,|\, c)  \, p(c) \, dc \, . 
\end{aligned}%
\end{equation}

Under standard identification assumptions (see Appendix~\ref{app:proofs_identification}), $\psi(Q)$ corresponds to the mediation identification functional for $\E(Y(a_1, M(a_0)))$, while $\gamma(Q)$ corresponds to the front-door identification functional for $\E(Y(a_0))$, where $Y(a,m)$ denotes the potential outcome under $A=a$ and $M=m$ \citep{pearl01direct, pearl1995causal, guo2023flexible, guo2024average}. 

Given $n$ i.i.d. observations $(C_i, A_i, M_i, Y_i)_{i=1}^n$ and nuisance estimates collected in $\widehat Q$, plug-in estimators of $\psi(Q)$ and $\gamma(Q)$ are given by 
\begin{equation}\label{eq:causal_targets_plugin}
\begin{aligned}
\psi(\widehat{Q})
&= \frac{1}{n} \sum_{i = 1}^n \theta(\widehat{Q})(C_i) \, , \\
\gamma(\widehat{Q})
&= \frac{1}{n} \sum_{i = 1}^n \big\{ I(A_i = a_0)Y_i +   \theta(\widehat{Q})(C_i) \, \widehat{\pi}(a_1 \,|\, C_i) \big\} \, .
\end{aligned}%
\end{equation}
When $M$ is continuous, evaluation of $\theta(\widehat Q)(c)$ requires computing $\int \widehat{\mu}(m, a_1, c)\, \widehat{f}_{M|A,C}(m \,|\, a_0, c)\, dm$, which may involve numerical integration or Monte Carlo approximation under an estimated working model for the conditional density $f_{M \,|\, A,C}$. In many practical settings, this can be statistically and computationally challenging. A common simplification is to coarsen the mediator by discretization. 

Next, we formalize such coarsened functionals and analyze the resulting coarsening errors.

%%%%%%%%%%%%%%%%%%%%%%%%%%%%%%%%%%%%%%%%%%%%%%%%%%%%%%%%%%%%
\section{Coarsened estimands}
\label{sec:coarsened}

For clarity we present the development for a univariate mediator $M\in\mathbb{R}$; extensions to multivariate mediators are discussed in Appendix~\ref{app:sec_multiple_mediators}. For simplicity, we assume throughout that the support of $M$ is contained in a compact interval ${\cal M}=[\ell,u]\subset\mathbb{R}$.  

Let $h:\mathbb{R}\to\{1,\ldots,K\}$ be a measurable discretization map that partitions the support of $M$ into disjoint bins ${\cal B}_k=\{m:h(m)=k\}$, $k\in\{1,\ldots,K\}$. Let $w_k=\sup_{m_1,m_2\in{\cal B}_k}|m_1-m_2|$ denote the bin width and define $\widetilde M=h(M)$. The rule $h$ may correspond, for instance, to thresholding $M$ at a fixed or data-dependent cutoff, but more generally can encode any measurable partition. For example, if $M$ is BMI and the observed values lie in the interval $[\ell,u]$, one may define $K=4$ bins ${\cal B}_1=[\ell,18.5)$, ${\cal B}_2=[18.5,25)$, ${\cal B}_3=[25,30)$, and ${\cal B}_4=[30,u]$, with $\widetilde M=1,\ldots,4$ denoting underweight, healthy weight, overweight, and obesity, respectively.

For each bin $k$, define $\mu_k(a,c){=}\E(Y\,|\, \widetilde M{=}k, A{=}a, C{=}c)$ and $g_k(a,c)=p(\widetilde M{=}k\,|\, A{=}a, C{=}c)$.  
The coarsened analogue of \eqref{eq:estimand} is
\begin{align}
\theta_h(Q)(c)=\sum_{k=1}^K \mu_k(a_1,c)\, g_k(a_0,c) \, .
\label{eq:estimand_coarsened}
\end{align}
The coarsened versions of estimands in \eqref{eq:causal_targets}, denoted by $\psi_{h}(Q)$ and $\gamma_{h}(Q)$, are obtained by replacing $\theta$ with $\theta_h$. We primarily focus on $\theta$ since coarsening behavior propagates to $\psi$ and $\gamma$ through $\theta$ by averaging over $C$. 

 Given our notation, we can write $\theta(Q)(c)$ in \eqref{eq:estimand} as 
\begin{align*}
    \sum_{k=1}^K  \Big\{ \int \mu(m, a_1,c) \, p(m\,|\, \widetilde{M} = k, a_0, c) \, dm \Big\} \,  g_k(a_0, c) \, . 
\end{align*}
Let $\mu_{k,a_1}(a,c){=}\E(\mu(M,a_1,c)\,|\, \widetilde M{=}k, A{=}a, C{=}c)$. Thus, $\theta(Q)(c)$ in \eqref{eq:estimand} can be expressed as 
\begin{align}
    \theta(Q)(c) = \sum_{k=1}^K  \mu_{k, a_1}(a_0, c) \, g_k(a_0, c) \, . 
\end{align}

\begin{lemma}\label{lem:coarsening_error}
For each $c\in\mathcal{C}$, the coarsening error defined as $\Delta_h(Q)(c)=\theta_h(Q)(c)-\theta(Q)(c)$ equals
\begin{align}
\Delta_h(Q)(c)
\!=\! \sum_{k=1}^K \{\mu_k(a_1,c) \!-\! \mu_{k,a_1}(a_0,c)\}\, g_k(a_0,c) \, .
\label{eq:coarsening_error}
\end{align}
If $m \mapsto \mu(m,a_1,c)$ is continuously differentiable on each ${\cal B}_k$ and for each fixed $c$ there exists a finite constant $L(c) < \infty$ such that $\sup_{m\in{\cal B}_k}|\mu'_m(m,a_1,c)|\le L(c)$ for all $k\in\{1, \ldots, K\}$ and $L(C)$ is square-integrable under $P_C$, then $|\Delta_h(Q)(c)| = O\big(w_{\max,K}\big)$, where $w_{\max,K} \coloneqq \max_{k\in\{1,\ldots,K\}} \, w_k$ is the maximum bin width. In particular, if $M$ has bounded support and the $K$ bins are chosen to have equal width, then $\Delta_h(Q)(c)=O(1/K)$.
\end{lemma}

We refer to $\Delta_h(Q)$ as the coarsening error. Since inference is ultimately aimed at $\psi(Q)$, this error may also be interpreted as a coarsening bias induced by replacing the original estimand with its coarsened analogue.

When one bin is much wider than the others, e.g., $\{0\}$ versus $(0,M_{\max}]$, the worst-case bound based on $w_{\max,K}$ can be conservative; a sharper bound uses the weighted average $\sum_{k=1}^K w_k g_k(a_0,c)$. See a proof in Appendix~\ref{app:proofs_lem:coarsening_error}.

To understand the magnitude of coarsening error in \eqref{eq:coarsening_error} and develop corrections for the resulting coarsening bias, we take a closer look at the difference between $\mu_k(a_1, c)$ and $\mu_{k, a_1}(a_0, c)$. 

Assume $\mu$ is twice continuously differentiable on each ${\cal B}_k$, and define the within-bin mean 
\begin{align}
    m_k(a,c) = \E(M \,|\, A=a, C=c, \widetilde M=k) \ . 
    \label{eq:within_bin_mean}
\end{align}%
Under the conditions of Lemma~\ref{lem:coarsening_error}, a Taylor expansion of $\mu(M, a_1, c)$ around $m_k$ yields  
\begin{equation}\label{eq:first_order_scaling}
\mu_k(a_1,c)-\mu_{k,a_1}(a_0,c) = \mu'_m(m_k(a_1,c),a_1,c)\{m_k(a_1,c) \!-\! m_k(a_0,c)\} \!+\! O(w_k^2) \, , 
\end{equation}
where $\mu_m'$ is the derivative of $\mu$ w.r.t $m$; see Appendix~\ref{app:proofs_lem:coarsening_error}. 

Equation \eqref{eq:first_order_scaling} reveals that the coarsening error in \eqref{eq:coarsening_error} scales linearly with the shift in within-bin mediator means between treatment levels and the local slope of the outcome regression in $M$. The higher-order terms are proportional to the curvature of $\mu$ and the within-bin spread of $M$; under mild smoothness conditions, these remainder terms are of second order in the bin width. Therefore $\Delta_h(Q)(c)$ admits a first-order approximation with remainder $O(w_{\max,K}^2)$. In the following remark, we re-express the coarsening error via an exact covariance representation. 

\begin{remark}\label{remark:covariance}
$\Delta_h(Q)(c)$ admits an exact representation in terms of the within-bin covariance between (i) the conditional outcome surface $\mu(M,a_1,c)$ and (ii) the treatment-induced density ratio $r_k(M|c) = p(M\,|\, \widetilde M=k, A=a_0,C=c)/p(M\,|\, \widetilde M=k, A=a_1,C=c)$.  
The coarsening error vanishes if within every bin, these two quantities are uncorrelated under the distribution of $M$ given $(\widetilde M=k, A=a_1,C=c)$.  
This occurs, for instance, if $\mu$ is locally constant in $M$ within bins, or if the conditional distribution of $M$ does not differ between $A=a_0$ and $A=a_1$ inside each bin; see Appendix~\ref{app:proofs_remark:covariance}. 
\end{remark}

The plug-in estimate of the coarsened estimand $\theta_h(Q)(c)$ is
\begin{align}
\theta_h(\widehat Q)(c)=\sum_{k=1}^K \widehat\mu_k(a_1,c)\,\widehat g_k(a_0,c) \, ,
\label{eq:plug-in-estimator}
\end{align}
where $\widehat\mu_k(a,c)$ and $\widehat g_k(a,c)$ estimate $\mu_k(a,c)$ and $g_k(a,c)$, e.g., via regressions of $Y$ on $(\widetilde M,A,C)$ and $\widetilde M$ on $(A,C)$. 
The plug-in estimators $\psi_h(\widehat Q)$ and $\gamma_h(\widehat Q)$ are obtained by replacing $\theta(\widehat{Q})$ with $\theta_h(\widehat{Q})$ in \eqref{eq:causal_targets_plugin}.

We decompose $\theta_{h}(\widehat{Q})(c) {-} \theta(Q)(c)$ as: 
\begin{align}
   &\underbrace{ \big\{ \theta_{h}(\widehat{Q})(c) {-}  \theta_{h}(Q)(c) \big\} }_{\substack{\text{estimation error}}}  {+} \underbrace{ \big\{ \theta_{h}(Q)(c) {-} \theta(Q)(c) \big\} }_{\text{coarsening error}} \, . 
\end{align}
Assume that for fixed $h$, the estimator $\theta_h(\widehat Q)$ satisfies
\begin{align*}
    \|\theta_h(\widehat Q)-\theta_h(Q)\|_{L_2(P_C)}=O_p(n^{-1/2}) \, , 
\end{align*}%
where $\|g\|_{L_2(P_C)}^2=\E(g(C)^2)$. 
Then, by Lemma~\ref{lem:coarsening_error},
\begin{align*}
\|\theta_h(\widehat Q)-\theta(Q)\|_{L_2(P_C)}
=
O_p(n^{-1/2}) + O(w_{\max,K}) \, . 
\end{align*}%
Hence the asymptotic bias is driven by the coarsening error $\Delta_h(Q)$, which is first order in the bin width. 

Let $K=K_n$ increase with $n$. If $M$ has bounded support and equal-width bins are used, then $w_{\max,K_n}=O(1/K_n)$. If $K_n\to\infty$ and $w_{\max,K_n}=o(n^{-1/2})$,  
% (e.g., $K_n=o(n^{1/2})$), 
the coarsening error is negligible relative to sampling error.
% and $\theta_h(\widehat Q)$ is asymptotically equivalent to $\theta$ in $L_2(P_C)$. 
Under equal-width binning, this requires $K_n/\sqrt n \to \infty$. 

%%%%%%%%%%%%%%%%%%%%%%%%%%%%%%%%%%%%%%%%%%%%%%%%%%%%%%%%%%%%
\section{Debiased coarsened estimands} 
\label{sec:debiased-bin}

We introduce a simple correction that removes the leading coarsening error by evaluating the outcome regression $\mu(., a_1, c)$ at within-bin representatives of the mediator under $A=a_0$. 
Given $m_k(a,c)$ in \eqref{eq:within_bin_mean}, define the debiased coarsened functional
\begin{align}
\widetilde\theta_h(Q)(c) = \sum_{k=1}^K \mu\big(m_k(a_0,c), a_1, c\big)\, g_k(a_0,c) \, . 
\label{eq:tilde-theta}
\end{align}

Taking a first-order Taylor expansion of $\mu(M,a_1,c)$ around $m_k(a_0,c)$ and then taking expectations conditional on $(\widetilde M = k, A = a_0, C = c)$ motivates the approximation of $\mu_{k,a_1}(a_0,c)$ by $\mu\big(m_k(a_0,c),a_1,c\big)$; see Appendix~\ref{app:proofs_lem:first-order-removal}. Thus \eqref{eq:tilde-theta} replaces the in-bin conditional expectation of $\mu(M,a_1,c)$ under $A=a_0$ by its evaluation at the within-bin mean under the same distribution.

The debiased coarsened causal parameters $\widetilde\psi_h(Q)$ and $\widetilde\gamma_h(Q)$ are obtained from \eqref{eq:causal_targets} by replacing $\theta$ with $\widetilde\theta_h$.

\begin{lemma}
\label{lem:first-order-removal} 
For each $c \in \mathcal{C}$, the coarsening error defined as $\widetilde\Delta_h(Q)(c)=\widetilde\theta_h(Q)(c)-\theta(Q)(c)$ equals
\begin{align}
\widetilde\Delta_h(Q)(c) \!=\!\! \sum_{k=1}^K \{ \mu(m_k(a_0,c),a_1,c) \!-\! \mu_{k,a_1}(a_0,c) \!\} g_k(a_0,c).
\label{eq:coarsening_error_tilde_delta_h}
\end{align}%
If $m \mapsto \mu(m,a_1,c)$ is twice continuously differentiable and $R_k(c) = \mu\big(m_k(a_0,c),a_1,c\big) - \mu_{k,a_1}(a_0,c)$, then 
{
\begin{align}
|R_k(c)|
\le
\frac12
\sup_{m\in{\cal B}_k}
\!\! |\mu''_m(m,a_1,c)|
\,
\E\big((M-m_k(a_0,c))^2 \,|\, a_0,c,k\big).
\label{eq:remainder-bound}
\end{align}
}%
If $\sup_{m\in{\cal B}_k}|\mu''_m(m,a_1,c)| \leq L(c)$, where $L(C)$ is square-integrable under $P_C$, then  $\widetilde\Delta_h(Q)(c) = O(w_{\max,K}^2)$. If $M$ has bounded support and the $K$ bins are chosen to have equal width, then $\widetilde\Delta_h(Q)(c)=O(1/K^2)$. 
\end{lemma}
See Appendix~\ref{app:proofs_lem:first-order-removal} for a proof. 

Lemma~\ref{lem:first-order-removal} shows that the leading
first-order component of the coarsening error in Lemma~\ref{lem:coarsening_error} is eliminated. The remaining error is proportional to the curvature of $\mu(m,a_1,c)$ and the within-bin variance of $M$, and is thus second order in the bin width.

The plug-in estimate of the debiased coarsened estimand $\widetilde\theta_h(Q)(c)$ in \eqref{eq:tilde-theta} is
\begin{align}
\widetilde\theta_h(\widehat Q)(c)
=
\sum_{k=1}^K
\widehat\mu\big(\widehat m_k(a_0,c), a_1, c\big)\,
\widehat g_k(a_0,c) \, ,
\label{eq:plug-in-estimator-tilde}
\end{align}%
where $\widehat m_k(a_0,c) = \widehat \E(M \,|\, A=a_0, C=c, \widetilde M=k)$. The estimators $\widetilde\psi_h(\widehat Q)$ and $\widetilde\gamma_h(\widehat Q)$ are obtained by replacing $\theta(\widehat{Q})$ with $\widetilde\theta_h(\widehat{Q})$ in \eqref{eq:causal_targets_plugin}.

We decompose $\widetilde\theta_{h}(\widehat{Q})(c) {-} \theta(Q)(c)$ as 
{ 
\begin{align}
    &\underbrace{ \big\{ \widetilde\theta_{h}(\widehat{Q})(c) -  \widetilde\theta_{h}(Q)(c) \big\} }_{\substack{\text{estimation error}}}
    {+}
    \underbrace{ \big\{ \widetilde\theta_{h}(Q)(c) {-} \theta(Q)(c) \big\} }_{\text{coarsening error}} \, .
\end{align}
}%

Assume that for fixed $h$,
\begin{align*}
  \|\widetilde\theta_h(\widehat Q)-\widetilde\theta_h(Q)\|_{L_2(P_C)} = O_p(n^{-1/2}).  
\end{align*}%
Then, by Lemma~\ref{lem:first-order-removal},
\begin{align*}
\|\widetilde\theta_h(\widehat Q)-\theta(Q)\|_{L_2(P_C)} = O_p(n^{-1/2}) + O(w_{\max,K}^2).
\end{align*}%

For fixed $K$, the asymptotic bias is driven by the coarsening error $\widetilde{\Delta}_h(Q)$, which is second order in the bin width. 

Let $K=K_n$ increase with $n$. If $M$ has bounded support and equal-width bins are used, then $w_{\max,K_n}=O(1/K_n)$, and hence $\widetilde\Delta_h(Q)(c)=O(1/K_n^2)$. If $K_n\to\infty$ and $w_{\max,K_n}^2=o(n^{-1/2})$, 
% (e.g., $K_n=o(n^{1/4})$), 
then the coarsening error is negligible relative to sampling error.
% and $\widetilde\theta_h(\widehat Q)$ is asymptotically equivalent to $\theta(Q)$ in $L_2(P_C)$. 
Under equal-width binning, this requires $K_n/n^{1/4}\to\infty$.

Compared to the naive coarsened functional $\theta_h(Q)(c)$, which incurs first-order coarsening error of order $O(w_{\max,K})$, the debiased coarsened functional $\widetilde\theta_h(Q)(c)$ achieves a second-order approximation of order $O(w_{\max,K}^2)$ under the stated smoothness conditions. Under equal-width binning, this corresponds to $O(1/K)$ versus $O(1/K^2)$.

%%%%%%%%%%%%%%%%%%%%%%%%%%%%%%%%%%%%%%%%%%%%%%%%%%%%%%%%%%%%
\section{Smoothed debiased coarsened estimands}
\label{sec:smoothed}

Sections~\ref{sec:coarsened} and~\ref{sec:debiased-bin} analyzed the approximation error induced by coarsening. We now turn to statistical estimation.

Throughout the previous sections we assumed that, for fixed $h$, the estimation error of the plug-in estimator is $O_p(n^{-1/2})$ in $L_2(P_C)$. Such behavior typically requires sufficiently fast convergence of all nuisance estimators. When flexible machine learning methods are used to estimate $\mu$, $\pi$, and $g_k$, plug-in estimators can exhibit non-negligible first-order estimation bias.

A von Mises expansion shows that for a smooth target parameter $\psi(Q)$, $\psi(\widehat Q) - \psi(Q) = - P\{\phi(\widehat Q)\} + R_2(\widehat Q,Q)$, where $\phi(Q)$ is the canonical gradient, efficient influence function (EIF), and $R_2$ is a second-order remainder \citep{van2000asymptotic}. The leading term $-P\{\phi(\widehat Q)\}$ represents first-order estimation bias and depends directly on the quality of nuisance estimation. Achieving asymptotic linearity therefore requires sufficiently fast nuisance convergence.

Influence-function–based estimators, such as one-step and targeted minimum loss estimators, remove the first-order bias term and attain asymptotic linearity under weaker conditions. However, to construct such estimators, the target parameter must be pathwise differentiable under the statistical model. The debiased coarsened functional $\widetilde\theta_h(Q)(c)$ fails this requirement when $M$ is continuous, because it involves point evaluation of $\mu(m,a,c)$ at $m_k(a_0,c)$. Point evaluation is not a continuous linear functional on $L_2(P)$ unless $M$ is discrete, and therefore $\widetilde\psi_h$ and $\widetilde\gamma_h$ are not pathwise differentiable under the statistical model.
 
\subsection{Smoothed estimands and plug-ins} 

To restore pathwise differentiability and enable influence-function–based estimation, we introduce a smoothed version of the debiased coarsened functional. We regularize the point evaluation by replacing it with a localized average. Let $\mathcal K$ be a symmetric kernel and let $b>0$ be a bandwidth. Define $\mathcal K_b(u)=b^{-1}\mathcal K(u/b)$ and the normalized weight
{ 
\begin{align*}
\omega_{b,k}(m\,|\, a_1,c)
=
\frac{\mathcal K_b(m-m_k(a_0,c))}
{\E(\mathcal K_b(M-m_k(a_0,c)) \,|\, A=a_1,C=c)}. 
\end{align*}
}

Define the localized mean
\begin{align*}
    \mu_{b,k}(a_1,c) \!=\!
    \E( \mu(M,a_1,c) \, \omega_{b,k}(M | a_1,c)
    \,|\, A \!=\!a_1, C\!=\!c), 
\end{align*}
and the smoothed debiased coarsened estimand
\begin{align}
    \widetilde\theta_{h, b}(Q)(c)
    =
    \sum_{k=1}^K
    \mu_{b,k}(a_1,c)\,
    g_k(a_0,c).
\end{align}%
As $b \to 0$, the kernel $\mathcal K_b$ concentrates at $m_k(a_0,c)$ and $\widetilde\theta_{h, b}(Q)(c) \to \widetilde\theta_h(Q)(c)$. 

The smoothed debiased coarsened parameters $\widetilde\psi_{h, b}(Q)$ and $\widetilde\gamma_{h, b}(Q)$ are obtained from \eqref{eq:causal_targets} by replacing $\theta$ with $\widetilde\theta_{h, b}$.

\begin{lemma}
\label{lem:loc-bias}
For each $c \in \mathcal{C}$, the smoothing error $\widetilde\Delta_{h,b}^s(Q)(c)=\widetilde\theta_{h,b}(Q)(c)-\widetilde\theta_h(Q)(c)$ equals 
{
\begin{align}
   \widetilde\Delta_{h,b}^{s}(Q)(c) 
   {=} \sum_{k=1}^K \big\{\mu_{b,k}(a_1,c) {-} \mu(m_k(a_0,c),a_1,c)\big\} \, g_k(a_0,c). 
\end{align}
}%
Suppose:
(i) $m\mapsto\mu(m,a_1,c)$ is twice continuously differentiable near $m_k(a_0,c)$ with $\sup_{m,c}|\mu''_m(m,a_1,c)|<\infty$;
(ii) $\mathcal K$ is symmetric with $\int u \,\mathcal K(u)\,du=0$ and $\int u^2\mathcal K(u)\,du<\infty$; and 
(iii) $f_{M\,|\,A,C}(\cdot\,|\,a_1,c)$ is continuous and bounded away from zero near $m_k(a_0,c)$.
Then, as $b \to 0$, $\widetilde\Delta_{h,b}^s(Q)(c)=O(b^2)$. Combining with Lemma~\ref{lem:first-order-removal}, the total approximation error $\widetilde\Delta_{h,b}(Q)(c) = \widetilde\theta_{h,b}(Q)(c) - \theta(Q)(c)$ satisfies $\widetilde\Delta_{h,b}(Q)(c) = O\big(w_{\max,K}^2 + b^2\big)$. 
\end{lemma}
See a proof in Appendix~\ref{app:proofs_loc-bias}.   

The plug-in estimator for $\widetilde\theta_{h, b}(Q)(c)$ is 
\begin{align}
\widetilde\theta_{h,b}(\widehat{Q})(c)
&= \sum_{k=1}^K 
\widehat\mu_{b, k}(a_1, c)\,
\widehat{g}_k(a_0, c)\, , 
\label{eq:plug-in-estimator_hb}
\end{align}
where $\widehat\mu_{b, k}(a_1, c) = \widehat \E( \widehat \mu(M, a_1, c) \, \widehat \omega_{b, k}(M \,|\, a_1, c) \,|\, a_1, c)$. 

We can decompose $\widetilde\theta_{h,b}(\widehat{Q})(c) {-} \theta(Q)(c)$ as 

{ 
\begin{align}
&\underbrace{\big\{\widetilde\theta_{h,b}(\widehat{Q})(c) {-} \widetilde\theta_{h,b}(Q)(c)\big\}}_{\substack{\text{estimation error}}}
{+}
\underbrace{\big\{\widetilde\theta_{h,b}(Q)(c) {-} \theta(Q)(c)\big\}}_{\text{total approximation error}} \, .
\end{align}
}%

Assume that for fixed $h$ and $b$,
\begin{align*}
\|\widetilde\theta_{h,b}(\widehat Q)-\widetilde\theta_{h,b}(Q)\|_{L_2(P_C)} = O_p(n^{-1/2}).
\end{align*}
Then, by Lemma~\ref{lem:loc-bias},
\begin{align*}
\|\widetilde\theta_{h,b}(\widehat Q)-\theta(Q)\|_{L_2(P_C)} = O_p(n^{-1/2}) + O(w_{\max,K}^2+b^2).
\end{align*}%
Thus, for fixed $K$ and $b$, the asymptotic bias is driven by $\widetilde\Delta_{h,b}(Q)(c)$, which is second order in the bin width and quadratic in the smoothing bandwidth. 

Let $K=K_n$ and $b=b_n$ depend on $n$. Under equal-width binning, $w_{\max,K_n}=O(1/K_n)$, so
\begin{align*}
\widetilde\Delta_{h,b_n}(Q)(c) 
= \widetilde\theta_{h,b_n}(Q)(c)-\theta(Q)(c)
= O(1/K_n^2+b_n^2).
\end{align*}%
If $K_n\to\infty$, $b_n\to0$, and $1/K_n^2 + b_n^2 = o(n^{-1/2})$, for example $K_n=o(n^{1/4})$ and $b_n=o(n^{1/4})$, then $\widetilde\theta_{h,b_n}(\widehat Q)$ is asymptotically equivalent to $\theta(Q)$ in $L_2(P_C)$.

\subsection{Influence function based estimation}  
\label{subsec:EIF} 

As an example, consider the smoothed mediation functional 
\begin{align*}
    \widetilde\psi_{h, b}(Q) &= \int \widetilde\theta_{h,b}(Q)(c) \, p(c) \, dc \ . 
\end{align*}%
To present the influence function in a transparent form, we first treat the bin centers $m_k(a_0,c)$ as fixed quantities (e.g., midpoints or pre-specified representatives). 

\begin{theorem}\label{thm:eif}
The nonparametric efficient influence function for $\widetilde\psi_{h, b}(Q)$ when $m_k(a_0, k)$ is fixed, is
{
\begin{align*}
\widetilde\phi_{h, b}^\text{fixed}(Q)(O) 
&=
\sum_{k=1}^K 
\frac{\mathbb I(A=a_1)}{\pi(a_1\,|\, C)}\,
g_k(a_0,C)\,
\big\{ Y \omega_{b,k}(M\,|\, a_1,C)
 - \mu_{b,k}(a_1,C)\big\} \\[0.3em]
&\quad
+
\sum_{k=1}^K 
\frac{\mathbb I(A=a_0)}{\pi(a_0\,|\, C)}\,
\mu_{b,k}(a_1,C)\,
\big\{\mathbb I(\widetilde M=k) - g_k(a_0,C)\big\} \\[0.3em]
&\quad
+
\big\{\widetilde\theta_{h,b}(Q)(C) - \widetilde\psi_{h, b}(Q) \big\} \, .
\end{align*}
}
\end{theorem}
See a proof in Appendix~\ref{app:proofs:eif}. 

Let $\widehat{Q}$ collect all nuisance estimates. The one-step estimator of $\widetilde \psi_{h, b}(Q)$, when $m_k(a_0, k)$ is pre-specified, is 
\begin{align*}
\widetilde\psi_{h, b}^{\text{one-step}}(\widehat{Q})
=\widetilde\psi_{h, b}^\text{fixed}(\widehat{Q}) + 
\frac{1}{n}\sum_{i=1}^n 
\widetilde\phi_{h, b}^\text{fixed}(\widehat{Q})(O_i) \, , 
\end{align*}
where $\widetilde\phi_{h, b}^\text{fixed}(\widehat{Q})(O_i)$ is the evaluation of the influence function at $\widehat{Q}$ and $O_i$, and 
\begin{align*}
\widetilde\psi_{h, b}^\text{fixed}(\widehat{Q})
&{=} \frac{1}{n} \sum_{i = 1}^n \sum_{k=1}^{K} \widehat\mu_{b,k}^\text{fixed}(a_1,C_i)\,\widehat g_k(a_0,C_i) \, , \\
\widehat\mu_{b,k}^\text{fixed}(a_1,c)
& {=} \widehat{\E}\big[\widehat\mu(M,a_1,c)\,\omega_{b,k}(M\,|\, a_1,c)\,|\, A{=}a_1,C{=}c\big] \, . 
\end{align*}

Under sufficient regularity conditions, $n^{-1/4}$ rates of the nuisance estimators, and cross-fitting, the one-step estimator is asymptotically linear with asymptotic variance of $\E(\widetilde\phi_{h, b}^\text{fixed}(Q)(O)^2)$. 

When $m_k(a_0,c)$ is estimated, an additional influence function contribution arises, denoted by $\widetilde \phi^\omega_{h, b}(Q)(O)$: 

{ 
\begin{align*}
&\widetilde \phi^\omega_{h, b}(Q)(O) 
\!=\!\!
\sum_{k=1}^K \!
\alpha_k(C)\,
\frac{\mathbb I(A=a_0, \widetilde M=k)}{\pi(a_0\,|\, C)}
\big\{M \!-\! m_k(a_0,C)\big\} \, ,  \notag 
% \label{eq:EIF_extra_mk_simplified}
\end{align*}
}%

where $\alpha_k(c) \coloneqq \frac{\partial}{\partial m}\mu_{b,k}(a_1,c)$. The full EIF for $\widetilde\psi_{h, b}(Q)$ is therefore $\widetilde \phi_{h, b}(Q)(O) = \widetilde \phi_{h, b}^\text{fixed}(Q)(O) + \widetilde \phi^\omega_{h, b}(Q)(O)$. See a proof in Appendix~\ref{app:proofs:eif}.

%%%%%%%%%%%%%%%%%%%%%%%%%%%%%%%%%%%%%%%%%%%%%%%%%%%%%%%%%%%%
\section{Simulation studies}
\label{sec:sim}

The R code is available on
\href{https://github.com/xxou/Debias_Mediator_Discretization}{GitHub}. 

We study estimation of the mediation functional $\psi(Q)$ in  \eqref{eq:causal_targets}, with $a_1 = 1$ and $a_0 = 0$. Our goals are threefold: (i) quantify the population-level coarsening error of the discretized estimand $\theta_h(Q)(c)$ and verify the first-order scaling in Lemma~\ref{lem:coarsening_error}; (ii) evaluate the finite-sample performance of the plug-in and the one-step estimators for the coarsened estimand $\psi_h(Q)$, and the debiased coarsened estimand $\widetilde{\psi}_h(Q)$, under equal-frequency discretization; (iii) examine the sensitivity of the plug-in estimators to different discretization schemes including equal-width and predefined-width discretizations. 

All simulations are conducted under the following data-generating process (DGP). The baseline covariate $C$ takes values in $\{-2, -1, 0, 1, 2\}$ with probability mass $(0.15, 0.20, 0.18, 0.30, 0.17)$ respectively;
\begin{align*} 
A \,|\, C  & \sim \mathrm{Bernoulli}\big(\mathrm{expit}(0.5C)\big),\\ 
M \,|\, A,C  & \sim N\big(-0.6C {+} 2A {+} 0.5AC,\,1\big), \\
Y \,|\, M,A,C & \sim N\big( \mu(M, A, C), \, 1\big),
\end{align*}

\noindent 
where $\mu(M, A, C) = 0.8C + 1.5A + 0.75AM + 0.20 MC^2 + 0.1 M^3 +  0.55 AC$, nonlinear in $M$, ensuring nonzero curvature and hence non-negligible second-order terms in the Taylor expansions underlying Lemmas~\ref{lem:coarsening_error} and~\ref{lem:first-order-removal}. 

\vspace{0.4cm}
\noindent\textbf{Simulation \#1: Population-level coarsening error.}  

We first isolate the coarsening error characterized in Lemma~\ref{lem:coarsening_error}, $\Delta_{h}(Q)(c){=}\theta_h(Q)(c){-}\theta(Q)(c)$. To approximate the theoretical quantities appearing in $\Delta_h(Q)(c)$, we generate a Monte Carlo sample of size $n{=}10^6$. No nuisance estimation is involved in this experiment. Instead, we exploit the known parametric forms to compute: $g_k(a, c), m_k(a, c), \mu_k(a, c)$. We consider equal-frequency discretizations of $M$ with $K{=2}$ and $K{=}6$ bins. For $K{=}2$, the bins are $\mathcal{B}_1{=}\{M \le 1.21\}$ and $\mathcal{B}_2{=}\{M > 1.21\}$. For $K{=}6$, we define:  $\mathcal{B}_1{=}\{M \le -0.38\}$, $\mathcal{B}_2{=}(-0.38, 0.53]$, $\mathcal{B}_3{=}(0.53,1.21]$, $\mathcal{B}_4{=}(1.21,1.81]$, $\mathcal{B}_5{=}(1.81, 2.49]$, and $\mathcal{B}_6{=}\{M > 2.49\}$.

Table~\ref{tab:Delta_h_P_c} reports $\Delta_h(Q)(c)$ across levels of $C$. For all levels, the coarsening error decreases substantially as $K$ increases. However, even at $K=6$, the error remains non-negligible, illustrating that discretization can alter the target parameter even with a relatively fine partition. 

Equation~\eqref{eq:first_order_scaling} shows that the leading term of $\Delta_h(Q)(c)$ is proportional to $ \mu'_m(m_k(a_1,c),a_1,c)\{m_k(a_1,c) \!-\! m_k(a_0,c)\}$. Figure~\ref{fig:within-bin-means} plots the within-bin shifts $m_k(a_1,c) \!-\! m_k(a_0,c)$ for discretizations with $K=2$ and $K=6$. Increasing $K$ reduces these shifts across bins and covariate strata, thereby attenuating the dominant first-order component of the coarsening error. The empirical behavior aligns closely with the Taylor expansion derived in Lemma~\ref{lem:coarsening_error}. 

\begin{table}[!t]
\caption{Coarsening error $\Delta_h(Q)(c)$ for equal-frequency discretizations with $K=2$ and $K=6$.}
\label{tab:Delta_h_P_c}
\begin{tabular*}{\linewidth}{@{\extracolsep{\fill}}l|rrrrr}
\toprule
 & C = -2 & C = -1 & C = 0 & C = 1 & C = 2 \\ 
\midrule\addlinespace[2.5pt]
K = 2 & 1.103 & 0.871 & 0.876 & 1.522 & 3.390 \\ 
K = 6 & 0.195 & 0.149 & 0.216 & 0.550 & 1.562 \\ 
\bottomrule
\end{tabular*}
\end{table}

\begin{figure}[!t]
\centering
\includegraphics[width=0.7\linewidth]{pics/compare_mk_K2_K6_freq.png}
\vspace{-0.2cm}
% \caption{ Within-bin differences in conditional mediator means, $m_k(a_1,c){-}m_k(a_0,c)$, under two separate discretizations with $K{=}2$ (red) and $K{=}6$ (blue). Each horizontal segment represents one mediator bin, with bin indices shown on the upper axis for $K{=}2$ and on the lower axis for $K{=}6$. Panels correspond to different values of the covariate $C$.} 
 \caption{Within-bin differences $m_k(a_1,c)-m_k(a_0,c)$ under $K=2$ (red, upper axis) and $K=6$ (blue, lower axis) discretizations. Each horizontal segment represents one mediator bin; panels correspond to values of $C$.} 
\label{fig:within-bin-means}
\end{figure}

\begin{figure}[t]
\centering
\includegraphics[width=.8\linewidth]{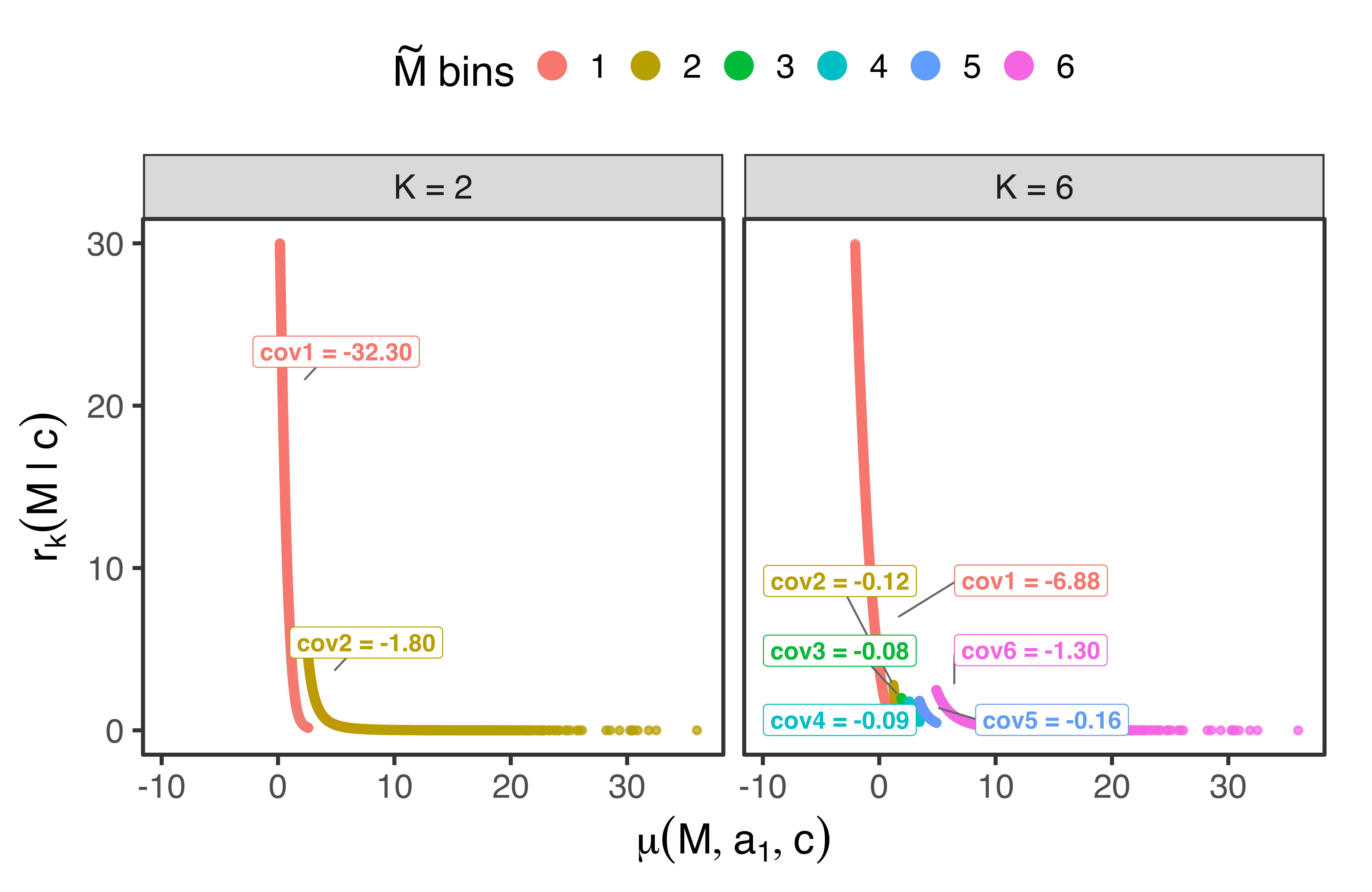}
\caption{Within-bin covariance between $\mu(M,a_1,C)$ and $r_k(M \,|\, C)$ at $C=0$, as formalized in Remark~\ref{remark:covariance}. Finer discretization reduces the magnitude of this covariance.}
\label{fig:covariance_C0}
\end{figure}

\begin{figure}[t]
\centering
\includegraphics[width=0.9\linewidth]{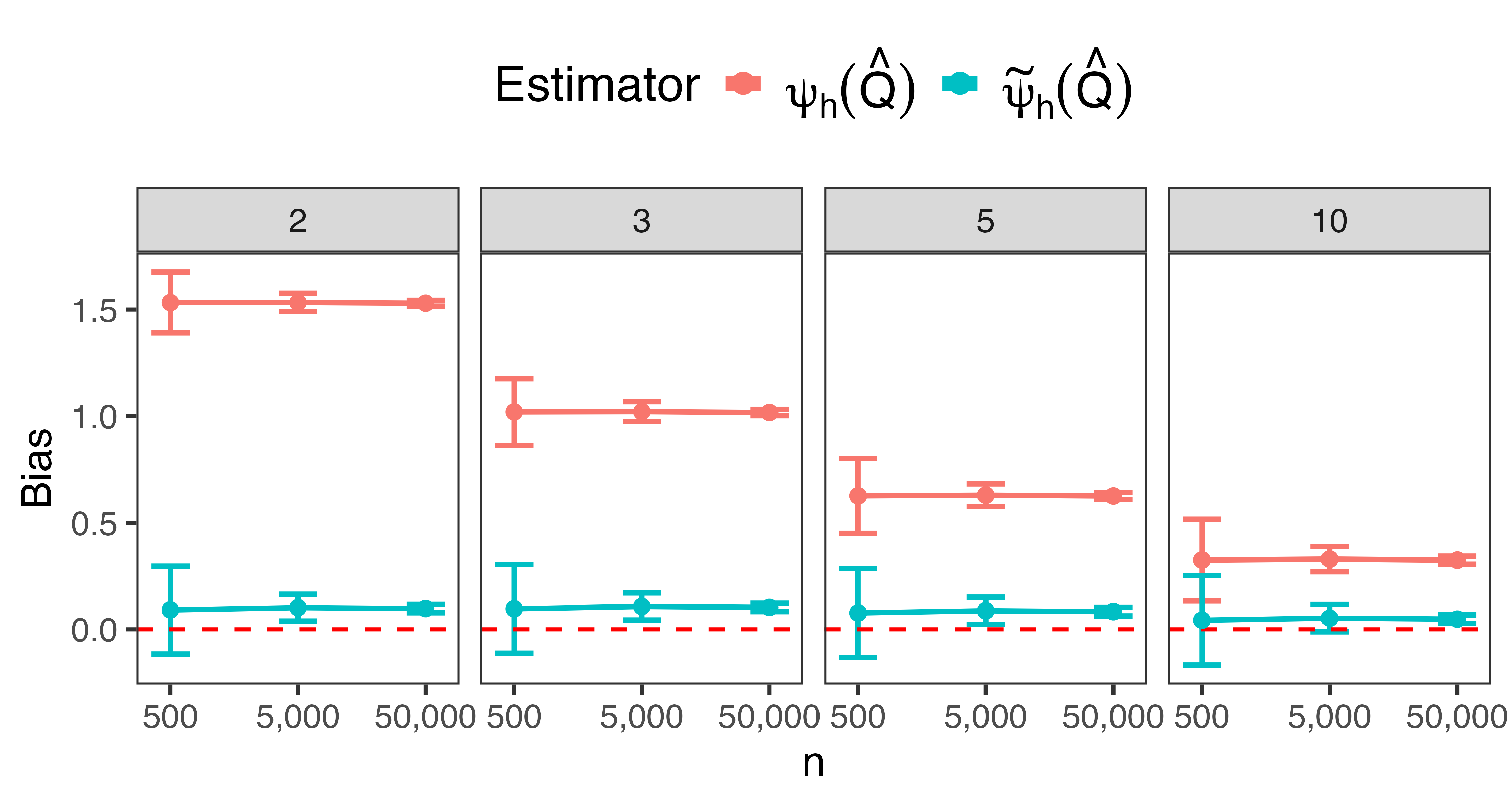}
\caption{Bias of the plug-in estimates of the coarsened estimand,  $\psi_h(\widehat{Q})$, and the debiased coarsened estimand, $\widetilde{\psi}_h(\widehat{Q})$, as functions of sample size for several discretization levels $K$. Points denote the Monte Carlo mean error (bias) across 1,000 simulation replicates. Error bars show the Monte Carlo standard deviation of the estimation error.  }
\label{fig:trends_n_freq}
\end{figure}

Remark~\ref{remark:covariance} expresses the coarsening error as a within-bin covariance between $\mu(M,a_1,c)$ and the ratio $r_k(M\,|\, c)$. Figure~\ref{fig:covariance_C0} illustrates this at $C=0$. With finer discretization, the covariance magnitude drops substantially, giving a complementary view of the reduced coarsening error.

Together, these results empirically validate the $O(1/K)$ first-order behavior stated in Lemma~\ref{lem:coarsening_error}.

\vspace{0.4cm}
\noindent\textbf{Simulation \#2: Finite-sample performance.}

We now study statistical estimation via simulations with $n \in \{500, 5{,}000, 50{,}000\}$, using 1,000 Monte Carlo replications for each $n$. We consider equal-frequency discretizations of the mediator over a broad range of bin counts $K$.

For each discretization level $K$, we compare the  plug-in and one-step estimates of the coarsened estimand $\psi_h(Q)$ and the debiased coarsened estimand $\widetilde{\psi}_h(Q)$. All nuisance functions are estimated using correctly specified generalized linear models. The within-bin means $m_k(a_0,c)$ and bin probabilities $g_k(a,c)$ are computed under the truncated normal distribution implied by the DGP.

Figure~\ref{fig:trends_n_freq} reports the results. 
Two distinct phenomena are evident: (i) increasing $n$ reduces Monte Carlo variability, and (ii) for fixed $K$, the estimation bias of $\psi_h(\widehat Q)$ does not vanish as $n$ increases. The persistence of bias reflects population coarsening error, not estimation error. In contrast, $\widetilde{\psi}_h(\widehat Q)$ exhibits dramatically reduced bias across all sample sizes, consistent with the second-order scaling $\widetilde \Delta_h(Q)(c) = O(w_{\max,K}^2)$ from Lemma~\ref{lem:first-order-removal}.  

Because bias patterns are qualitatively similar across sample sizes, we focus on $n=5{,}000$ and report analogous results for $n=500$ in Appendix~\ref{app:simu_small_n}. Figure~\ref{fig:plugin} summarizes absolute bias, variance, mean squared error (MSE), and coverage of 95\% confidence intervals (CIs) as functions of $K$.

The coarsened plug-in estimator, $\psi_h(\widehat Q)$, exhibits substantial bias and poor coverage for small $K$, improving gradually as $K$ increases, reflecting the $O(w_{\max,K})$ decay of the coarsening error. In contrast, the debiased coarsened estimator, $\widetilde{\psi}_h(\widehat Q)$, achieves near-zero bias even for relatively small $K$, smaller MSE, and coverage close to the nominal level across a broad range of discretization levels. Variance is only modestly affected by the debiasing step. These findings demonstrate that evaluating $\mu$ at the within-bin mean $m_k(a_0,c)$ effectively removes the leading coarsening error without inducing meaningful variance inflation.

\begin{figure}[t]
\centering
\includegraphics[width=.8\linewidth]{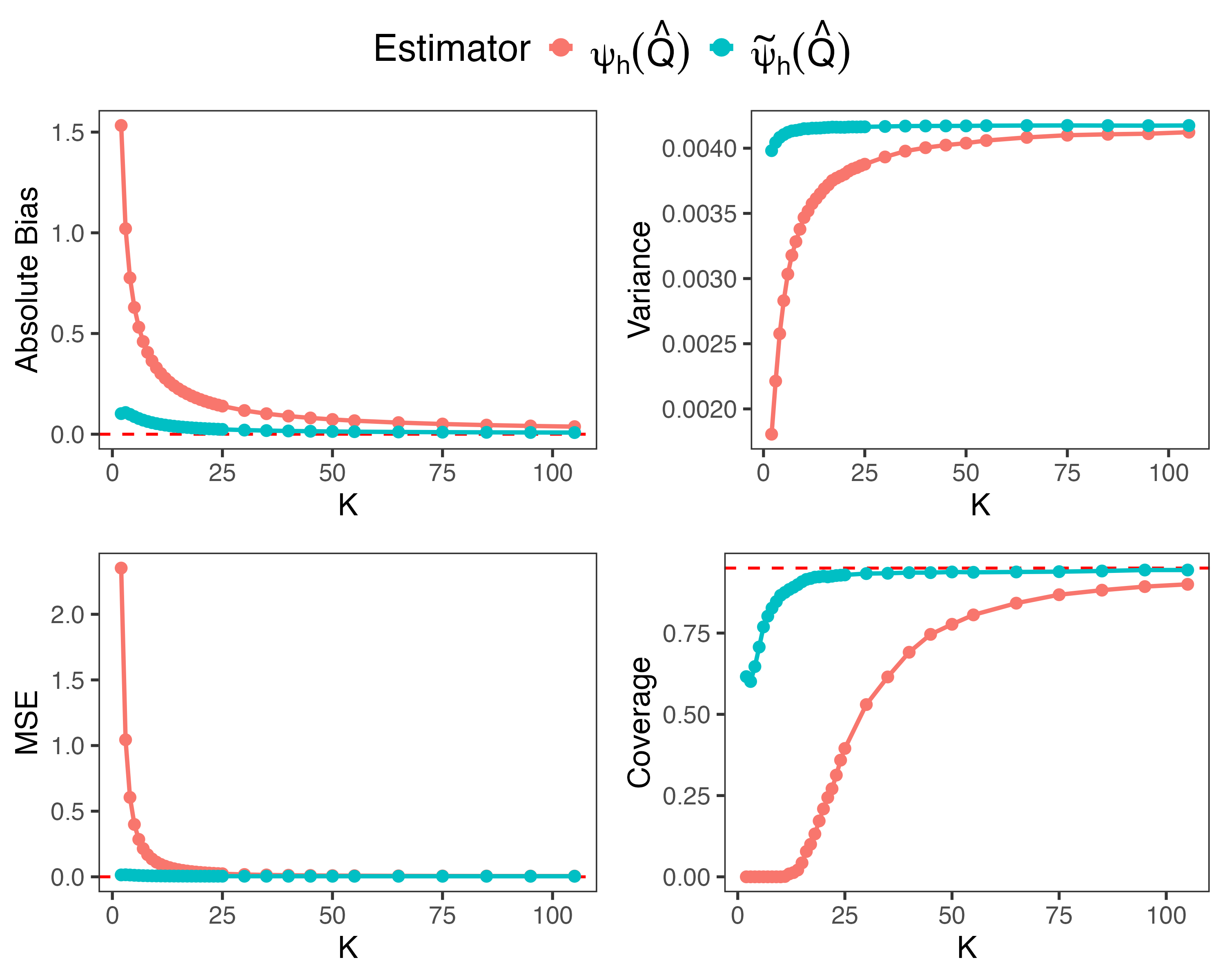}
\caption{The performance of the coarsened plug-in estimator $\psi_h(\widehat{Q})$ and the debiased estimator $\widetilde{\psi}_h(\widehat{Q})$ as functions of the number of mediator bins $K$ for sample size $n=5{,}000$.}
\label{fig:plugin}
\end{figure}

\begin{figure}[h]
\centering
\includegraphics[width=1\textwidth]{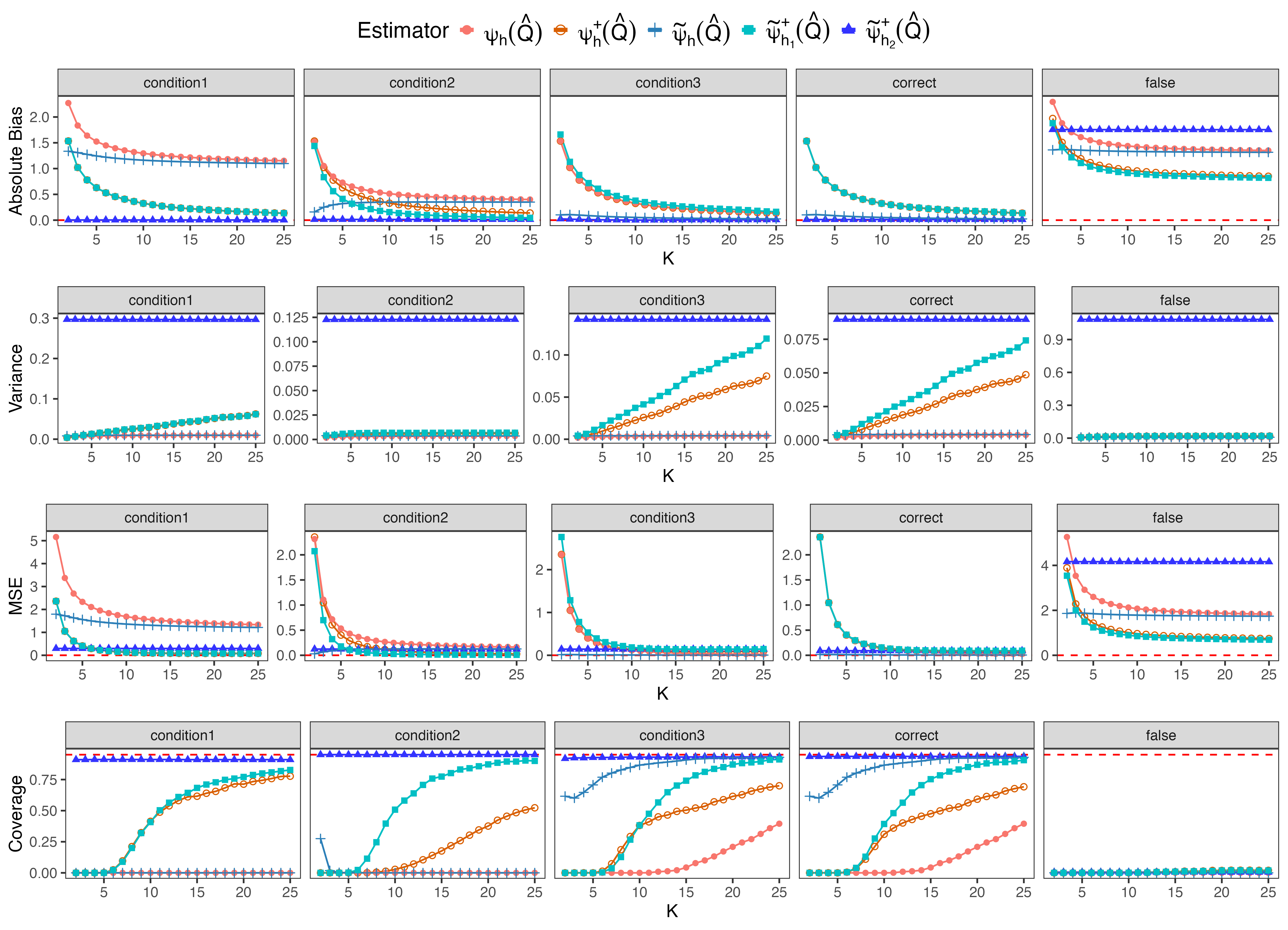}
\caption{Comparison of plug-in and one-step estimators for $n=5{,}000$ under nuisance-model misspecification. ``Correct'' and ``False'' indicate correctly specified and misspecified nuisance models, respectively. Conditions 1–3 correspond to misspecification of $\{\mu, \mu_k\}$, $\{g, g_k\}$, and $\pi$, respectively. Performance is summarized in terms of absolute bias, variance, MSE, and 95\% confidence interval coverage.}
\label{fig:all}
\end{figure}

One-step estimators are often used to reduce estimation bias by incorporating influence-function corrections. Define the generalized propensity score $g(a|m, c) = p(A = a| M = m, C = c)$. The one-step estimators for $\psi(Q)$ and $\psi_h(Q)$, denoted $\psi^+(\widehat Q)$ and $\psi_h^{+}(\widehat Q)$ are  
{
\begin{align}
\psi^{+}(\widehat Q) =& \frac{1}{n} \sum_{i=1}^n  \bigg[
\frac{\mathbb{I}(A_i = a_1)}{\widehat{\pi}(a_0 \,|\, C_i)}
\frac{\widehat{g}(a_0|M_i, C_i)}{\widehat{g}(a_1|M_i, C_i)}
\big( Y {-}\widehat{\mu}(M_i, a_1, C_i) \big) \label{eq:one_step_theta} \\
& \hspace{-0.1cm} + \frac{\mathbb{I}(A_i = a_0)}{\hat{\pi}(a_0 \,|\, C_i)}
\Big(\widehat{\mu}(M_i, a_1, C_i) {-} \theta(\widehat{Q})(C_i) \Big) {+} \theta(\widehat{Q})(C_i)
\bigg], \notag 
\end{align}
\begin{align}
\psi_h^+(\widehat{Q})  =& \frac{1}{n} \sum_{i=1}^n  \bigg[
\frac{\mathbb{I}(A_i = a_1)}{\widehat{\pi}(a_1 \,|\, C_i)}
\frac{\widehat{g}_k(a_0, C_i)}{\widehat{g}_k(a_1, C_i)}
\Big( Y -\widehat{\mu}_k(a_1, C_i) \Big) \label{eq:one_step_theta_h} \\
& \hspace{-0.1cm} + \frac{\mathbb{I}(A_i = a_0)}{\widehat{\pi}(a_0 \,|\, C_i)}
\Big(\widehat{\mu}_k(a_1, C_i) -\theta_h(\widehat{Q})(C_i) \Big) + \theta_h(\widehat{Q})(C_i)
\bigg]. \notag 
\end{align}
}
The estimator in \eqref{eq:one_step_theta_h} is consistent for $\psi_h(Q)$ if either (i) $\widehat{\pi}$ and $\widehat{g}_k$, (ii) $\widehat{\pi}$ and $\widehat{\mu}_k$, or (iii) $\widehat{g}_k$ and $\widehat{\mu}_k$ are consistent. 

Note that $\psi_h^{+}(\widehat Q)$ still targets the coarsened functional $\psi_h(Q)$ rather than the original parameter $\psi(Q)$. To remedy this limitation, we compare: 
(i) $\psi_h^{+}(\widehat Q)$, the one-step estimator for $\psi_h(Q)$,
(ii) $\widetilde \psi^{+}_{h_1}(\widehat Q)$, a naive derivative-based correction that replaces $\widehat{\mu}_k$ by $\widehat{\mu}(m_k(a_0,c),a_1,c)$ and $\theta_h(\widehat{Q})(c)$ with $\widetilde{\theta}_h(\widehat{Q})(c)$ in  \eqref{eq:one_step_theta_h}, 
(iii) $\widetilde \psi^{+}_{h_2}(\widehat Q)$, the one-step estimator that replaces $\theta(\widehat{Q})(c)$ with $\widetilde{\theta}_h(\widehat{Q})(c)$  in \eqref{eq:one_step_theta}, 
(iv) the coarsened plug-in estimator $\psi_h(\widehat Q)$, (v) the debiased plug-in estimator $\widetilde{\psi}_h(\widehat Q)$. 

Figure~\ref{fig:all} presents results under correct specification and a range of misspecification for nuisance models. Details  of the misspecification configurations are given in Appendix~\ref{app:simu_misspecification_cases}.

The one-step estimator $\psi_h^{+}(\widehat Q)$ yields modest improvements in bias and coverage relative to $\psi_h(\widehat Q)$, but does not eliminate discretization bias when $K$ is small, since the target remains $\psi_h(Q)$. The first derivative estimator $\widetilde \psi^{+}_{h_1}(\widehat Q)$ behaves similarly to $\psi_h^{+}(\widehat Q)$, with improved coverage, but does not meaningfully reduce discretization bias in the way that $\widetilde \psi_h(\widehat Q)$ does. In contrast, $\widetilde \psi^{+}_{h_2}(\widehat{Q})$ achieves uniformly low bias across different $K$ and near-nominal coverage across scenarios except when all nuisance functions are misspecified. This strong performance arises because $\widetilde \psi^{+}_{h_2}(\widehat Q)$ combines two bias-reduction mechanisms: its one-step correction targets the original parameter $\psi(Q)$, reducing first-order estimation bias due to nuisance misspecification, and the debiased coarsened functional $\widetilde\theta(Q)(c)$ removes the leading discretization bias. Notably, although the debiased plug-in estimator can still be affected by nuisance-model misspecification, under correct specification $\widetilde{\psi}_h(\widehat Q)$ performs closest to $\widetilde \psi^{+}_{h_2}(\widehat Q)$ among the other four estimators, with only minor differences in bias and coverage.

These findings demonstrate that the coarsening error arises at the level of the target functional and therefore cannot be eliminated by influence-function correction alone. Correcting the functional through the proposed debiased coarsened construction is essential for valid inference. The debiased plug-in estimator achieves substantial bias reduction while retaining computational simplicity and strong empirical performance when nuisance models are well estimated, which can be facilitated by flexible machine learning methods.

\vspace{0.4cm}
\noindent\textbf{Simulation \#3: Comparisons of discretization schemes.}
% \label{subsec:sim3}

Using the same DGP, we repeated the simulation procedure with 1,000 Monte Carlo replications for $n \in \{500, 5{,}000, 50{,}000\}$ and considered two alternatives to the equal-frequency discretization used above. First, we used empirical equal-width discretization, where the cut points are computed from the observed range of $M$ within each simulated dataset. Second, we used predefined-width discretization, in which the interval $[-1.5,3.3]$ is partitioned into $K$ subintervals of equal length and the two boundary cut points are set to observed minimum and maximum values. For example, when $K = 3$, the resulting bins are $ \leq 0.1$, $(0.1, 1.7]$, and $>1.7$. Thus, for a fixed $K$, all Monte Carlo replications  use the same cut points.

Figure~\ref{fig:trend_n_discretization} displays bias trends as $n$ increases for specific $K$, and Figure~\ref{fig:trend_K_discretization} summarizes absolute bias, variance, MSE, and coverage as functions of $K$ for $n = 5{,}000$. The behavior of the coarsened estimator $\psi_h(\widehat Q)$ depends meaningfully on the discretization rule. Under empirical equal-width discretization, the bias of $\psi_h(\widehat Q)$ can increase with $n$ for some fixed values of $K$, particularly when $K$ is small, except for $K{=}2$. This occurs because the mediator has unbounded support, and $M$ is unimodal. When the sample size increases, the empirical range tends to expand, leading to wider equal-width bins. In particular, wider central bins can contain a larger proportion of observations and amplify the within-bin differences in conditional mediator means, $m_k(a_1,c)-m_k(a_0,c)$. In contrast, under predefined-width discretization, the interior bin widths are fixed for each $K$. Consequently, the bias of $\psi_h(\widehat Q)$ remains stable as $n$ increases for a fixed $K$, consistent with the behavior observed under equal-frequency discretization. However, as shown in Figure~\ref{fig:trend_K_fixed}, although the bias decreases as $K$ increases, it does not converge to zero. This residual bias arises because the two tail bins continue to contribute approximation error even as the interior discretization becomes finer; in particular, these tail bins always contain the regions $ \leq -1.5$ and $\geq 3.3$. Across both schemes, the proposed debiased estimator $\widetilde{\psi}_h(\widehat Q)$ substantially reduces bias and remains close to the true target even for relatively coarse discretizations.  Appendix~\ref{app:simu_discretization} provides additional details.

These results clarify the practical role of bin construction. Equal-frequency bins stabilize empirical bin masses; empirical equal-width bins are sensitive to sample extremes when $M$ has unbounded support; and predefined-width bins reflect externally fixed categories, such as clinically meaningful thresholds, but they lack the flexibility to adapt to a specific dataset, and coarsening error may remain even when $K$ is large. The debiased estimator is considerably less sensitive to these choices, supporting its use when the discretization is dictated by interpretability, scientific convention, or data-analytic convenience.

%%%%%%%%%%%%%%%%%%%%%%%%%%%%%%%%%%%%%%%%%%%%%%%%%%%%%%%%%%%%
\section{A real data application}
\label{sec:real_data}

We apply our proposed estimation framework to data from the B\_PROUD study, an observational investigation assessing the impact of Mobile Stroke Unit (MSU) dispatch on post-stroke outcomes in Berlin \citep{ebinger2017berlin} conducted between February 2017 and May 2019. It can be used to evaluate the effect of additional MSU care on 3-month functional outcomes among patients for whom an MSU was received.

We analyzed 768 patients eligible for reperfusion therapy, of whom 588 (76.6\%) received MSU care ($A=1$) and 180 (23.4\%) received conventional emergency services ($A=0$). The outcome $Y$ is the 3-month modified Rankin Scale (mRS), an ordinal measure ranging from 0 (no symptoms) to 6 (death). In our study, we treat mRS as continuous. The full mediator $M$ is defined as the time from ambulance dispatch to thrombolysis and is set to zero for patients who did not receive thrombolytic therapy. We adjust for two baseline continuous covariates: systolic blood pressure and stroke severity. Missing data are handled using multiple imputation.

This dataset was previously analyzed  by  \cite{piccininni2023effect} employing a front-door approach to estimate the causal effect of MSU dispatch on 3-month mRS using a three-category mediator.  Subsequently, \cite{guo2023flexible} considered both continuous and binary versions of the mediator, exploring different cutoffs and implementing one-step and TMLE estimators of the front-door functional. In this analysis, we compare three estimators of the front-door functional $\gamma(Q)$ for $\E(Y(a_0))$ in \eqref{eq:causal_targets} with $a_0 = 0$:
(i) the coarsened plug-in estimator $\gamma_h(\widehat{Q})$,
(ii) the  debiased plug-in estimator $\widetilde{\gamma}_h(\widehat{Q})$, and 
(iii)  the plug-in estimator $\gamma_s(\widehat{Q})$ based on sequential regression, which avoids explicit estimation of the mediator density by rewriting $\theta(Q)(c)$ as
\begin{align}
\theta(Q)(c)
= \E(\tilde{Y} \,|\, a_0, c) , \quad \text{where} \quad  \widetilde{Y} = \E(Y \,|\, M, a_1, c) \ . \label{eq:theta_seq}
\end{align}
All nuisance functions were estimated using super learner with candidates including \texttt{glm.interaction}, \texttt{glmnet}, \texttt{random forests} (1000 trees), \texttt{xgboost}, and \texttt{dbarts}.

\begin{figure}[t]
    \centering
    \includegraphics[width=0.6\linewidth]{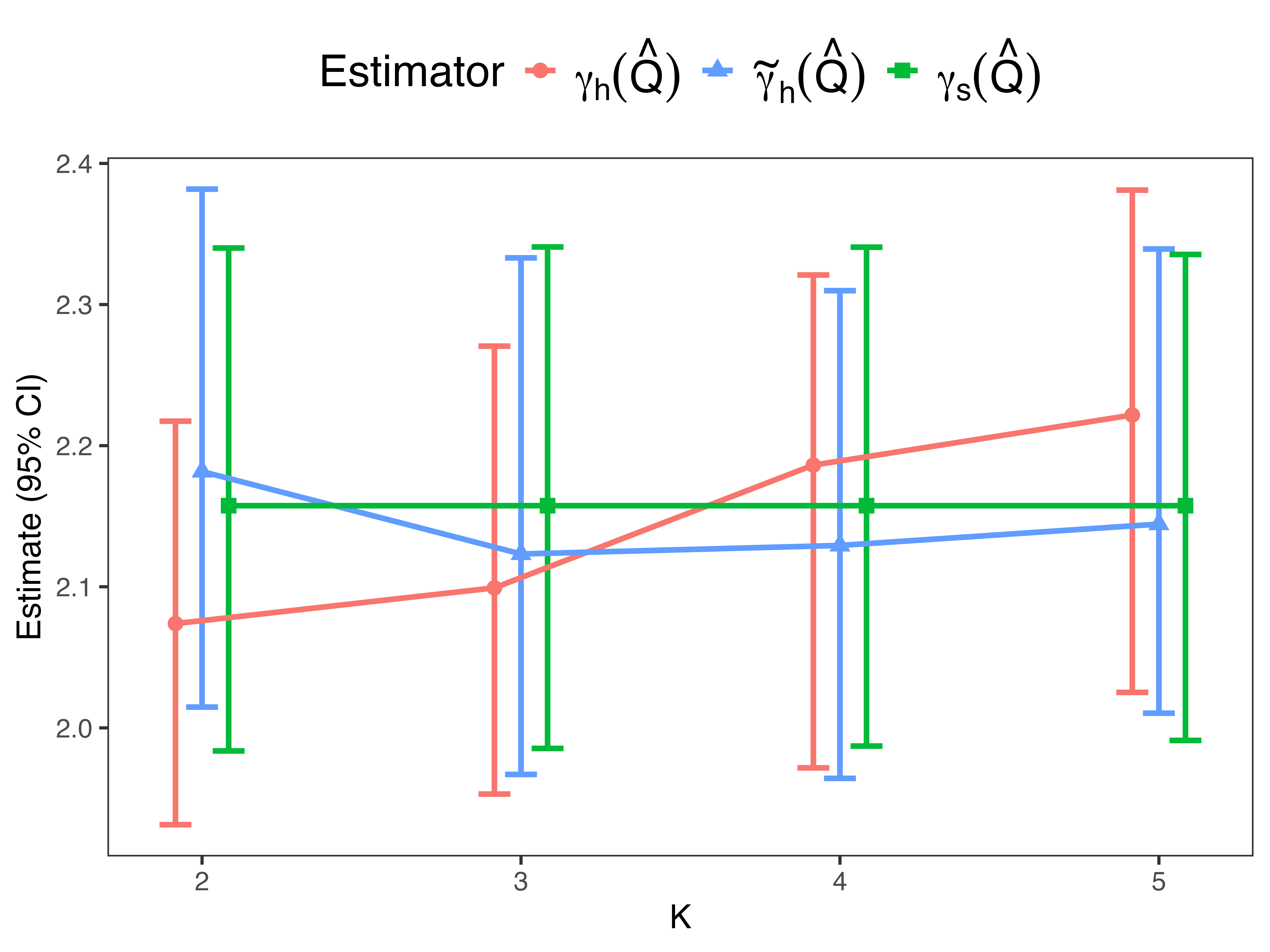}
    \caption{Comparison of plug-in estimators $\gamma_h(\widehat{Q})$,  $\widetilde\gamma_h(\widehat{Q})$  and $\gamma_s(\widehat{Q})$ in B\_PROUD study for estimating $\gamma(Q) = \E [Y(a_0)]$ }
    \label{fig:read_data_est}
\end{figure}

\begin{figure}[!ht]
    \centering
    \includegraphics[width=0.9\linewidth]{pics/fig_B_data_M_shifts.png}
    \caption{Estimated within-bin differences in conditional mediator means, $\widehat{m}_k(a_1, C_i)-\widehat{m}_k(a_0, C_i)$, under discretizations with $K=2$. The red line denotes zero and the blue dashed lines indicate the median of within-bin differences.}
    \label{fig:read_data_M_shits}
\end{figure}

Figure~\ref{fig:read_data_est} presents point estimates along with 95\% CIs obtained via bootstrapping.  In terms of point estimates, the debiased estimator $\widetilde\gamma_h(\widehat{Q})$ and the sequential estimator $\gamma_s(\widehat{Q})$ are closer to each other than either is to the coarsened estimator $\gamma_h(\widehat{Q})$.  When $K=2$, the CIs of the three estimators exhibit limited overlap, and $\gamma_h(\widehat{Q})$ (2.074; 95\% CI: 1.931–2.217) appears somewhat farther from the debiased estimator $\widetilde\gamma_h(\widehat{Q})$ (2.182; 95\% CI: 2.015–2.382) and the sequential estimator $\gamma_s(\widehat{Q})$ (2.157; 95\% CI: 1.984–2.340). For $K = 3, 4,$ and $5$, the  CIs largely overlap. In particular, the estimate of $\widetilde\gamma_h(\widehat{Q})$ with $K = 5$  (2.144; 95\% CI: 2.010–2.340) is nearly identical to the estimate of $\gamma_s(\widehat{Q})$, indicating substantial agreement under finer discretization.

The coarsened estimator $\gamma_h(\widehat{Q})$ does not perform as poorly as might be anticipated. This pattern is interpretable upon examining the within-bin mediator mean shifts across treatment levels, as illustrated in Figure~\ref{fig:read_data_M_shits}. For $K=2$, the estimated shifts in the two bins cluster around opposite signs: one predominantly positive and the other predominantly negative. As indicated in \eqref{eq:coarsening_error}, the coarsening bias depends on both the mediator shifts within bins and the local derivative structure of $\mu(m, a_1, c)$. When mediator shifts differ in sign across bins, their weighted contributions may partially offset. This cancellation effect provides a plausible explanation for the relatively small empirical deviation of the coarsened estimator compared with the other estimators in this dataset.

However, such offsets depend on specific features of the data structure and cannot  always be guaranteed in practice. Therefore, although the  coarsened plug-in estimator may perform well in certain empirical settings, relying on it without bias correction remains potentially risky. In contrast, the proposed debiased estimator offers a more reliable alternative, mitigating first-order coarsening bias without imposing restrictive assumptions on the mediator–outcome relationship.

%%%%%%%%%%%%%%%%%%%%%%%%%%%%%%%%%%%%%%%%%%%%%%%%%%%%%%%%%%%%
\section{Discussion}
\label{sec:discussion}

We study the bias induced by discretizing a continuous mediator in causal functionals. Even under correct identification and consistent nuisance estimation, naive discretization alters the target parameter and introduces a first-order coarsening error that scales with the bin width. Evaluating the outcome regression at within-bin means under the appropriate treatment-reference distribution removes this leading error term and reduces the approximation error to second order under standard smoothness conditions.
A key distinction is between functional approximation (coarsening) error and statistical estimation error. Influence-function–based estimators address the latter but do not correct the bias induced by discretization when the estimand itself has been modified. Correcting the functional is therefore essential; once this is done, standard semiparametric tools can be applied to control estimation error and conduct valid inference.
In practice, discretization is often used to simplify computation or avoid direct conditional density estimation. We show that the resulting coarsening error can be formally characterized and substantially reduced through a simple within-bin correction that preserves computational tractability. Although we focus on mediation and front-door functionals, the same ideas extend to causal parameters involving integration of a regression surface against a conditional distribution under suitable smoothness conditions.

%%%%%%%%%%%%%%%%%%%%%%%%%%%%%%%%%%%%%%%%%%%%%%%%%%%%%%%%%%%

% \vspace{0.75cm}
% \section{Back Matter}
% There are a some final, special sections that come at the back of the paper, in the following order:
% \begin{itemize}
%   \item Author Contributions (optional)
%   \item Acknowledgements (optional)
%   \item References
% \end{itemize}
% They all use an unnumbered \verb|\subsubsection|.

% For the first two special environments are provided.
% (These sections are automatically removed for the anonymous submission version of your paper.)
% The third is the ‘References’ section.
% (See below.)

% (This ‘Back Matter’ section itself should not be included in your paper.)

% \begin{contributions} % will be removed in pdf for initial submission 
% 					  % (without ‘accepted’ option in \documentclass)
%                       % so you can already fill it to test with the
%                       % ‘accepted’ class option
%     Briefly list author contributions. 
%     This is a nice way of making clear who did what and to give proper credit.
%     This section is optional.

%     H.~Q.~Bovik conceived the idea and wrote the paper.
%     Coauthor One created the code.
%     Coauthor Two created the figures.
% \end{contributions}

% \begin{acknowledgements} % will be removed in pdf for initial submission,
% 						 % (without ‘accepted’ option in \documentclass)
%                          % so you can already fill it to test with the
%                          % ‘accepted’ class option
%     Briefly acknowledge people and organizations here.

%     \emph{All} acknowledgements go in this section.
% \end{acknowledgements}
\newpage
\noindent \textbf{Acknowledgments} 

\noindent 
This work is supported by NIH grant R21ES036795. 

%%%%%%%%%%%%%%%%%%%%%%%%%%%%%%%%%%%%%%%%%%%%%%%%%%%%%%%%%%%
% References

\bibliographystyle{unsrt} 
\bibliography{Debias} 

%%%%%%%%%%%%%%%%%%%%%%%%%%%%%%%%%%%%%%%%%%%%%%%%%%%%%%%%%%%

\newpage
\appendix

\setcounter{figure}{0}
\renewcommand{\thefigure}{S\arabic{figure}}

\setcounter{table}{0}
\renewcommand{\thetable}{S\arabic{table}}

\setcounter{equation}{0}
\renewcommand{\theequation}{S\arabic{equation}}

\begin{center}
{\Large \bf Coarsening Bias from Variable Discretization in Causal Functionals (Supplementary Material)}
\end{center}

This Supplementary Material is organized as follows. 
Appendix~\ref{app:glossary} provides a glossary of the main terms and notation.
Appendix~\ref{app:sec_multiple_mediators} presents extensions to settings with multiple mediators. 
Appendix~\ref{app:proofs} contains all technical proofs. 
Appendix~\ref{app:simulations} provides additional details for the simulation studies. 

\section{Glossary of terms and notation}
\label{app:glossary}

Table~\ref{tab:glossary} summarizes the main terms and notation used throughout the paper.

{
\setlength{\tabcolsep}{5pt}
\begin{longtable}{p{0.22\linewidth} p{0.75\linewidth}}
\caption{Glossary of terms and notation.}
\label{tab:glossary}\\
\toprule
\textbf{Term or notation} & \textbf{Definition} \\
\midrule
\endfirsthead
\multicolumn{2}{c}{\tablename\ \thetable\ (continued)}\\
\toprule
\textbf{Term or notation} & \textbf{Meaning} \\
\midrule
\endhead
\midrule
\multicolumn{2}{r}{Continued on next page}\\
\endfoot
\bottomrule
\endlastfoot

\multicolumn{2}{l}{\textbf{Observed data and nuisance functions}}\\
\addlinespace
$O=(C,A,M,Y)$ & Observed data: baseline covariates $C$, binary exposure $A$, mediator $M$, and outcome $Y$, respectively; $a_0$ and $a_1$ denote the two exposure levels.\\
$P_C$, $p(c)$ & Marginal distribution and density of the baseline covariates $C$.\\
$\mu(m,a,c)$ & Outcome regression, $\E(Y\mid M=m,A=a,C=c)$.\\
$\pi(a\mid c)$ & Propensity score, $p(A=a\mid C=c)$.\\
$g(a\mid m,c)$ & Generalized propensity score, $p(A=a\mid M=m,C=c)$.\\
$Q$ & Collection of nuisance functions required by a target functional, such as $\mu$, $f_{M\mid A,C}$, and $\pi$.\\

\addlinespace
\multicolumn{2}{l}{\textbf{Original target functionals}}\\
\addlinespace
$\theta(Q)(c)$ & Conditional mediation functional, $\int \mu(m,a_1,c)f_{M\mid A,C}(m\mid a_0,c)\,dm$.\\
$\psi(Q)$ & Marginal mediation functional obtained by averaging $\theta(Q)(C)$ over $P_C$.\\
$\gamma(Q)$ & Front-door functional.\\

\addlinespace
\multicolumn{2}{l}{\textbf{Coarsening and discretized functionals}}\\
\addlinespace
$h$, $K$, $\mathcal{B}_k$ & Coarsening map, number of bins, and the $k$th mediator bin, respectively.\\
$\widetilde M=h(M)$ & Coarsened mediator taking values in $\{1,\ldots,K\}$.\\
$w_k$, $w_{\max,K}$ & Width of bin $\mathcal{B}_k$ and the maximum bin width.\\
$\mu_k(a,c)$ & Coarsened-mediator outcome regression, $\E(Y\mid \widetilde M=k,A=a,C=c)$.\\
$\mu_{k,a_1}(a,c)$ & Conditional mean of $\mu(M,a_1,c)$, $\E(\mu(M, a_1, c) \mid \widetilde M=k, A=a, C=c)$.\\
$m_k(a,c)$ & Conditional mediator mean, $\E(M\mid \widetilde M=k,A=a,C=c)$.\\
$g_k(a,c)$ & Conditional bin probability, $p(\widetilde M=k\mid A=a,C=c)$.\\
$p_{k,a}(m \mid c)$ &  Conditional mediator density, $p(m \mid \widetilde M=k, A=a,C=c)$.\\
$r_k(m\mid c)$ & Within-bin conditional density ratio, $\frac{p_{k,a_0}(m \,|\, c)}{p_{k,a_1}(m \,|\, c)}$.\\
$\theta_h(Q)(c)$ & Coarsened conditional mediation functional, $\sum_{k=1}^K\mu_k(a_1,c)g_k(a_0,c)$.\\
$\psi_h(Q)$, $\gamma_h(Q)$ & Coarsened versions of the marginal mediation and front-door functionals.\\
$\Delta_h(Q)(c)$ & Coarsening error, $\theta_h(Q)(c)-\theta(Q)(c)$.\\

\addlinespace
\multicolumn{2}{l}{\textbf{Debiasing and smoothing}}\\
\addlinespace
$\widetilde\theta_h(Q)(c)$ & Debiased coarsened functional, $\sum_{k=1}^K\mu\{m_k(a_0,c),a_1,c\}g_k(a_0,c)$.\\
$\widetilde\psi_h(Q)$, $\widetilde\gamma_h(Q)$ & Debiased coarsened versions of the marginal mediation and front-door functionals.\\
$R_k(c)$ & Within-bin debiasing remainder, $\mu\{m_k(a_0,c),a_1,c\}-\mu_{k,a_1}(a_0,c)$.\\
$\widetilde\Delta_h(Q)(c)$ & Approximation error of the debiased coarsened functional, $\widetilde\theta_h(Q)(c)-\theta(Q)(c)$.\\
$\mathcal{K}$, $b$, $\mathcal{K}_b$ & Kernel function, bandwidth, and rescaled kernel $\mathcal{K}_b(u)=b^{-1}\mathcal{K}(u/b)$.\\
$\omega_{b,k}(m\mid a_1,c)$ & Normalized localized kernel weight centered at $m_k(a_0,c)$.\\
$\mu_{b,k}(a_1,c)$ & Kernel-localized outcome mean for the $k$th bin.\\
$\widetilde\theta_{h,b}(Q)(c)$ & Smoothed, debiased coarsened functional, $\sum_{k=1}^K\mu_{b,k}(a_1,c)g_k(a_0,c)$.\\
$\widetilde\psi_{h,b}(Q)$, $\widetilde\gamma_{h,b}(Q)$ & Smoothed, debiased versions of the marginal mediation and front-door functionals.\\
$\widetilde\Delta^s_{h,b}(Q)(c)$ & Smoothing error, $\widetilde\theta_{h,b}(Q)(c)-\widetilde\theta_h(Q)(c)$.\\
$\widetilde\Delta_{h,b}(Q)(c)$ & Total approximation error, $\widetilde\theta_{h,b}(Q)(c)-\theta(Q)(c)$.\\

\addlinespace
\multicolumn{2}{l}{\textbf{Estimators}}\\
\addlinespace
$\theta(\widehat Q)(c)$ & Plug-in estimator of the original conditional mediation functional.\\
$\psi(\widehat Q)$, $\gamma(\widehat Q)$ & Plug-in estimators of the original mediation and front-door functionals.\\
$\theta_h(\widehat Q)(c)$ & Coarsened plug-in estimator, $\sum_{k=1}^K\widehat\mu_k(a_1,c)\widehat g_k(a_0,c)$.\\
$\psi_h(\widehat Q)$, $\gamma_h(\widehat Q)$ & Coarsened plug-in estimators obtained by replacing $\theta(\widehat Q)$ with $\theta_h(\widehat Q)$.\\
$\widetilde\theta_h(\widehat Q)(c)$ & Debiased coarsened plug-in estimator, $\sum_{k=1}^K\widehat\mu\{\widehat m_k(a_0,c),a_1,c\}\widehat g_k(a_0,c)$.\\
$\widetilde\psi_h(\widehat Q)$, $\widetilde\gamma_h(\widehat Q)$ & Debiased coarsened plug-in estimators obtained by replacing $\theta(\widehat Q)$ with $\widetilde\theta_h(\widehat Q)$.\\
$\widetilde\theta_{h,b}(\widehat Q)(c)$ & Smoothed, debiased coarsened plug-in estimator based on $\widehat\mu_{b,k}$ and $\widehat g_k$.\\
$\widetilde\psi_{h,b}^{\text{fixed}}(\widehat Q)$, $\widetilde\psi_{h,b}^{\text{one-step}}(\widehat Q)$ & Fixed-center plug-in and one-step estimators of the smoothed, debiased mediation functional.\\
$\psi^+(\widehat Q)$, $\psi_h^+(\widehat Q)$ & One-step estimators targeting the original and coarsened mediation functionals, respectively.\\
$\widetilde\psi_{h_1}^+(\widehat Q)$ & Derivative one-step estimator that replaces $\widehat{\mu}_k$ by $\widehat{\mu}(m_k(a_0,c),a_1,c)$ and $\theta_h(\widehat{Q})(c)$ with $\widetilde{\theta}_h(\widehat{Q})(c)$ in  $\psi_h^+(\widehat Q)$.\\
$\widetilde\psi_{h_2}^+(\widehat Q)$ & Derivative one-step estimator that replaces $\theta(\widehat{Q})(c)$ with $\widetilde{\theta}_h(\widehat{Q})(c)$  in $\psi^+(\widehat Q)$.\\
$\gamma_s(\widehat Q)$ & Sequential-regression plug-in estimator of the front-door functional.\\
\end{longtable}
}

% \vspace{1cm}
\section{Extensions to  multiple mediators}
\label{app:sec_multiple_mediators}

We discuss the extensions with two (sets of) mediators. Let $M_1$ and $M_2$ denote two ordered continuous or multi-dimensional mediators. Define the outcome regression $ \mu( m_1, m_2, a, c) = \E(Y | M_1=m_1, M_2=m_2, A=a, C=c)$, the conditional mediator density $f_{M_1|A, C}(m_1|a, c) = p(M_1 = m_1| A= a, C =c)$ and $f_{M_2|M_1, A, C}(m_2|m_1, a, c) = p(M_2 = m_2| M_1 = m_1, A= a, C =c)$. Let $Q = \{\mu, f_{M_1|A, C}, f_{M_2|M_1, A, C} \}$ collects the nuisance functions.  Let $(a_y, a_1, a_2) \in \{0, 1\}^3$, and for each $c \in \mathcal{C}$,
\begin{align}
    \theta(Q)(c) = \int\!\!\!\int  \mu(m_2, m_1, a_y, c)\, f_{M_2|M_1, A,C}(m_2 \,|\, m_1, a_2, c)\, f_{M_1| A,C}(m_1 \,|\, a_1, c) \, dm_2 \, dm_1 \,  . 
    \label{eq:estimand_2M}
\end{align}
Similarly, for $i=1, 2$, let $h_i: \mathbb{R} \mapsto \{1, \ldots, K_i\} $ denote the measurable discretization maps that partitions the support of $M_i$ into $K_i$ disjoint bins: ${\cal B}_{i, k_i} = \{m_i: h_i(m_i)=k_i\}, k_i\in \{1, \ldots, K_i\}$. Let $w_{i, k_i} = \sup_{m_{i, 1}, m_{i, 2} \in {\cal B}_{i, k_i} }|m_{i, 1} - m_{i, 2}|$ denote the bin width and define the coarsened mediator $\widetilde{M_i} = h_i(M_i), i=1,2$.

For each bin  $k_i \in \{1, \ldots, K_i\}$, define $g_{ k_1}(a_1, c) = p(\widetilde{M}_1 = k_1| A=a_1, C=c)$, $g_{k_1, k_2}(a_2, c) = p(\widetilde{M}_2 = k_2| \widetilde{M}_1 = k_1, A=a_2, C=c)$, and $\mu_{k_1, k_2}(a_y, c) = \E(Y | \widetilde{M}_2 = k_2, \widetilde{M}_1 = k_1, A = a_y, C=c)$. The coarsened analogue of \eqref{eq:estimand_2M} is: 
\begin{align}
    \theta_{h}(Q)(c) = 
     \sum_{k_1=1}^{K_1} \sum_{k_2=1}^{K_2}   
    \mu_{k_1, k_2}(a_y, c)\,  g_{k_1, k_2}(a_2, c) \, g_{ k_1}(a_1, c) .
\end{align}
To introduce the similar correction as one mediator in main manuscript, we first define
\begin{align*}
g_{k_2}(m_1,  a_2, c) & = p(\widetilde{M}_2 = k_2| M_1 = m_1, A=a_2, C=c) \, , \\
\mu_{k_2, a_y}(m_1, a_2, c) &= \E(\mu(M_1, M_2, a_y, C)\,|\, \widetilde{M}_2=k_2, M_1=m_1, A=a_2, C=c) , \\
    \eta_{k_2, a_y}(m_1, a_2, c) & = \mu_{k_2, a_y}(m_1, a_2, c) p(\widetilde{M}_2 = k_2| M_1 = m_1, A=a_2, C=c),   \\
\mu_{\eta, k_1, k_2, a_2, a_y}(a_1, c) &= \E\big(\eta_{k_2, a_y}(M_1, a_2, C)\,|\, \widetilde{M}_1=k_1,  A=a_1, C=c \big) \ . 
\end{align*}
Then, the conditional functional $ \theta(Q)(c)$ could be written as
\begin{align}
  \theta(Q)(c)  &=   \sum_{k_1=1}^{K_1} \sum_{k_2=1}^{K_2}  \Big \{ \int \big\{ \int 
    \mu(m_1, m_2, a_y,c) \, p(m_2\,|\, \widetilde{M}_2=k_2, m_1, a_2, c) dm_2 \big \} \notag \\
   & \hspace{2cm} \times p(\widetilde{M}_2=k_2 \,|\, m_1, a_2, c) \, p(m_1\,|\, \widetilde{M}_1=k_1, a_1, c)  \,  dm_1 \Big\} \, \notag \\
   & \hspace{2cm} \times  p(\widetilde{M}_1=k_1 \,|\,  a_1, c)   \notag \\
   & = \sum_{k_1=1}^{K_1} \sum_{k_2=1}^{K_2} \mu_{\eta, k_1, k_2, a_2, a_y}(a_1, c) g_{ k_1}(a_1, c) \, \ .
\end{align}
Define the within-bin mean
\begin{align*}
 m_{1, k_1}(a_1,c) &= \E[M_1\,|\, \widetilde M_1=k_1, A=a_1, C=c ], \\
 m_{2, k_2}(m_1, a_2,c) &= \E[M_2\,|\, \widetilde M_2=k_2, M_1=m_1, A=a_2, C=c ] \, . 
\end{align*}
Thus,  we define the debiased coarsened functional as 
\begin{align}
\widetilde \theta_h(Q)(c) 
= \sum_{k_1=1}^{K_1} \sum_{k_2=1}^{K_2} \mu \Big(m_{1, k_1}(a_1,c), m_{2, k_2}\big(m_{1, k_1}(a_1,c), a_2, c \big), a_y,c \Big) g_{k_2}\big(m_{1, k_1}(a_1,c),  a_2, c \big)    g_{ k_1}(a_1, c) \, \ .
\end{align}
For each $c \in \mathcal{C}$, the coarsening error $\widetilde \Delta_h(Q)(c) =  \widetilde \theta_h(Q)(c)  -  \theta(Q)(c)$ satisfies
\begin{equation}
    \begin{aligned}
    \widetilde \Delta_h(Q)(c)  = 
    &\sum_{k_1=1}^{K_1} \sum_{k_2=1}^{K_2} \Big\{\mu \Big(m_{1, k_1}(a_1,c), m_{2, k_2}\big(m_{1, k_1}(a_1,c), a_2, c \big), a_y,c \Big) g_{k_2}\big(m_{1, k_1}(a_1,c),  a_2, c \big) \\
    & \hspace{2cm} - \mu_{\eta, k_1, k_2, a_2, a_y}(a_1, c) \Big\}    g_{ k_1}(a_1, c) \, \ .
\end{aligned}
\end{equation}
Assume $\mu(m_1, m_2 , a_y , c)$ is twice continuously differentiable on each $\mathcal{B}_{i, k_i}$. Inside each bin $\mathcal{B}_{2, k_2}$, apply a second-order Taylor expansion of $\mu(m_1, m_2 , a_y , c)$ around $m_{2, k_2}(m_1, a_2,c)$:
\begin{align}
    \mu(m_1, M_2, a_y,c) = 
     & \mu(m_1, m_{2, k_2}(m_1, a_2, c), a_y,c)  + \mu'_{m_2}(m_1, m_{2, k_2}(m_1, a_2, c), a_y,c) (M_2 -  m_{2, k_2}(m_1, a_2, c)) \notag \\
     & + \frac{1}{2}\mu''_{m_2}(m_1, M_2^*, a_y,c)(M_2 -  m_{2, k_2}(m_1, a_2, c) )^2 \ ,  \label{eq:exact_mu} 
\end{align}
for some $M_2^* \in \mathcal{B}_{2, k_2}$.

Assume $\eta_{k_2, a_y}(m_1, a_2, c)$ is twice continuously differentiable on each $\mathcal{B}_{1, k_1}$. Inside each bin $\mathcal{B}_{1, k_1}$ for $M_1$, apply a second-order Taylor expansion of $\eta_{k_2, a_y}(M_1, a_2, c)$ around $m_{1, k_1}(a_1,c)$:
\begin{align}
    \eta_{k_2, a_y}(M_1, a_2, c) = 
    &  \eta_{k_2, a_y}(m_{1, k_1}(a_1,c), a_2, c) +    (\eta_{k_2, a_y})'_{m_1}(m_{1, k_1}(a_1,c), a_2, c) (M_1 -m_{1, k_1}(a_1,c) ) \notag \\
    & + \frac{1}{2} (\eta_{k_2, a_y})''_{m_1}(M^*_1, a_2, c) (M_1 -m_{1, k_1}(a_1,c) )^2\label{eq:exact_pi_k2_ay} \ .
\end{align}
for some $M_1^* \in \mathcal{B}_{1, k_1}$.
%\begin{align}
%    \mu(m_1, M_2, a_y,c) 
%     &\approx \mu(m_1, m_{2, k_2}(m_1, a_2, c), a_y,c)  + \mu'_{m_2}(m_1, m_{2, k_2}(m_1, a_2, c), a_y,c) (M_2 -  m_{2, k_2}(m_1, a_2, c) \ ,  \label{eq:approx_mu} \\
     %----
 %   \mu_{k_2, a_y}(M_1, a_2,c) 
 %   &\approx  \mu_{k_2, a_y}(m_{1, k_1}(a_1,c), a_2,c)   + \mu'_{m_1}(m_{1, k_1}(a_1,c), a_2,c) (M_1 -  m_{1, k_1}(a_1,c) \ , \label{eq:approx_mu_k2_ay}  \\
    %----
 %   \eta_{k_2, a_y}(M_1, a_2, c) 
%    &\approx  \eta_{k_2, a_y}(m_{1, k_1}(a_1,c), a_2, c) + \frac{\partial{\pi^{*}}_{k_2, a_y}(M_1, a_2, c)}{\partial m_1}\bigg|_{M_1 = m_{1, k_1}(a_1,c) }(M_1 -m_{1, k_1}(a_1,c) )  \label{eq:approx_pi_k2_ay} \ .
%\end{align}
%\begin{align*}
%\mu_{\eta, k_1, k_2, a_2, a_y}(a_1, c) 
%&= \E(\eta_{k_2, a_y}(M_1, a_2, c)\,|\, \widetilde{M}_1{=}k_1,  A{=}a_1, C{=}c)  \\
%& \overset{\ref{eq:approx_pi_k2_ay}}{\approx} \, \eta_{k_2, a_y}(m_{1, k_1}(a_1,c), a_2, c) \\
%& = \mu_{k_2, a_y}(m_{1, k_1}(a_1,c), a_2,c) \, p(\widetilde{M}_2=k_2 \,|\, m_{1, k_1}(a_1,c), a_2, c)  \\
%& = \E(\mu(m_{1, k_1}(a_1,c), M_2, a_y, c)\,|\, \widetilde{M}_2{=}k_2, m_{1, k_1}(a_1,c), A{=}a_2, C{=}c) \, p(\widetilde{M}_2=k_2 \,|\, m_{1, k_1}(a_1,c), a_2, c) \\
%& \overset{\ref{eq:approx_mu}}{\approx}  \mu(m_{1, k_1}(a_1,c), m_{2, k_2}(m_{1, k_1}(a_1,c), a_2, c), a_y,c) \, p(\widetilde{M}_2=k_2 \,|\, m_{1, k_1}(a_1,c), a_2, c)
%\end{align*}

Take conditional expectation sequentially :
{
\begin{align*}
&\mu_{\eta, k_1, k_2, a_2, a_y}(a_1, c)\\ 
&= \E \big(\eta_{k_2, a_y}(M_1, a_2, c)\,|\, \widetilde{M}_1{=}k_1,  A{=}a_1, C{=}c \big)  \\
& \overset{\ref{eq:exact_pi_k2_ay}}{=} \, \eta_{k_2, a_y}(m_{1, k_1}(a_1,c), a_2, c) + \frac{1}{2} \E \big( (\eta_{k_2, a_y})''_{m_1}(M^*_1, a_2, c) (M_1 -m_{1, k_1}(a_1,c) )^2 \big | \widetilde{M}_1{=}k_1,  A{=}a_1, C{=}c \big)\\
& = \mu_{k_2, a_y}(m_{1, k_1}(a_1,c), a_2,c) \, g_{k_2}\big(m_{1, k_1}(a_1,c),  a_2, c \big)  \\
& \hspace{0.5cm} +  \frac{1}{2} \E \big( (\eta_{k_2, a_y})''_{m_1}(M^*_1, a_2, c) (M_1 -m_{1, k_1}(a_1,c) )^2 \big | \widetilde{M}_1{=}k_1,  A{=}a_1, C{=}c \big)\\
& = \E \big(\mu(m_{1, k_1}(a_1,c), M_2, a_y, c)\,|\, \widetilde{M}_2{=}k_2, m_{1, k_1}(a_1,c), A{=}a_2, C{=}c \big) \, g_{k_2}\big(m_{1, k_1}(a_1,c),  a_2, c \big) \\
& \hspace{0.5cm}+ \frac{1}{2} \E \big( (\eta_{k_2, a_y})''_{m_1}(M^*_1, a_2, c) (M_1 -m_{1, k_1}(a_1,c) )^2 \big | \widetilde{M}_1{=}k_1,  A{=}a_1, C{=}c \big) \\
& \overset{\ref{eq:exact_mu}}{=}  \Big\{\mu \big(m_{1, k_1}(a_1,c), m_{2, k_2}(m_{1, k_1}(a_1,c), a_2, c), a_y,c \big)  \\
& \hspace{0.5cm} + \frac{1}{2} \E \big(\mu''_{m_2}(m_{1, k_1}(a_1,c), M_2^*, a_y,c)(M_2 -  m_{2, k_2}(m_{1, k_1}(a_1,c), a_2, c) )^2 \,|\, \widetilde{M}_2{=}k_2, m_{1, k_1}(a_1,c), A{=}a_2, C{=}c \big) \Big\} \\
& \hspace{0.75cm} \times g_{k_2}\big(m_{1, k_1}(a_1,c),  a_2, c \big) \\
& \hspace{0.5cm} + \frac{1}{2} \E \big( (\eta_{k_2, a_y})''_{m_1}(M^*_1, a_2, c) (M_1 -m_{1, k_1}(a_1,c) )^2 \big | \widetilde{M}_1{=}k_1,  A{=}a_1, C{=}c \big) \ .
\end{align*}
}

Thus
{\small
\begin{align*}
R_{k_1, k_2}(c) 
& = \mu \big(m_{1, k_1}(a_1,c), m_{2, k_2}(m_{1, k_1}(a_1,c), a_2, c), a_y,c \big)   \, g_{k_2}\big(m_{1, k_1}(a_1,c),  a_2, c \big) - \mu_{\eta, k_1, k_2, a_2, a_y}(a_1, c) \\
& = - \frac{1}{2} \E \Big(\mu''_{m_2}(m_{1, k_1}(a_1,c), M_2^*, a_y,c)(M_2 -  m_{2, k_2}(m_{1, k_1}(a_1,c), a_2, c) )^2 \,|\, \widetilde{M}_2{=}k_2, m_{1, k_1}(a_1,c), A{=}a_2, C{=}c \Big) \\
& \hspace{0.75cm} \times g_{k_2}\big(m_{1, k_1}(a_1,c),  a_2, c \big) \\
& \hspace{0.5cm} - \frac{1}{2} \E \big( (\eta_{k_2, a_y})''_{m_1}(M^*_1, a_2, c) (M_1 -m_{1, k_1}(a_1,c) )^2 \big | \widetilde{M}_1{=}k_1,  A{=}a_1, C{=}c \big) \, .
\end{align*}
}

Given $|g_{k_2}\big(m_{1, k_1}(a_1,c),  a_2, c \big)| < 1$, the monotonicity and linearity of expectations, taking absolute values, we obtain the following upper bound on $|R_{k_1, k_2}(c)|$:
{\footnotesize
\begin{align*}
    |R_{k_1, k_2}(c) | 
    &\leq \frac{1}{2} \E \Big(|\mu''_{m_2}(m_{1, k_1}(a_1,c), M_2^*, a_y,c)|(M_2 -  m_{2, k_2}(m_{1, k_1}(a_1,c), a_2, c) )^2 \,|\, \widetilde{M}_2{=}k_2, m_{1, k_1}(a_1,c), A{=}a_2, C{=}c \Big)  \\
    & \hspace{0.5cm}+ \frac{1}{2} \E \big( |(\eta_{k_2, a_y})''_{m_1}(M^*_1, a_2, c)| (M_1 -m_{1, k_1}(a_1,c) )^2 \big | \widetilde{M}_1{=}k_1,  A{=}a_1, C{=}c \big) \\
    &\leq \frac{1}{2} \,  \sup_{m_2 \in \mathcal{B}_{2, k_2}(c)}|\mu''_{m_2}(m_{1, k_1}(a_1,c), m_2, a_y,c)| \E \big((M_2 -  m_{2, k_2}(m_{1, k_1}(a_1,c), a_2, c) )^2 \,|\, \widetilde{M}_2{=}k_2, m_{1, k_1}(a_1,c), A{=}a_2, C{=}c \big) \\
    & \hspace{0.5cm} + \frac{1}{2} \sup_{m_1 \in \mathcal{B}_{1, k_1}(c)} |(\eta_{k_2, a_y})''_{m_1}(m_1, a_2, c)|  \E \big( (M_1 -m_{1, k_1}(a_1,c) )^2  \,|\, \widetilde{M}_1{=}k_1,  A{=}a_1, C{=}c \big)\ . 
\end{align*}
}
For any $(a_1, a_2, c)$, the conditional law $p(m_1\,|\, \widetilde M_1=k_1, A=a_1, C=c)$ is supported on $B_{1, k_1}(c)$ and $p(m_2\,|\, \widetilde M_2=k_2,  M_1 = m_{1, k_1}(a_1,c), A=a_2, C=c)$ is supported on $B_{2, k_2}(c)$, hence 
\begin{align*}
 \text{Var}(M_1 \,|\, k_1, a_1, c) & \le  \frac{ ( \sup_{m_{1, 1}, m_{1, 2} \in \mathcal{B}_{1, k_1}(c)} |m_{1, 1} - m_{1, 2}|)^2}{4}  = \frac{w_{1,k_1}(c)^2}{4} \\
 \text{Var}(M_2 \,|\, k_2, m_{1, k_1}(a_1,c) , a_2, c) & \le  \frac{ ( \sup_{m_{2, 1}, m_{2, 2} \in \mathcal{B}_{2, k_2}(c)} |m_{2, 1} - m_{2, 2}|)^2}{4}  = \frac{w_{2,k_2}(c)^2}{4}  \ .
\end{align*}
For $\sup_{m_1 \in \mathcal{B}_{1, k_1}(c)} |(\eta_{k_2, a_y})''_{m_1}(m_1, a_2, c)| \le L_1(c)$ and $\sup_{m_2 \in \mathcal{B}_{2, k_2}(c)}|\mu''_{m_2}(m_{1, k_1}(a_1,c), m_2, a_y,c)| \le L_2(c)$, and the fact that $\sum_{k=1}^K g_{k_1}(a_1,c)=1$,  we have $\widetilde{\Delta}_h(Q)(c) = O(w_{\text{max}, 1, K_1}(c)^2+w_{\text{max}, 2, K_2}(c)^2)$.

%%%%%%%%%%%%%%%%%%%%%%%%%%%%%%%%%%%%%%%%%%%%%%%%%%%%%%%%%%%%
\clearpage
\section{Proofs}
\label{app:proofs} 

\subsection{Identification proofs} 
\label{app:proofs_identification} 

For the identification of mediation functional $\E(Y(a_1, M(a_0)))$ and the front-door functional $\E(Y(a_0))$,  we first introduce general assumptions that apply throughout. For any $a$ and  $m$, we assume
\begin{enumerate}
    \item[(A)] \textit{Consistency}, which indicates that observed outcome and mediators match their counterfactuals when treatment and mediator values are set at observed values; i.e. $Y(a, m) = Y$ if $A =a$ and $M=m$, and $M(a) = M$ if $A=a$.
    \item[(B)] \textit{Positivity}, which states that $P(A = 1 \,|\, C = c) > 0$ when $P(C = c) > 0$, and $P(A = 1 \,|\, M = m, C = c) > 0$ when $P(M = m, C = c) > 0$.
\end{enumerate}

To identify the mediation functional $\E(Y(a_1, M(a_0)))$, we additionally need:
\begin{enumerate}
\item[(C1)] \textit{Conditional ignorability}, which assumes the absence of unmeasured  confounders between the treatment-mediator, treatment-outcome and mediator-outcome pairs; i.e. for any $a, a_0, a_1$ and  $m$: (i) $Y(a_1, m), M(a_0) \indep A \,|\, C$; (ii)  $Y(a_1, m) \indep M(a) \,|\, A, C$ .
\end{enumerate}
Under (A)–(C1), the mediation functional is identified as
\begin{equation}
\begin{aligned}
& \hspace{-0.5cm} \E(Y(a_1, M(a_0))) \\
& = \int \!\!\! \int \E(Y(a_1, m) | M(a_0) = m, C = c) \, p(M(a_0) = m|C = c) \, p(C = c) \, dm \, dc \\
& =  \int \!\!\! \int \E(Y(a_1, m) | M(a_0) = m, A = a_1, C = c) \, p(M = m| A = a_0, C = c) \, p(C = c) \, dm \, dc \\
& =  \int \!\!\! \int \E(Y(a_1, m) | M(a_1) = m, A = a_1, C = c) \, p(M = m| A = a_0, C = c) \, p(C = c) \, dm \, dc \\
& =  \int \!\!\! \int \E(Y | M = m, A = a_1, C = c) \, p(M = m| A = a_0, C = c) \, p(C = c) \, dm \, dc 
\end{aligned}
\end{equation}

To identify the front-door functional $\E(Y(a_0))$, in addition to (A) and (B), we require:
\begin{enumerate}
\item[(C2)] \textit{Conditional ignorability}, which assumes the absence of unmeasured  confounders between the treatment-mediator, and mediator-outcome pairs; i.e. (i) $M(a_0) \indep A \,|\, C$; (ii)  $Y(m) \indep M \,|\, A, C$; (iii)  $Y(m) \indep M(a_0) \,|\, C$.
\item[(D)] \textit{No direct effect}, which assumes that $M$ blocks all directed paths from  $A$ to $Y$, i.e., $Y(a,m) {=} Y(m)$ for any $a$ and $m$.
\end{enumerate}
Under (A), (B), (C2) and (D), the front-door functional is identified as
\begin{equation}
\begin{aligned}
& \hspace{-0.15cm} \E(Y(a_0))  \\
&= \int \!\!\! \int \E(Y(a_0)|M(a_0) = m, C = c) \, p(M(a_0) = m|C = c) \, p(C = c) \, dm \, dc \\
&= \int \!\!\! \int \E(Y(a_0, m)|M(a_0) = m, C = c) \, p(M(a_0) = m| A = a_0, C = c) \, p(C = c) \, dm \, dc \\
&= \int \!\!\! \int \E(Y(m)|M(a_0) = m, C = c) \, p(M = m| A = a_0, C = c) \, p(C = c) \, dm \, dc \\
&= \int \!\!\! \int \E(Y(m)|C = c) \, p(M = m| A = a_0, C = c) \, p(C = c) \, dm \, dc \\
&= \int \!\!\! \int \Big\{ \sum_{a \in \{0, 1\}} \E(Y|M=m, A =a, C = c) \, p(A = a|C=c) \Big\} \, p(M = m| A = a_0, C = c) \, p(C = c) \, dm \, dc \\
&= p(A {=} a_0) \, \E(Y | A {=} a_0) +  \int \!\!\! \int \E(Y | M{=} m, A {=}a_1, C {=} c) \, p(M {=} m| A {=} a_0, C {=} c) \, p(A {=} a_1,  C {=} c) \, dm \,  dc \, . 
\end{aligned}
\end{equation}

\subsection{Lemma~\ref{lem:coarsening_error}} 
\label{app:proofs_lem:coarsening_error} 

Given our notations, we can rewrite $\theta(Q)(C)$ in \eqref{eq:estimand} as: 
\begin{align}
   \theta(Q)(C) 
   &= \sum_{k=1}^K   \Big\{ \int \mu(m, a_1,c) \, p(m\,|\, \widetilde{M} = k, a_0, c) \, dm \Big\} \,  g_k(a_0, c) 
    = \sum_{k=1}^K  \mu_{k, a_1}(a_0, c) \, g_k(a_0, c)  \ ,  
\end{align}

By definition of the coarsened estimand,
\begin{align*}
\theta_h(Q)(c)
=
\sum_{k=1}^K \mu_k(a_1,c)\, g_k(a_0,c).
\end{align*}
Therefore,
\begin{align}
\Delta_h(Q)(c)
&= \theta_h(Q)(c) - \theta(Q)(c) \\
&= \sum_{k=1}^K \mu_k(a_1,c)\, g_k(a_0,c)
    - \sum_{k=1}^K \mu_{k,a_1}(a_0,c)\, g_k(a_0,c) \\
&= \sum_{k=1}^K \big\{\mu_k(a_1,c) - \mu_{k,a_1}(a_0,c)\big\}\, g_k(a_0,c),
\end{align}
which proves \eqref{eq:coarsening_error}.

Further, suppose that $\sup_{m \in \mathcal{B}_k(c)} |\mu_m'(m,a_1,c)| \le L(c)$ for some
square-integrable function under $P_C$. Then from \eqref{eq:coarsening_error},
\begin{align}
\big|\Delta_h(Q)(c)\big|
&\le
\sum_{k=1}^K \big|\mu_k(a_1,c) - \mu_{k,a_1}(a_0,c)\big|\, g_k(a_0,c) \\
&\le
L(c) \sum_{k=1}^K \big|m_k(a_1,c) - m_k(a_0,c)\big|\, g_k(a_0,c).
\end{align}
Since both conditional means $m_k(a_1,c)$ and $m_k(a_0,c)$ lie in the same bin
$B_k(c)$, their difference is bounded by the bin width,
\begin{align*}
\big|m_k(a_1,c) - m_k(a_0,c)\big| \le w_k(c),
\end{align*}
so
\begin{align}
\big|\Delta_h(Q)(c)\big|
&\le
L(c) \sum_{k=1}^K w_k(c)\, g_k(a_0,c).
\end{align}
The weights $g_k(a_0,c)$ form a probability distribution over $k$, hence
$\sum_{k=1}^K w_k(c)\, g_k(a_0,c)$ is a weighted average of the bin widths and is
bounded by the maximum bin width,
\begin{align*}
\sum_{k=1}^K w_k(c)\, g_k(a_0,c)
\le \max_k w_k(c)
=: w_{\max,K}(c).
\end{align*}
Therefore
\begin{align*}
\Delta_h(Q)(c) = O\big(w_{\max,K}(c)\big).
\end{align*}
When the mediator has bounded support and the $K$ bins are chosen to have equal
width, $w_{\max,K}(c) = O(1/K)$, so $\Delta_h(Q)(c) = O(1/K)$. 

To prove \eqref{eq:first_order_scaling}, we assume $\mu$ is twice continuously differentiable and define the within-bin mean 
\begin{align}
    m_k(a,c) = \E(M \,|\, A=a, C=c, \widetilde M=k) \ . 
\end{align}
Applying a first-order Taylor expansion of $\mu(M,a_1,c)$ around $m_k(a,c)$ yields
\begin{align*}
\mu(M,a_1,c) 
\approx & \mu(m_k(a,c),a_1,c)
+ \mu_m'(m_k(a,c),a_1,c)\,(M - m_k(a,c)) \ , 
\end{align*}
where $\mu_m'$ denotes the partial derivative of $\mu$ with respect to $m$. Taking expectations w.r.t. $p(m \,|\, \widetilde{M}=k, a, c)$ shows that 
$\E(\mu(M,a_1,c)\,|\, \widetilde M=k, a,c) \approx \mu(m_k(a,c),a_1,c)$ up to second-order remainder terms.  
Hence, $\mu_k(a_1,c) - \mu_{k,a_1}(a_0,c)$ can be approximated by
\begin{align}
 \mu_k(a_1,c) - \mu_{k,a_1}(a_0,c) 
&\approx \mu(m_k(a_1,c),a_1,c) - \mu(m_k(a_0,c),a_1,c)  \nonumber\\
&\approx \mu_m'(m_k(a_1,c),a_1,c)\,\big(m_k(a_1,c) - m_k(a_0,c)\big) \, ,
\label{eq:first_order_bias}
\end{align}
where the second line follows another first-order Taylor expansion of $\mu$ around $m_k(a_1,c)$.

\subsection{Remark~\ref{remark:covariance}} 
\label{app:proofs_remark:covariance}

For each bin index $k$ and covariate value $c$, define
\begin{align}
    p_{k,a}(m \,|\, c) = p(m \,|\, \widetilde M=k, A=a, C=c) \ , \ r_k(m \,|\, c) = \frac{p_{k,a_0}(m \,|\, c)}{p_{k,a_1}(m \,|\, c)} \ . 
\end{align}
We can write
\begin{align*}
\mu_{k,a_1}(a_0,c) - \mu_k(a_1,c)
&= \int \mu(m,a_1,c)\, p_{k, a_0}(m \,|\, c)\, dm 
   - \int \mu(m,a_1,c)\, p_{k, a_1}(m \,|\, c)\, dm \\
&=  \int \mu(m,a_1,c)\, \big\{ r_k(m \,|\, c) - 1 \big\} \, p_{k, a_1}(m \,|\, c)\, dm \\ 
&= \E_{p_{k,a_1}}\Big(\mu(M,a_1,c)\, \big( r_k(M \,|\, c) - 1\big)  \Big) \\
&= \E_{p_{k,a_1}}\Big( \big( \mu(M,a_1,c)\, - \, \E_{p_{k, a_1}}\big(\mu(M,a_1,c)\big) \big) \, \big( r_k(M \,|\, c) - 1\big)  \Big)  \\
&= \Cov_{p_{k, a_1}}\!\big(\mu(M,a_1,c),\, r_k(M \,|\, c)\big) \ ,
\end{align*}
since $E_{p_{k, a_1}}\big(r_k(M \,|\, c)\big) = 1$.

\subsection{Lemma~\ref{lem:first-order-removal}}
\label{app:proofs_lem:first-order-removal}

In the proof of Lemma~\ref{lem:coarsening_error}, we showed $\theta(Q)(c) =\sum_{k=1}^K \mu_{k,a_1}(a_0,c)\, g_k(a_0,c)$. By definition of the debiased coarsened functional, $\widetilde\theta_h(Q)(c)=
\sum_{k=1}^K \mu\big(m_k(a_0,c),a_1,c\big)\, g_k(a_0,c)$. Subtracting the two displays yields \eqref{eq:coarsening_error_tilde_delta_h}.

We now bound $R_k(c)$ for a fixed bin index $k$. Inside each bin $\mathcal B_k$, apply a second-order Taylor expansion of $\mu(M,a_1,c)$ around $m_k(a_0,c)$: 
{\small 
\begin{align*}
    \mu(M,a_1,c) 
    &= \mu(m_k(a_0,c), a_1,c) + \mu'_m(m_k(a_0,c), a_1,c) (M - m_k(a_0,c)) + \frac{1}{2}\mu''_m(M^*,a_1,c)(M - m_k(a_0,c))^2 \ ,
\end{align*}
}%
for some $M^* \in \mathcal B_k$ (by the mean-value form of the remainder). Taking the conditional expectation under $(\widetilde M{=}k, A{=}a_0, C{=}c)$ gives 
{ 
\begin{align*}
     \E(\mu(M,a_1,c) \,|\, k, a_0, c)
    &= \mu(m_k(a_0,c), a_1,c) + \frac{1}{2} \E\big(\mu''_m(M^*,a_1,c)(M - m_k(a_0,c))^2 \,|\, k, a_0, c\big) \ .
\end{align*}
}
Thus
{ 
\begin{align*}
    R_k(c) 
    &= \mu(m_k(a_0,c), a_1, c) - \mu_{k, a_1}(a_0, c) 
    = - \frac{1}{2} \E\big(\mu''_m(M^*,a_1,c)(M - m_k(a_0,c))^2 \,|\, k, a_0, c\big) \ .
\end{align*}
}
Taking absolute values and given the monotonicity and linearity of expectations, we obtain the following upper bound on $| R_k(c) |$: 
\begin{align*}
    | R_k(c) | 
    &\leq \frac{1}{2} \, \E\big( | \mu''_m(M^*,a_1,c)| \, (M - m_k(a_0,c))^2 \,|\, k, a_0, c\big)  \\
    &\leq \frac{1}{2} \,  \sup_{m \in \mathcal{B}_k(c)} | \mu''_m(m,a_1,c)| \, \E\big( (M - m_k(a_0,c))^2 \,|\, k, a_0, c\big) \ . 
\end{align*}

Finally, since the conditional law of $M$ given $(A=a_0,C=c,\widetilde M=k)$ is supported on ${\cal B}_k$, its range is at most $w_k$, and hence 
\begin{align*}
 \E\!\left((M-m_k(a_0,c))^2 \mid A=a_0,C=c,\widetilde M=k\right)
&= \Var(M\mid A=a_0,C=c,\widetilde M=k) \\
&\le \frac{ ( \sup_{m_1, m_2 \in \mathcal{B}_k} |m_1 - m_2|)^2}{4} \\ 
&=w_k^2/4
\end{align*}

Combining with $\sup_{m\in{\cal B}_k}|\mu''_m(m,a_1,c)| \le L(c)$ yields
$|R_k(c)|\le L(c)\, w_k^2/8$. Furthermore, using the fact that $\sum_{k=1}^K g_k(a_0,c)=1$, we obtain
\begin{align*}
|\widetilde\Delta_h(Q)(c)|
&\le
\sum_{k=1}^K |R_k(c)|\, g_k(a_0,c)
\le
\frac{L(c)}{8}\sum_{k=1}^K w_k^2 g_k(a_0,c)
\le
\frac{L(c)}{8}\, w_{\max,K}^2,
\end{align*}
which implies $\widetilde\Delta_h(Q)(c)=O(w_{\max,K}^2)$.

\subsection{Lemma~\ref{lem:loc-bias}} 
\label{app:proofs_loc-bias} 

By definition: 
\begin{align*}
\widetilde\theta_{h,b}(Q)(c)
=
\sum_{k=1}^K \mu_{b,k}(a_1,c)\, g_k(a_0,c),
\qquad
\widetilde\theta_h(Q)(c)
=
\sum_{k=1}^K \mu\big(m_k(a_0,c),a_1,c\big)\, g_k(a_0,c),
\end{align*}
so
\begin{align*}
   \widetilde\Delta_{h,b}^{s}(Q)(c)
   &= \widetilde\theta_{h,b}(Q)(c) - \widetilde\theta_h(Q)(c) \\
   &= \sum_{k=1}^K \big\{\mu_{b,k}(a_1,c) - \mu\big(m_k(a_0,c),a_1,c\big)\big\}\, g_k(a_0,c),
\end{align*}
which establishes the first display.

It remains to show that each term in braces is of order $b^2$ uniformly in $k,c$ under the stated conditions. Fix $k$ and $c$, 
\begin{align*}
\mu_{b,k}(a_1,c)
&=
E\big\{\mu(m,a_1,c)\,\omega_{b,k}(M \,|\, a_1,c)\,|\, A=a_1,C=c\big\} \\
&=
\frac{\E\big\{\mu(m,a_1,c)\,\mathcal K_b(M-m_k(a_0,c))\,|\, A=a_1,C=c\big\}}
     {\E\big\{\mathcal K_b(M-m_k(a_0,c))\,|\, A=a_1,C=c\big\}} \\
&=
\frac{N_b}{D_b},
\end{align*}
where
\begin{align*}
N_b
&=
\int \mu(m,a_1,c)\,\mathcal K_b(m-m_k(a_0,c))\, f_{M\,|\, A,C}(m \,|\, a_1,c)\, dm, \\
\qquad
D_b
&=
\int \mathcal K_b(m-m_k(a_0,c))\, f_{M\,|\, A,C}(m \,|\, a_1,c)\, dm.
\end{align*}

Make the change of variables $m = m_k(a_0,c) + bu$, so that $dm = b\,du$ and
$\mathcal K_b(m-m_k(a_0,c)) = \frac{1}{b}\mathcal K\!\left(\frac{m-m_k(a_0,c)}{b}\right)
= \frac{1}{b}\mathcal K(u)$. Then
\begin{align*}
N_b
&=
\int \mu(m_k(a_0,c) + bu, a_1, c)\,\mathcal K_b\big(m_k(a_0,c)+bu-m_k(a_0,c)\big)\, f_{M\,|\, A,C}(m_k(a_0,c)+bu \,|\, a_1, c)\, b\,du \\
&=
\int \mu(m_k(a_0,c) + bu, a_1, c)\, f_{M\,|\, A,C}(m_k(a_0,c)+bu \,|\, a_1, c)\, \mathcal K(u)\, du, \\
D_b
&=
\int f_{M\,|\, A,C}(m_k(a_0,c)+bu \,|\, a_1, c)\,\mathcal K(u)\, du.
\end{align*}

By assumption (i), $\mu(\cdot)$ is twice continuously differentiable in a
neighborhood of $m_k(a_0,c)$ and its second derivative is uniformly bounded; similarly,
assumption (iii) ensures that $f_{M\,|\, A,C}(\cdot)$ is continuous and bounded away from zero
in a neighborhood of $m_k(a_0,c)$. Applying a second order Taylor expansion around $m_k(a_0,c)$,
\begin{align*}
\mu(m_k(a_0,c)+bu, a_1, c)
&=
\mu(m_k(a_0,c), a_1, c) + \mu'_m(m_k(a_0,c), a_1, c)\, bu + \tfrac12 \mu''_m(\widetilde M_1, a_1, c)\, b^2 u^2, \\
f_{M\,|\, A,C}(m_k(a_0,c)+bu \,|\, a_1, c)
&=
f_{M\,|\, A,C}(m_k(a_0,c) \,|\, a_1, c) + f'(m_k(a_0,c) \,|\, a_1, c)\, bu + \tfrac12 f''(\widetilde M_2 \,|\, a_1, c)\, b^2 u^2,
\end{align*}
for some $\widetilde M_1,\widetilde M_2$ between $m_k(a_0,c)$ and $m_k(a_0,c)+bu$. Multiplying these
expansions and using that the second derivatives are uniformly bounded yields
\begin{align*}
&\mu(m_k(a_0,c)+bu, a_1, c)\, f_{M\,|\, A,C}(m_k(a_0,c)+bu \,|\, a_1, c) \\
&\hspace{1cm}= \mu(m_k(a_0,c), a_1, c) f_{M\,|\, A,C}(m_k(a_0,c) \,|\, a_1, c) \\
&\hspace{1.5cm}+ \big\{\mu'_m(m_k(a_0,c), a_1, c) f_{M\,|\, A,C}(m_k(a_0,c) \,|\, a_1, c) + \mu(m_k(a_0,c), a_1, c) f'(m_k(a_0,c) \,|\, a_1, c)\big\} bu \\
&\hspace{1.5cm}+ O(b^2 u^2),
\end{align*}
where the $O(b^2 u^2)$ term is uniform in $u$ on the support of $\mathcal K$.

Substituting into the expression for $N_b$ and using assumption (ii) on the
kernel,
\begin{align*}
\int \mathcal K(u)\,du = 1,\qquad \int u\,\mathcal K(u)\,du = 0,\qquad
\int u^2 \mathcal K(u)\,du < \infty,
\end{align*}
we obtain
\begin{align*}
N_b
&=
\int \big[\mu(m_k(a_0,c), a_1, c) f_{M\,|\, A,C}(m_k(a_0,c) \,|\, a_1, c) \\
&\hspace{1cm}+ \big\{\mu'_m(m_k(a_0,c), a_1, c) f_{M\,|\, A,C}(m_k(a_0,c) \,|\, a_1, c) + \mu(m_k(a_0,c), a_1, c) f'(m_k(a_0,c) \,|\, a_1, c)\big\} bu \\ 
&\hspace{1cm}+ O(b^2 u^2)\big]\, \mathcal K(u)\, du \\
&=
\mu(m_k(a_0,c), a_1, c) f_{M\,|\, A,C}(m_k(a_0,c) \,|\, a_1, c)
+ O(b^2),
\end{align*}
since the term proportional to $bu$ integrates to zero by symmetry of
$\mathcal K$, and the $O(b^2 u^2)$ term integrates to $O(b^2)$ because
$\int u^2 \mathcal K(u)\,du < \infty$.

A similar argument for $D_b$ gives
\begin{align*}
D_b
&=
\int \big[f_{M\,|\, A,C}(m_k(a_0,c) \,|\, a_1, c) + f'(m_k(a_0,c) \,|\, a_1, c) bu + O(b^2 u^2)\big]\, \mathcal K(u)\, du \\
&=
f_{M\,|\, A,C}(m_k(a_0,c) \,|\, a_1, c) + O(b^2).
\end{align*}
Assumption (iii) ensures that $f_{M\,|\, A,C}(m_k(a_0,c) \,|\, a_1, c)$ is bounded away from zero, so for $b$ small
enough $D_b$ is also bounded away from zero and we can write
\begin{align*}
\mu_{b,k}(a_1,c)
=
\frac{N_b}{D_b}
&=
\frac{\mu(m_k(a_0,c), a_1, c) f_{M\,|\, A,C}(m_k(a_0,c) \,|\, a_1, c) + O(b^2)}{f_{M\,|\, A,C}(m_k(a_0,c) \,|\, a_1, c) + O(b^2)} \\
&=
\mu(m_k(a_0,c), a_1, c) + O(b^2),
\end{align*}
where the $O(b^2)$ term is uniform in $k$ and $c$ under the stated uniform
boundedness conditions.

Therefore, for each $k$,
\begin{align*}
\mu_{b,k}(a_1,c) - \mu\big(m_k(a_0,c),a_1,c\big) = O(b^2).
\end{align*}
Plugging this into the expression for $\widetilde\Delta_{h,b}^{s}(Q)(c)$ and using that
$\sum_{k=1}^K g_k(a_0,c) = 1$ yields
\begin{align*}
\big|\widetilde\Delta_{h,b}^{s}(Q)(c)\big|
&\le
\sum_{k=1}^K \big|\mu_{b,k}(a_1,c) - \mu\big(m_k(a_0,c),a_1,c\big)\big|\,
g_k(a_0,c) \\
&\le
C^* b^2 \sum_{k=1}^K g_k(a_0,c)
=
C^* b^2,
\end{align*}
for some constant $C^*<\infty$ not depending on $b$. Hence
$\widetilde\Delta_{h,b}^{s}(Q)(c) = O(b^2)$ as $b \to 0$, as claimed. 

Furthermore, we can write 
\begin{align*} 
\widetilde\Delta_{h,b}(Q)(c)
&\coloneqq \widetilde\theta_{h,b}(Q)(c) - \theta(Q)(c) \\
&=
\underbrace{\widetilde\theta_h(Q)(c) - \theta(Q)(c)}_{\text{coarsening error }}
+
\underbrace{\widetilde\theta_{h,b}(Q)(c) - \widetilde\theta_h(Q)(c)}_{\text{smoothing error }} 
\\
&=
O\big(w_{\max,K}^2 + b^2\big).
\end{align*}

\subsection{Theorem~\ref{thm:eif}}
\label{app:proofs:eif} 

We derive the EIF for the smoothed mediation functional
\begin{align}
 \widetilde\psi_{h, b}(Q)
=
\E\!\left[
\sum_{k=1}^K \mu_{b,k}(a_1,C) \, g_k(a_0,C)
\right],
\label{eq:psi_def}
\end{align}
where
\begin{align}
\mu_{b,k}(a_1,c)
=
\E\!\left[
Y\,\omega_{b,k}(M\,|\, a_1,c)
\,\middle|\,
A=a_1, C=c
\right],
\qquad
g_k(a_0,c)
=
p(\widetilde M=k\,|\, A=a_0,C=c).
\label{eq:mu_bk_gk_def}
\end{align}
Recall that
\begin{align}
\omega_{b,k}(m\,|\, a_1,c)
=
\frac{\mathcal K_b\!\left(m-m_k(a_0,c)\right)}
{\E\!\left(\mathcal K_b\!\left(M-m_k(a_0,c)\right)\,|\, A=a_1,C=c\right)},
\qquad
\mathcal K_b(u)=b^{-1}\mathcal K(u/b).
\label{eq:omega_def}
\end{align}

The functional $\widetilde\psi_{h,b}(Q)$ depends on $Q$ in two distinct ways: 
(i) through the conditional laws defining $g_k(a_0,c)$ and $\mu_{b,k}(a_1,c)$,
(ii) through the nuisance center $m_k(a_0,c)$ that appears inside the weight $\omega_{b,k}$. 

Along a regular parametric submodel $Q_\varepsilon$ with score $S$, the pathwise derivative decomposes into two contributions,
\[
\left.\frac{d}{d\varepsilon}\widetilde\psi_{h,b}(Q_\varepsilon)\right|_{\varepsilon=0}
=
\underbrace{\left.\frac{d}{d\varepsilon}\widetilde\psi_{h,b}\!\left(Q_\varepsilon; \{m_k\}\right)\right|_{\varepsilon=0}}_{\text{vary $Q$ holding $\{m_k\}$ fixed}}
+
\underbrace{\sum_{k=1}^K
\left.\frac{\partial \widetilde\psi_{h,b}}{\partial m_k}\right|_{Q}\,
\left.\frac{d}{d\varepsilon}m_k(Q_\varepsilon)\right|_{\varepsilon=0}}_{\text{chain rule through $m_k$}} .
\]

We compute these two contributions separately. 

%%%%%%%%%%%%%%%%%%%%%%%%%%%%%%%%%%%%%%%%%%%%%%%%%%%%%%%%%%%%%%%%%%%%%%%%%%%%%%
\paragraph{Part (I): Derivative with respect to the conditional laws (holding $m_k$ fixed).}
%%%%%%%%%%%%%%%%%%%%%%%%%%%%%%%%%%%%%%%%%%%%%%%%%%%%%%%%%%%%%%%%%%%%%%%%%%%%%%

In this first step we treat the function $m \mapsto \omega_{b,k}(m\mid a_1,c)$ as fixed, so that $\mu_{b,k}(a_1,c)=\E\{Y\omega_{b,k}(M\mid a_1,c)\mid A=a_1,C=c\}$ is a standard regression functional; the dependence of the normalization in \eqref{eq:omega_def} on the conditional law is accounted for when we later add the chain-rule correction through $m_k$.

Let $Z_k \equiv Z_k(O) := Y\,\omega_{b,k}(M\,|\, a_1,C)$. Then $\mu_{b,k}(a_1,c)=\E(Z_k\,|\, A=a_1,C=c)$.

\medskip
For fixed $\omega_{b,k}$, the EIF for the regression functional $c\mapsto\E(Z_k\,|\, A=a_1,C=c)$ is
\begin{align}
D_{\mu_{b,k}}^{\mathrm{fix}}(O)
&=
\frac{\mathbb I(A=a_1)}{\pi(a_1\,|\, C)}
\Big\{
Z_k - \mu_{b,k}(a_1,C)
\Big\}. \nonumber 
\\
&=\frac{\mathbb I(A=a_1)}{\pi(a_1\,|\, C)}
\Big\{
Y\,\omega_{b,k}(M\,|\, a_1,C) - \mu_{b,k}(a_1,C)
\Big\}.
\label{eq:IF_mu_bk_fixed}
\end{align}

This corresponds to the derivative of $\widetilde\psi_{h,b}(Q)$ with respect to the conditional distribution argument in the multivariate chain rule. 

\medskip
For $g_k(a_0,c)=\E(\mathbb I(\widetilde M=k)\,|\, A=a_0,C=c)$, the EIF is
\begin{align}
D_{g_k}(O)
=
\frac{\mathbb I(A=a_0)}{\pi(a_0\,|\, C)}
\Big\{
\mathbb I(\widetilde M=k) - g_k(a_0,C)
\Big\}.
\label{eq:IF_gk}
\end{align}

\medskip
Since $\widetilde\theta_{h,b}(C)=\sum_k g_k(a_0,C)\mu_{b,k}(a_1,C)$ is a product of two regression functionals, its derivative with respect to the conditional laws follows from the usual product rule applied pointwise in $C$.
The EIF for $ \widetilde\psi_{h, b}(Q)=\E(\widetilde\theta_{h,b}(Q)(C))$ under fixed $\omega_{b, k}$ is
\begin{align}
 \widetilde\phi_{h, b}^\text{fixed}(Q)
&=
\sum_{k=1}^K g_k(a_0,C)\,D_{\mu_{b,k}}^{\mathrm{fix}}(O)
+
\sum_{k=1}^K \mu_{b,k}(a_1,C)\,D_{g_k}(O)
+
\Big\{\widetilde\theta_{h,b}(Q)(C)- \widetilde\psi_{h, b}(Q)\Big\}.
\label{eq:EIF_fixed_omega}
\end{align}
Substituting \eqref{eq:IF_mu_bk_fixed} and \eqref{eq:IF_gk} yields the explicit displayed formula.

%%%%%%%%%%%%%%%%%%%%%%%%%%%%%%%%%%%%%%%%%%%%%%%%%%%%%%%%%%%%%%%%%%%%%%%%%%%%%%
\paragraph{Part (II): Derivative with respect to the nuisance centers $m_k(a_0,c)$.}
%%%%%%%%%%%%%%%%%%%%%%%%%%%%%%%%%%%%%%%%%%%%%%%%%%%%%%%%%%%%%%%%%%%%%%%%%%%%%%

Now consider the actual weight in \eqref{eq:omega_def}, which depends on the unknown center
\[
m_k(a_0,c)=\E(M\,|\, A=a_0,C=c,\widetilde M=k)
\]
and on the normalizing denominator
\[
\E\!\left\{\mathcal K_b\!\left(M-m_k(a_0,c)\right)\,|\, A=a_1,C=c\right\}.
\]
This induces an additional EIF component through the pathwise derivative of $\mu_{b,k}(a_1,c)$ with respect to the center $m_k(a_0,c)$.

\medskip
The EIF for the conditional mean $m_k(a_0,C)=\E(M\,|\, A=a_0,C,\widetilde M=k)$ is
\begin{align}
D_{m_k}(O)
=
\frac{\mathbb I(A=a_0)\,\mathbb I(\widetilde M=k)}{\pi(a_0\,|\, C)\,g_k(a_0,C)}
\Big\{M-m_k(a_0,C)\Big\}.
\label{eq:IF_mk}
\end{align}

To compute the second term in the multivariate chain rule, we differentiate $\mu_{b,k}(a_1,c)$ with respect to its second argument $m_k(a_0,c)$ while holding the conditional law of $(Y,M)\mid(A=a_1,C=c)$ fixed.

Writing $\mu_{b,k}(a_1,c)=N_k(c)/D_k(c)$ with
\[
N_k(c)=\E\!\left\{Y\,\mathcal K_b\!\left(M-m_k(a_0,c)\right)\,|\, A=a_1,C=c\right\},
\quad
D_k(c)=\E\!\left\{\mathcal K_b\!\left(M-m_k(a_0,c)\right)\,|\, A=a_1,C=c\right\}, 
\]
and applying the quotient rule yields (assuming the usual regularity conditions that justify differentiation under the conditional expectation)
\begin{align}
\alpha_k(c) \coloneqq \frac{\partial}{\partial m}\mu_{b,k}(a_1,c)
=
-
\frac{
\E\!\left[\{Y-\mu_{b,k}(a_1,c)\}\,\mathcal K_b'\!\left(M-m_k(a_0,c)\right)\,|\, A=a_1,C=c\right]
}{
\E\!\left[\mathcal K_b\!\left(M-m_k(a_0,c)\right)\,|\, A=a_1,C=c\right]
}.
\label{eq:dk_mu_bk_dm}
\end{align}

\medskip
By the multivariate chain rule, the contribution of the nuisance $m_k$ to the EIF is 
\begin{align}
\widetilde \phi^\omega_{h, b}(Q)(O)
=
\sum_{k=1}^K g_k(a_0,C)\,\alpha_k(C)\,D_{m_k}(O).
\label{eq:EIF_extra_mk}
\end{align}
Substituting \eqref{eq:IF_mk} into \eqref{eq:EIF_extra_mk} simplifies this term to
\begin{align}
\widetilde \phi^\omega_{h, b}(Q)(O)
=
\sum_{k=1}^K
\alpha_k(C)\,
\frac{\mathbb I(A=a_0)\,\mathbb I(\widetilde M=k)}{\pi(a_0\,|\, C)}
\Big\{M-m_k(a_0,C)\Big\}.
\label{eq:EIF_extra_mk_simplified}
\end{align}

\medskip
Combining the derivative with respect to the conditional laws and the derivative with respect to the nuisance centers yields the full efficient influence function 
\begin{align}
\widetilde \phi_{h, b}(Q)(O)
=
\widetilde \phi_{h, b}^\text{fixed}(Q)(O)
+
\widetilde \phi^\omega_{h, b}(Q)(O), 
\label{eq:EIF_full}
\end{align}
where $\widetilde \phi_{h, b}^\text{fixed}(Q)(O)$ is given in \eqref{eq:EIF_fixed_omega} and $\widetilde \phi^\omega_{h, b}(Q)(O)$ is given in \eqref{eq:EIF_extra_mk_simplified}.

%%%%%%%%%%%%%%%%%%%%%%%%%%%%%%%%%%%%%%%%%%%%%%%%%%%%%%%%%%%%
\section{Details in simulations}
\label{app:simulations}

% -------------------------------------
\subsection{Misspecification scenarios}
\label{app:simu_misspecification_cases}

In this section, we present a detailed explanation of the misspecification scenarios implemented in Simulation~\#2. For each of the estimators, the corresponding collection of nuisance functions $Q$ is given as follows: (i) $\psi_h(\widehat{Q})$: $\{\mu_k, g_k\}$; (ii) $\widetilde\psi_h(\widehat{Q})$: $\{\mu, m_k, g_k\}$; (iii) $\psi_h^{+}(\widehat{Q})$: $\{\mu_k, g_k, \pi\}$; (iv) $\widetilde\psi^{+}_{h_1}(\widehat{Q})$: $\{\mu, m_k, g_k, \pi\}$;
(v) $\widetilde\psi^{+}_{h_2}(\widehat{Q})$: $\{\mu, m_k, g_k, g, \pi\}$.

Table~\ref{tab:misspec_sim2} presents the specification and misspecification configurations of the nuisance functions.  The plug-in estimator $\psi_h(\widehat{Q})$ coincides with the specification under Condition~3 as well as the fully correctly specified setting; the same holds for $\widetilde\psi_h(\widehat{Q})$. Under the DGP described in Section~\ref{sec:sim}, correctly specified nuisance functions can be consistently estimated using GLMs. Misspecification is introduced by constructing models with nonlinear and irrelevant functional forms, including exponential, cosine, inverse, and higher-order polynomial terms.

% \ding{51}  % ✓ check mark
% \ding{55}   % ✗ cross mark

\begin{table}[t]
    \centering
    \caption{Misspecification scenarios for nuisance functions in Simulation \#2. 
A check mark (\ding{51}) indicates correct specification,  while a cross mark (\ding{55}) denotes model misspecification.}
\label{tab:misspec_sim2}
    \begin{tabular}{ l c c c c c }
    \toprule
        Function & Condition 1 & Condition 2 & Condition 3 & Correct & False \\ 
    \midrule 
        $m_k$   & \ding{51} & \ding{51} & \ding{51} & \ding{51} & \ding{55} \\ 
        $\mu$   & \ding{55} & \ding{51} & \ding{51} & \ding{51} & \ding{55} \\ 
        $\mu_k$ & \ding{55} & \ding{51} & \ding{51} & \ding{51} & \ding{55} \\ 
        $g$     & \ding{51} & \ding{55} & \ding{51} & \ding{51} & \ding{55} \\ 
        $g_k$   & \ding{51} & \ding{55} & \ding{51} & \ding{51} & \ding{55} \\
        $\pi$   & \ding{51} & \ding{51} & \ding{55} & \ding{51} & \ding{55} \\ 
    \bottomrule
    \end{tabular}
\end{table}

% ------------------------------------
\subsection{The performance in small sample sizes}
\label{app:simu_small_n}

\begin{figure}[t]
\centering
\includegraphics[width=.8\linewidth]{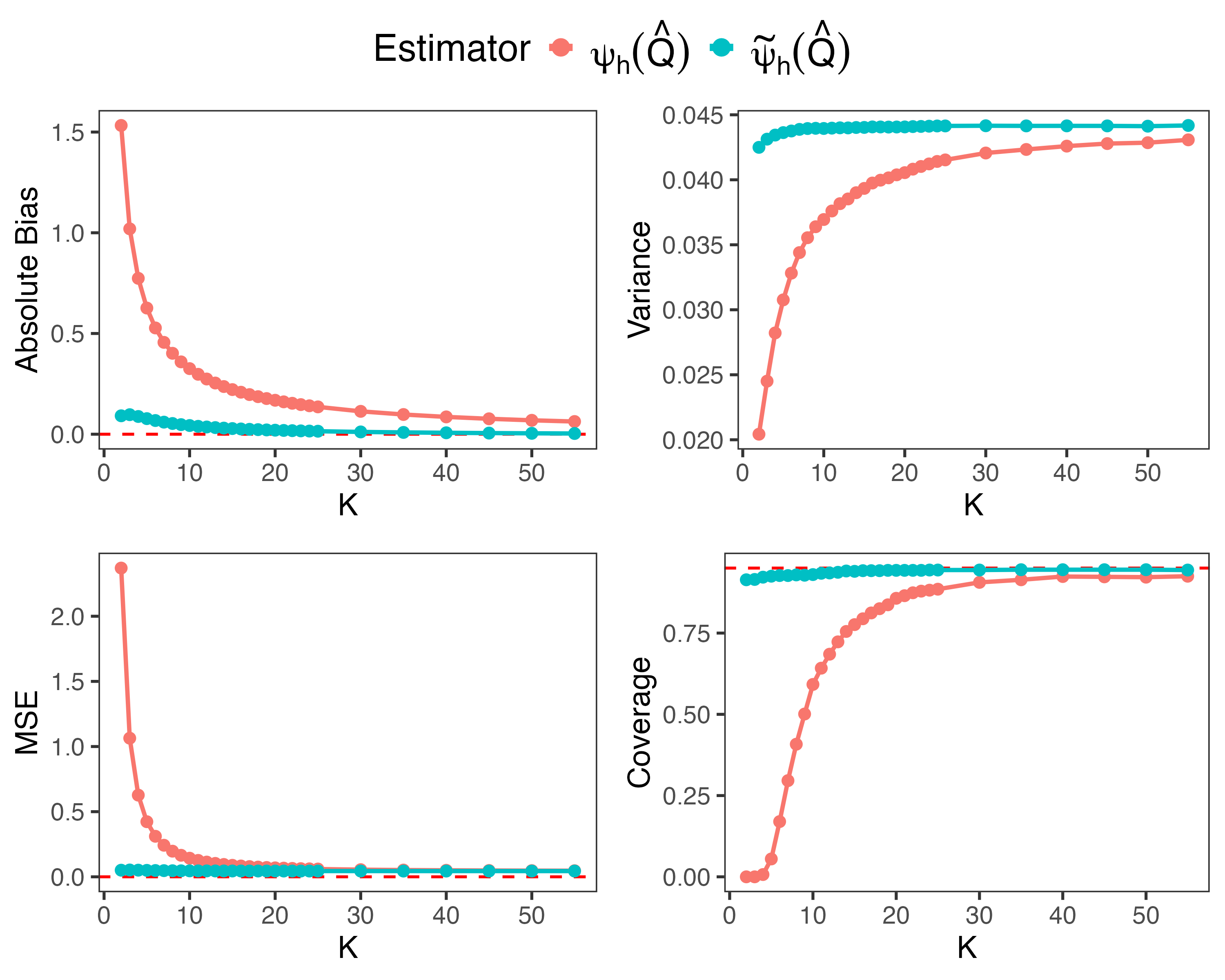}
\caption{The performance of the coarsened plug-in estimator $\psi_h(\widehat{Q})$ and the debiased estimator $\widetilde{\psi}_h(\widehat{Q})$ as functions of the number of mediator bins $K$ for sample size $n=500$.}
\label{fig:plugin_n500_freq}
\end{figure}

\begin{figure}[h]
\centering
\includegraphics[width=1\linewidth]{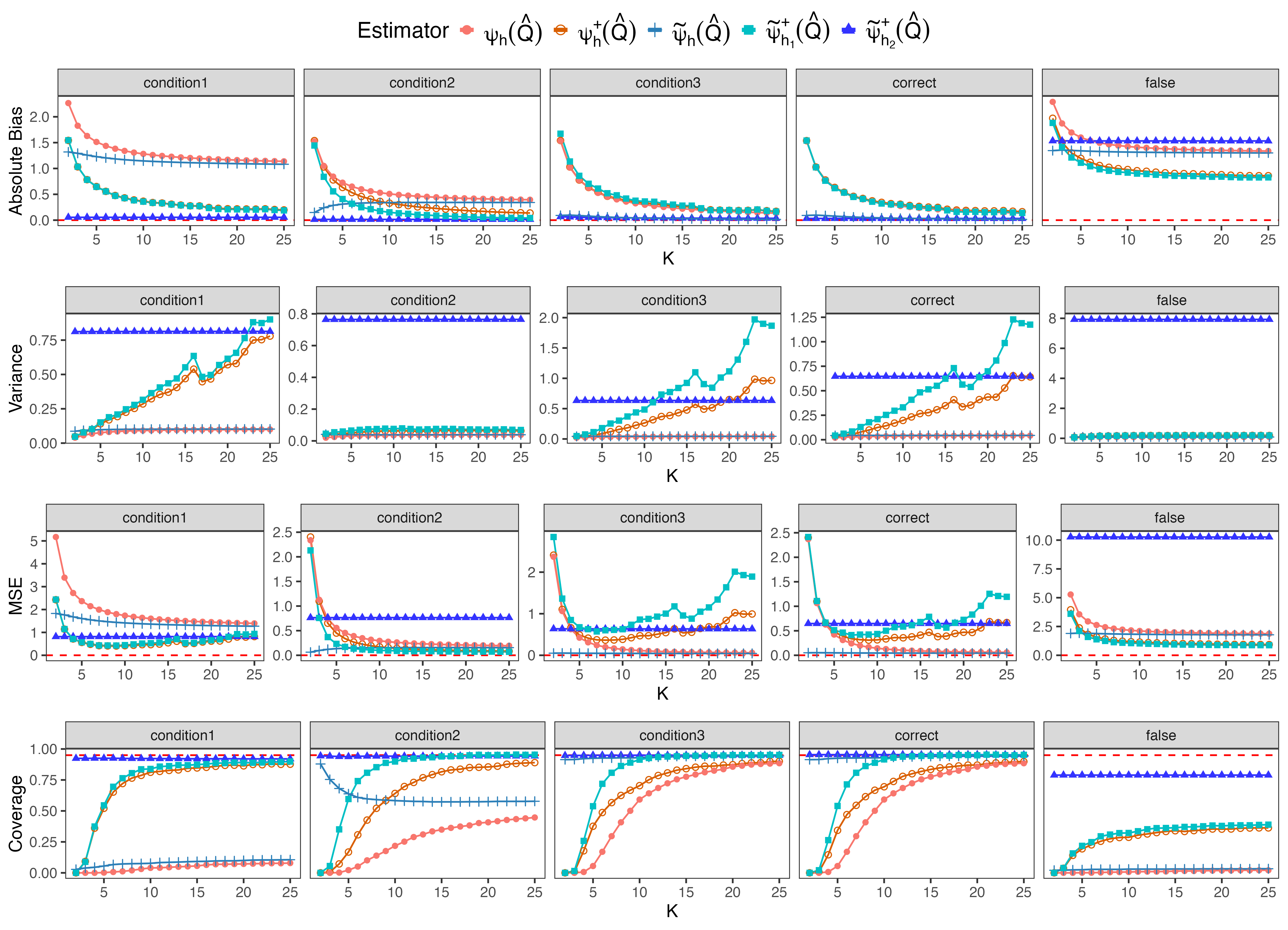}
\caption{Comparison of plug-in and one-step estimators under nuisance-model misspecification for $n=500$. ``Correct'' indicates all nuisance models are correctly specified, whereas ``False'' indicates all nuisance models are misspecified. Conditions 1–3 correspond to misspecification of $\{\mu, \mu_k\}$, $\{g, g_k\}$, and $\pi$, respectively. Performance is summarized in terms of bias, variance, MSE, and 95\% confidence interval coverage.}
\label{fig:all_n500_freq}
\end{figure}

Figure~\ref{fig:plugin_n500_freq} summarizes absolute bias, variance, MSE, and coverage of 95\% CIs as functions of $K$, and Figure~\ref{fig:all_n500_freq} presents results under correct specification and a range of misspecification scenarios for nuisance models in a smaller sample size ($n = 500$). The results are consistent with what we find in Simulation 2, where the debiased plug-in estimator $\widetilde \psi_h(\widehat{Q})$ shows substantial improvement compared with the coarsened estimator $\psi_h (\widehat Q)$. The one-step estimator $\psi_h^{+}(\widehat Q)$ improves bias and coverage relative to $\psi_h(\widehat Q)$, but does not eliminate discretization bias. The first derivative estimator $\widetilde \psi^{+}_{h_1}(\widehat Q)$ behaves similarly and does not meaningfully reduce discretization bias. In contrast, $\widetilde \psi^{+}_{h_2}(\widehat{Q})$ achieves  low bias and near-nominal coverage across scenarios. Thus, this simulation confirms the strong performance of our proposed debiased plug-in estimator even in small samples and supports the conclusions drawn in the main text.

% ------------------------------------
\subsection{Comparison of different discretization schemes}
\label{app:simu_discretization}

This section provides additional results for discretization schemes other than the equal-frequency discretization used in the main simulation analysis. 

\begin{figure*}[t]
\centering
\begin{subfigure}[t]{0.48\textwidth}
    \centering
    \includegraphics[width=\linewidth]{pics/compare_mk_K2_K6_width.png}
    \caption{Within-bin differences.}
    \label{fig:within-bin-means_width}
\end{subfigure}
\hfill
\begin{subfigure}[t]{0.48\textwidth}
    \centering
    \includegraphics[width=\linewidth]{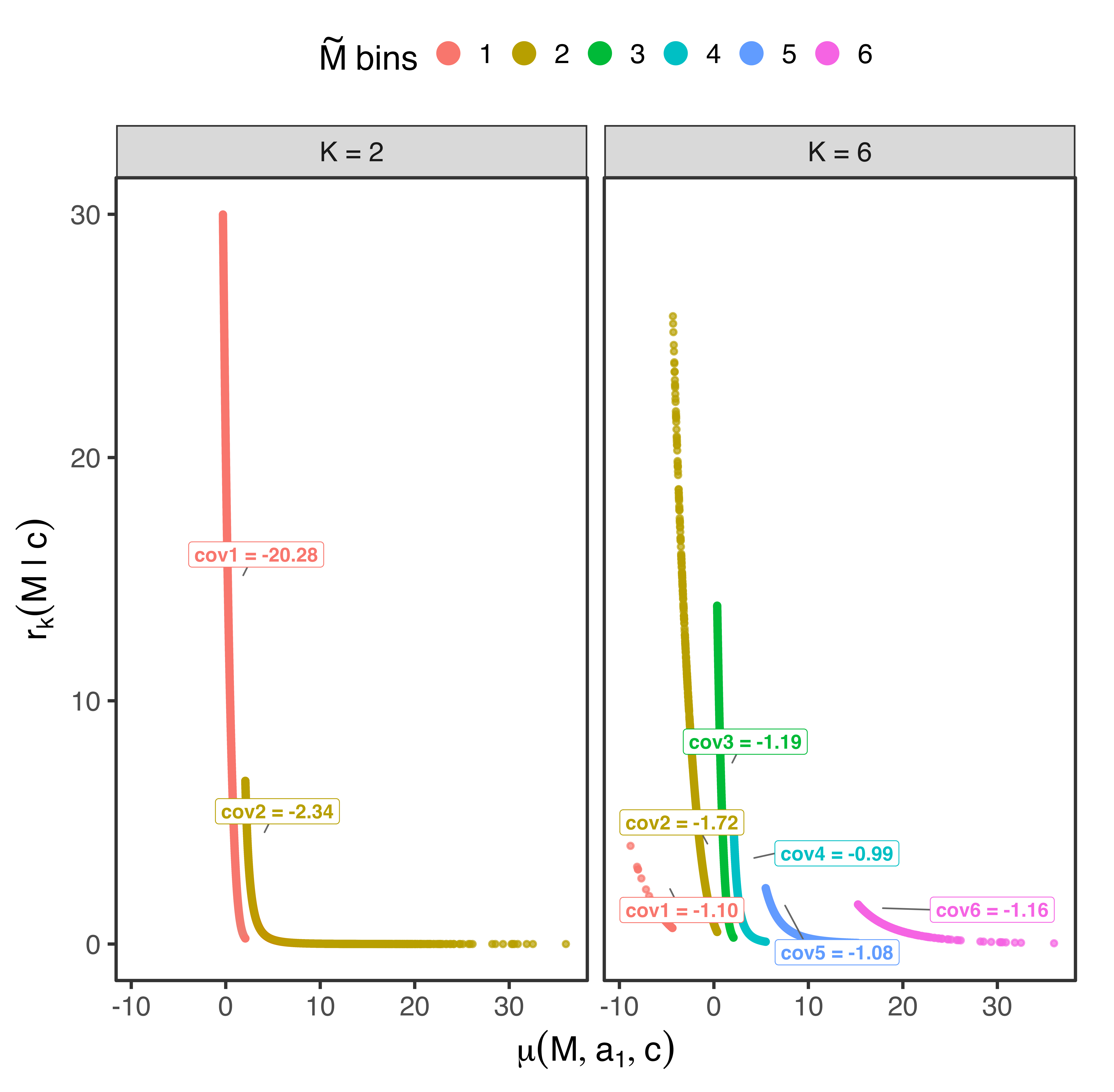}
    \caption{Within-bin covariance.}
    \label{fig::within-bin-covariance_width}
\end{subfigure}
\caption{Theoretical quantities approximation in large sample size. (a) Within-bin differences $m_k(a_1,c)-m_k(a_0,c)$ under $K=2$ (red, upper axis) and $K=6$ (blue, lower axis) equal-width discretizations. Each segment represents one mediator bin; panels correspond to values of $C$, and (b) Within-bin covariance between $\mu(M,a_1,C)$ and $r_k(M \,|\, C)$ at $C=0$. }
\label{fig:M_density_and_bin_difference}
\end{figure*}

We first examine the population-level coarsening error under equal-width discretization using the same large simulated sample as in Simulation \#1. The observed maximum and minimum of $M$ are 6.66 and $-5.23$, respectively, giving an empirical range of 11.89 and an equal-width bin length of $11.89/K$. For $K=2$, this yields the bins $\mathcal{B}_1=[-5.23,0.72]$ and $\mathcal{B}_2=(0.72,6.66]$. For $K=6$, the bins $\mathcal{B}_1,\ldots,\mathcal{B}_6$ are $[-5.23,-3.25]$, $(-3.25,-1.27]$, $(-1.27,0.72]$, $(0.72,2.70]$, $(2.70,4.68]$, and $(4.68,6.66]$. Figure~\ref{fig:within-bin-means_width} plots the within-bin shifts $m_k(a_1,c) - m_k(a_0,c)$ for these two discretization levels. Increasing $K$ reduces these shifts across bins and covariate strata, thereby attenuating the dominant first-order component of the discretization bias. Figure~\ref{fig::within-bin-covariance_width} illustrates that, at $C=0$, with finer discretization, the covariance magnitude drops substantially. These patterns are consistent with the main simulation results.

We then mainly focus on finite-sample estimator performance. Under the same DGP as in Section~\ref{sec:sim}, we compare the coarsened plug-in estimator $\psi_h(\widehat Q)$ with the proposed debiased plug-in estimator $\widetilde \psi_h(\widehat Q)$ using equal-width and predefined-width discretizations, and discuss the resulting patterns in relation to the equal-frequency results.

\begin{figure}[t]
\centering
\begin{subfigure}{0.49\textwidth}
    \centering
    \includegraphics[width=1\linewidth]{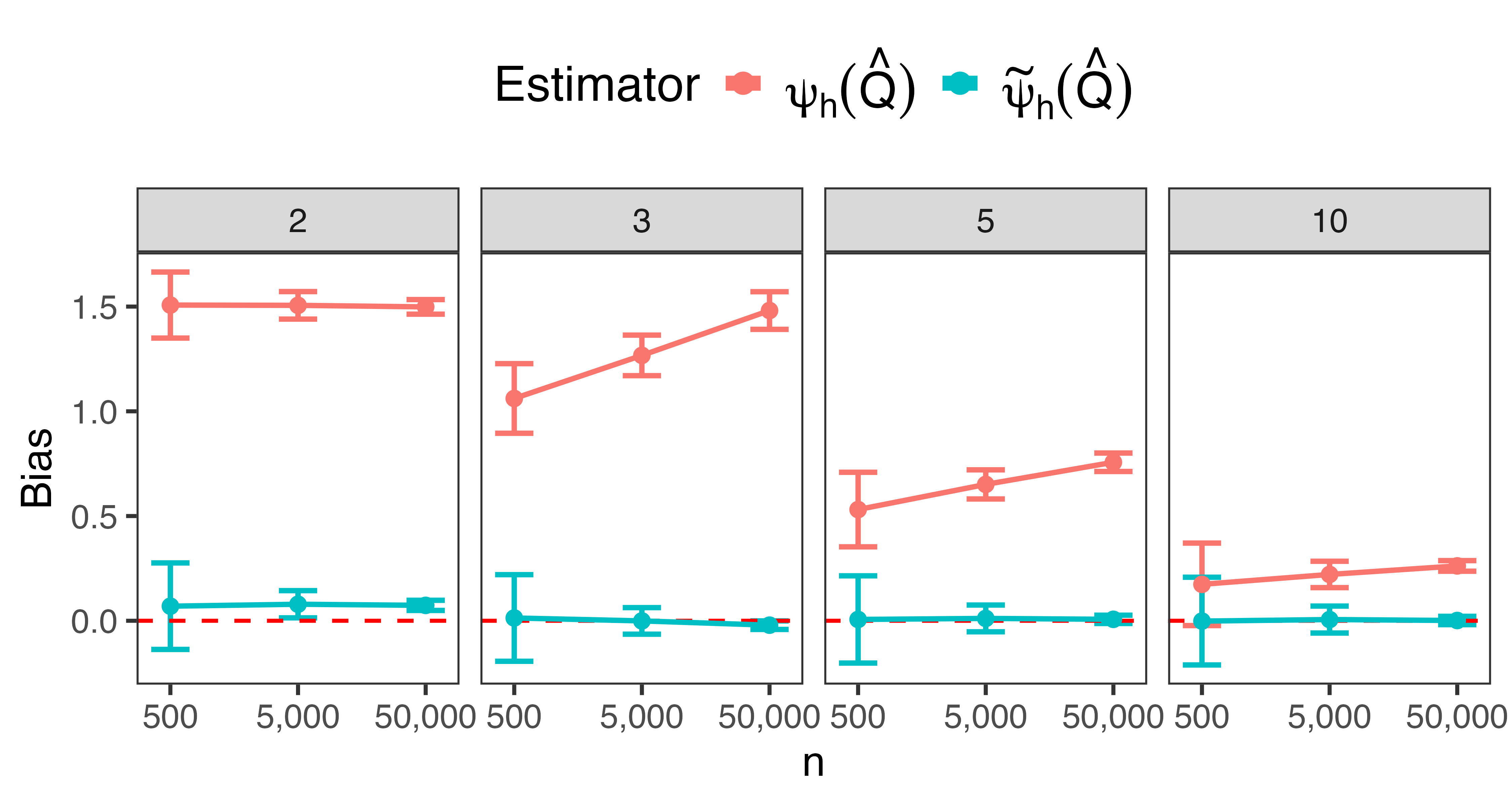}
    \caption{Equal width}
    \label{fig:trend_n_width}
\end{subfigure}
\hfill
\begin{subfigure}{0.49\textwidth}
    \centering
    \includegraphics[width=1\linewidth]{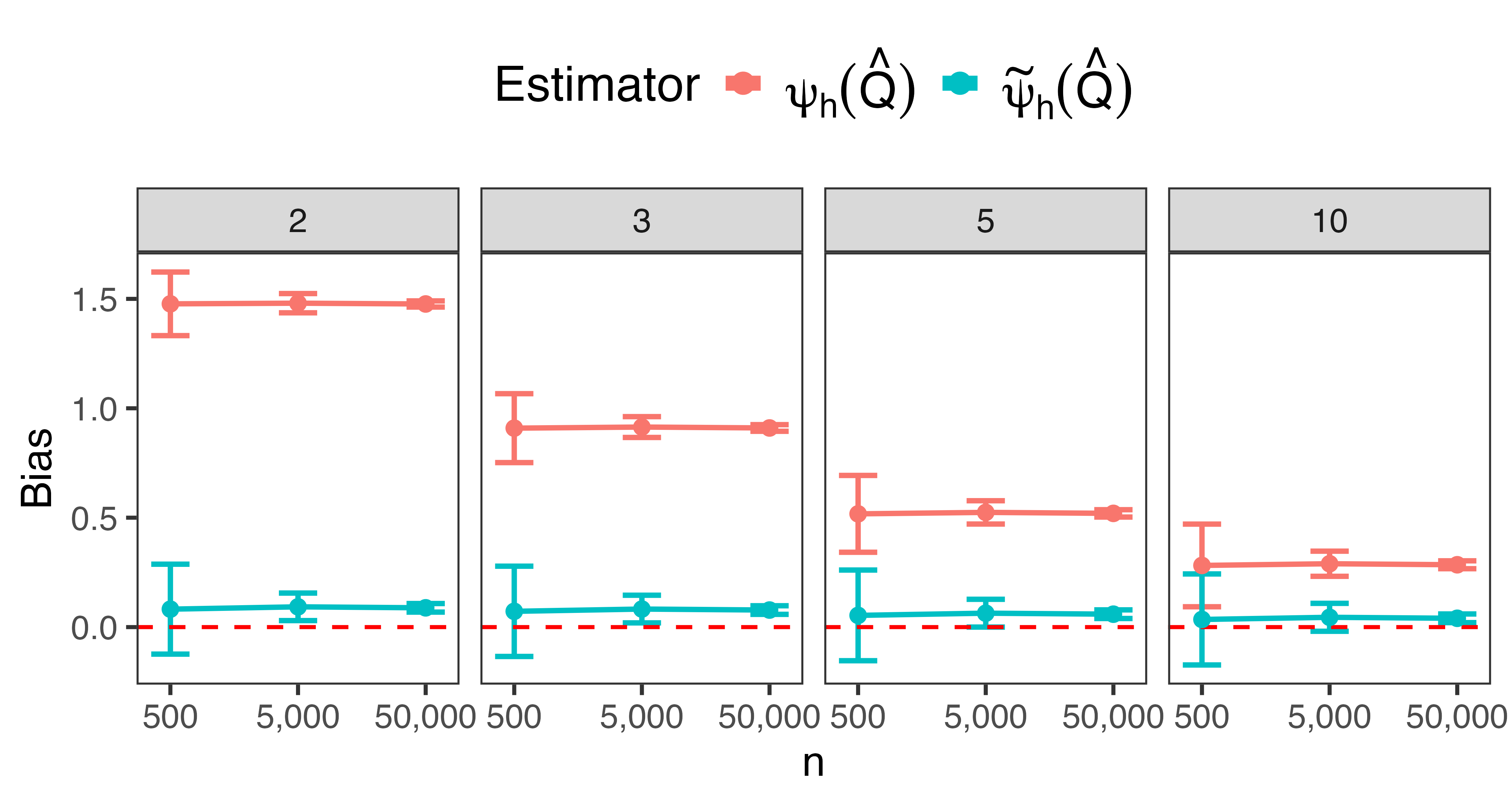}
    \caption{Predefined width}
    \label{fig:trend_n_fixed}
\end{subfigure}
\caption{Bias trends of the coarsened plug-in estimator $\psi_h(\widehat Q)$ and the debiased plug-in estimator $\widetilde\psi_h(\widehat Q)$ as the sample size $n$ increases under empirical equal-width and predefined-width discretizations.}
\label{fig:trend_n_discretization}
\end{figure}

\begin{figure}[!ht]
\centering
\begin{subfigure}{0.48\textwidth}
    \centering
    \includegraphics[width=1\linewidth]{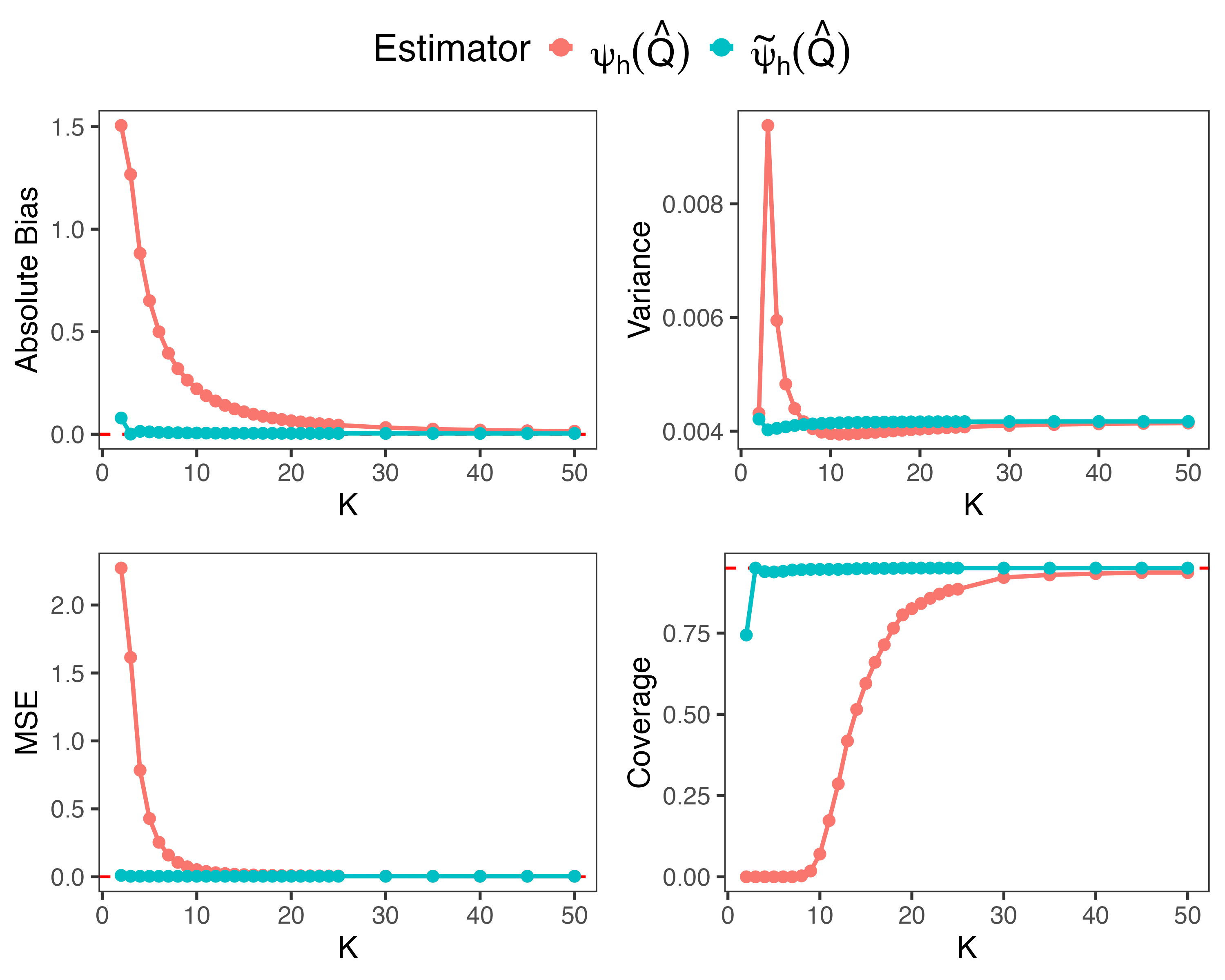}
    \caption{Equal width}
    \label{fig:trend_K_width}
\end{subfigure}
\hfill
\begin{subfigure}{0.48\textwidth}
    \centering
    \includegraphics[width=1\linewidth]{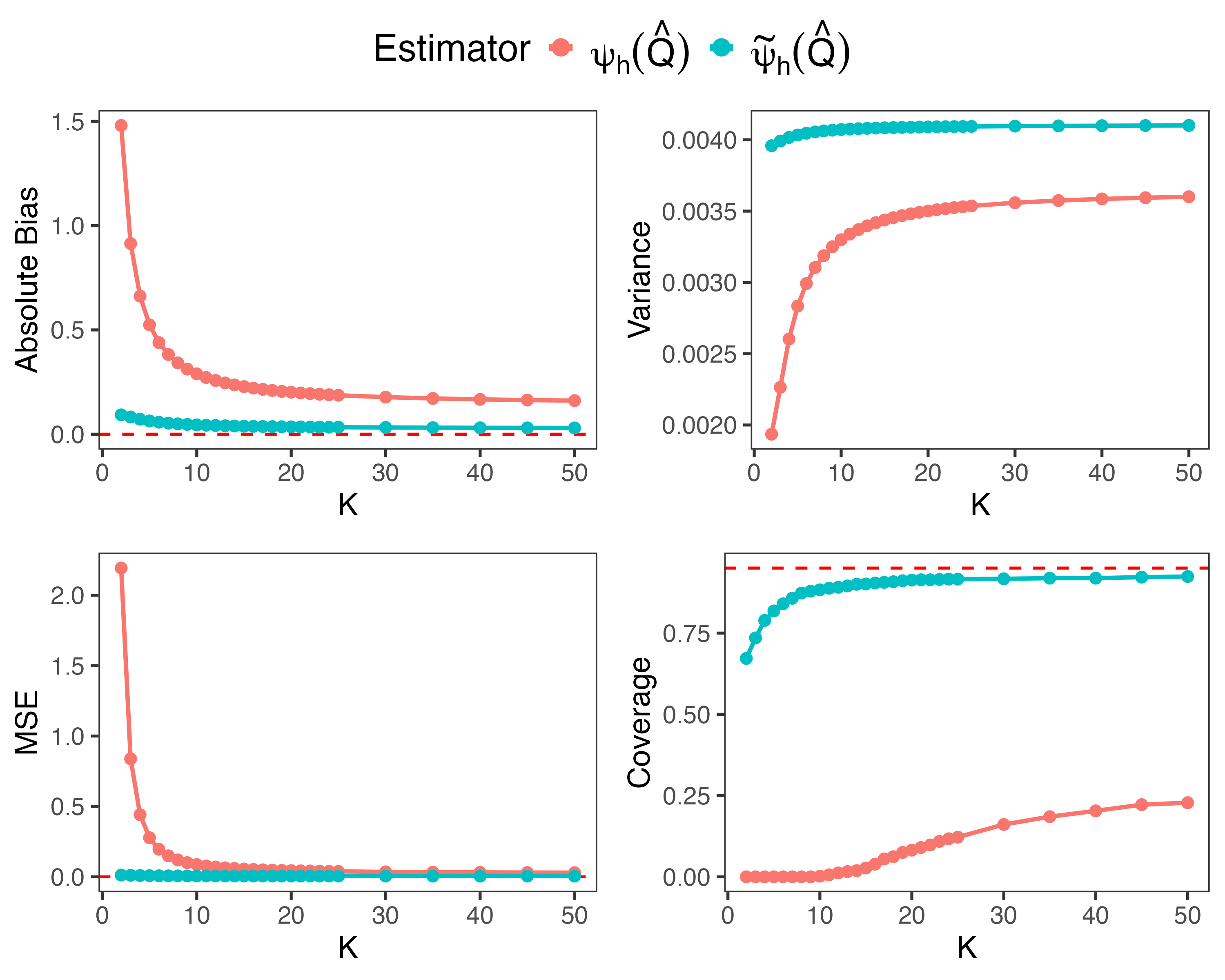}
    \caption{Predefined width}
    \label{fig:trend_K_fixed}
\end{subfigure}
\caption{Performance of the coarsened plug-in estimator $\psi_h(\widehat{Q})$ and the debiased estimator $\widetilde{\psi}_h(\widehat{Q})$ as functions of the number of mediator bins $K$ for sample size $n=5{,}000$ under different discretization schemes.}
\label{fig:trend_K_discretization}
\end{figure}

\textbf{Equal-frequency discretization.} The cut points are computed from the observed quantiles of $M$ in each simulated dataset, so each bin contains approximately the same number of observations. This rule makes the empirical bin probabilities stable by construction and makes the interior cut points converge to fixed population quantiles as $n$ increases. Table~\ref{tab:break_points_freq} illustrates the stability of the equal-frequency interior cut points. Consequently, for fixed $K$, the coarsening bias remains nearly unchanged across the sample sizes in Figure~\ref{fig:trends_n_freq}. As $K$ increases, the bins become narrower and the coarsened functional becomes more local. The within-bin differences in conditional mediator means, $m_k(a_1,c)-m_k(a_0,c)$, therefore tend to shrink, reducing the leading coarsening bias in $\psi_h(\widehat Q)$. The proposed debiased estimator further removes the first-order bias within each bin, leading to substantial bias reduction.

% The variance plots in Figure~\ref{fig:plugin} show that the variance increases with $K$ for both estimators, reflecting the usual \textit{bias-variance trade-off} induced by finer discretization. As $K$ increases, bins contain fewer observations, so bin-specific nuisance quantities such as $\widehat g_k(a,c)$, $\widehat m_k(a,c)$, and $\widehat\mu_k(a,c)$ are estimated with greater variability. The debiased estimator $\widetilde\psi_h(\widehat Q)$ generally has larger variance than the coarsened estimator $\psi_h(\widehat Q)$ because it achieves bias reduction through a more local construction. Specifically, $\widetilde\psi_h(\widehat Q)$ evaluates the outcome regression at the estimated within-bin conditional mean, $\widehat\mu(\widehat m_k(a_0,c),a_1,c)$. This removes the leading discretization bias, but also introduces additional variability from estimating $\widehat m_k(a_0,c)$ and propagating this uncertainty through the nonlinear outcome regression $\widehat\mu(m,a_1,c)$. As $K$ increases, the remaining coarsening bias becomes smaller and the correction is less pronounced; accordingly, the variance gap narrows between the two estimators.

The variance plots in Figure~\ref{fig:plugin} exhibit the usual bias-variance trade-off induced by finer discretization. As $K$ increases, bins contain fewer observations, so bin-specific quantities such as $\widehat g_k(a,c)$, $\widehat m_k(a,c)$, and $\widehat\mu_k(a,c)$ are estimated with greater variability, leading to increased variance for both estimators.
The debiased estimator $\widetilde\psi_h(\widehat Q)$ generally exhibits slightly larger variance for small $K$ because it relies on a more localized representation of the outcome regression, evaluating $\widehat\mu$ at the within-bin mean $\widehat m_k(a_0,c)$ rather than averaging outcomes over the entire bin. This localization removes the leading coarsening bias but provides less smoothing and therefore can increase variability.
As $K$ increases, the bins become narrower and $\mu_k(a_1,c)\approx \mu\big(m_k(a_0,c),a_1,c\big)$, so the two estimands become increasingly similar and the variance gap correspondingly disappears.

\begin{figure*}[h]
\centering
\begin{subfigure}[t]{0.30\textwidth}
    \centering
    \includegraphics[width=\linewidth]{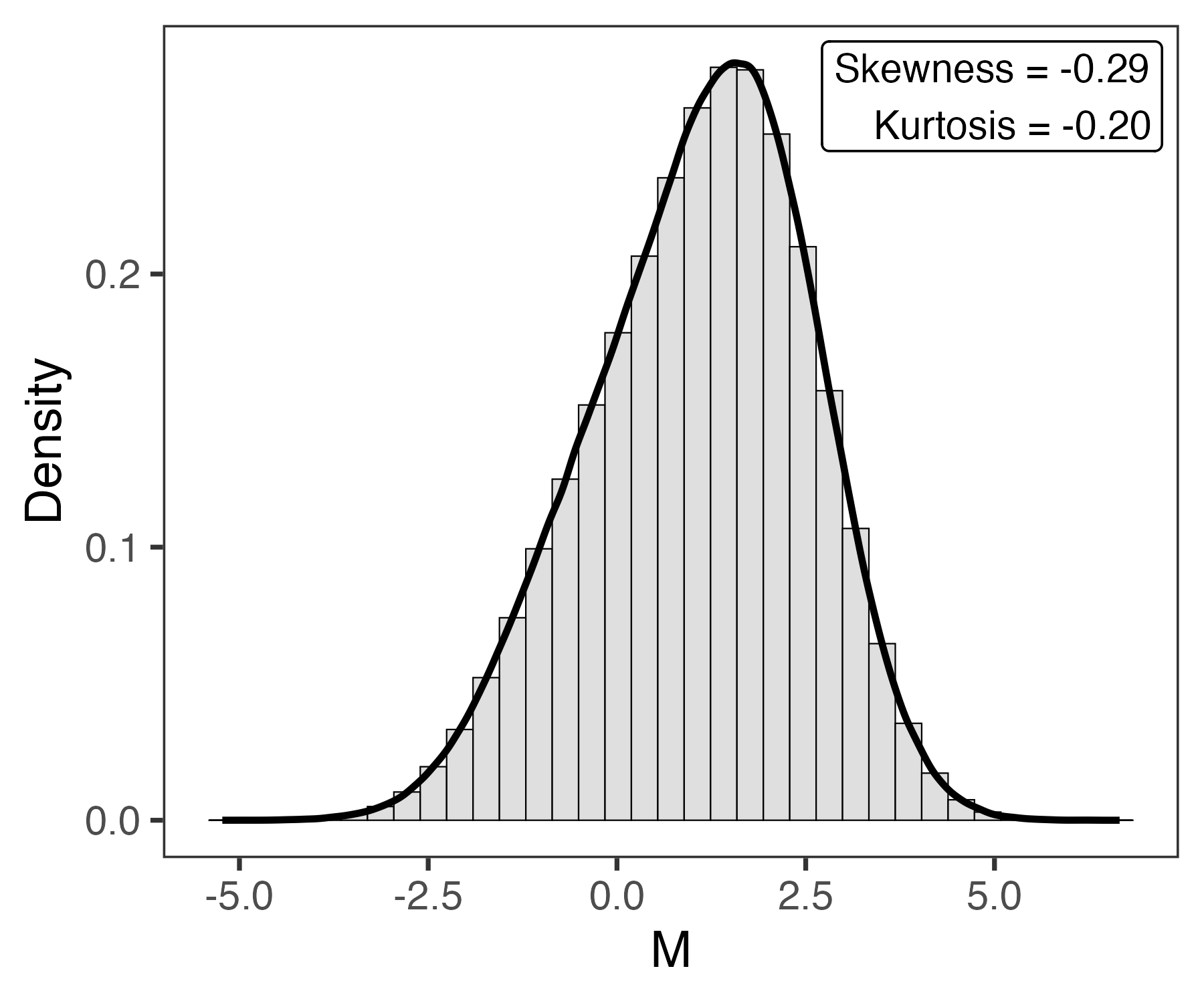}
    \caption{Empirical distribution of $M$.}
    \label{fig:M_hist_density}
\end{subfigure}
\hfill
\begin{subfigure}[t]{0.33\textwidth}
    \centering
    \includegraphics[width=\linewidth]{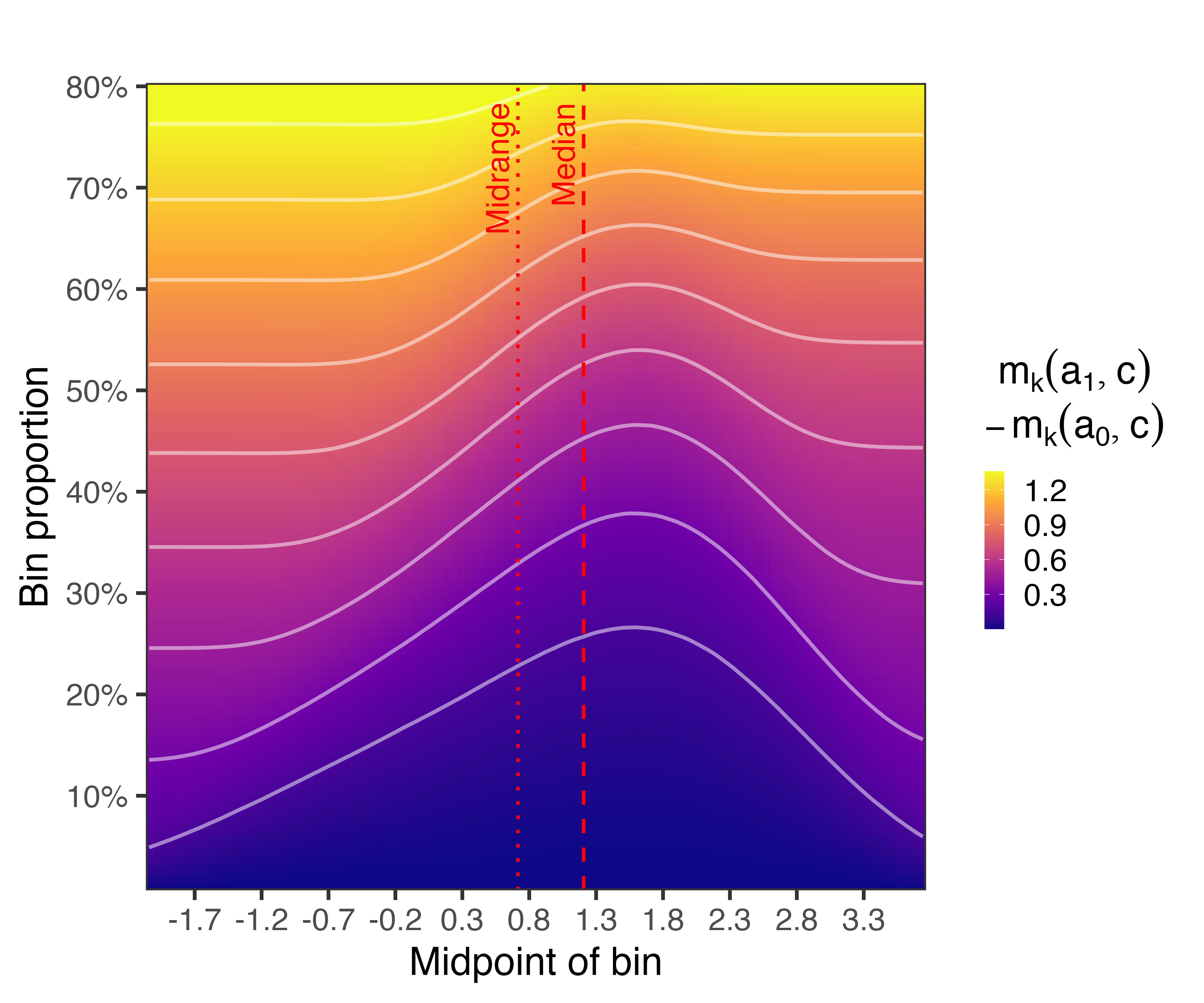}
    \caption{$m_k(a_1,c)-m_k(a_0,c)$.}
    \label{fig:bin_check_mk_difference}
\end{subfigure}
\hfill
\begin{subfigure}[t]{0.33\textwidth}
    \centering
    \includegraphics[width=\linewidth]{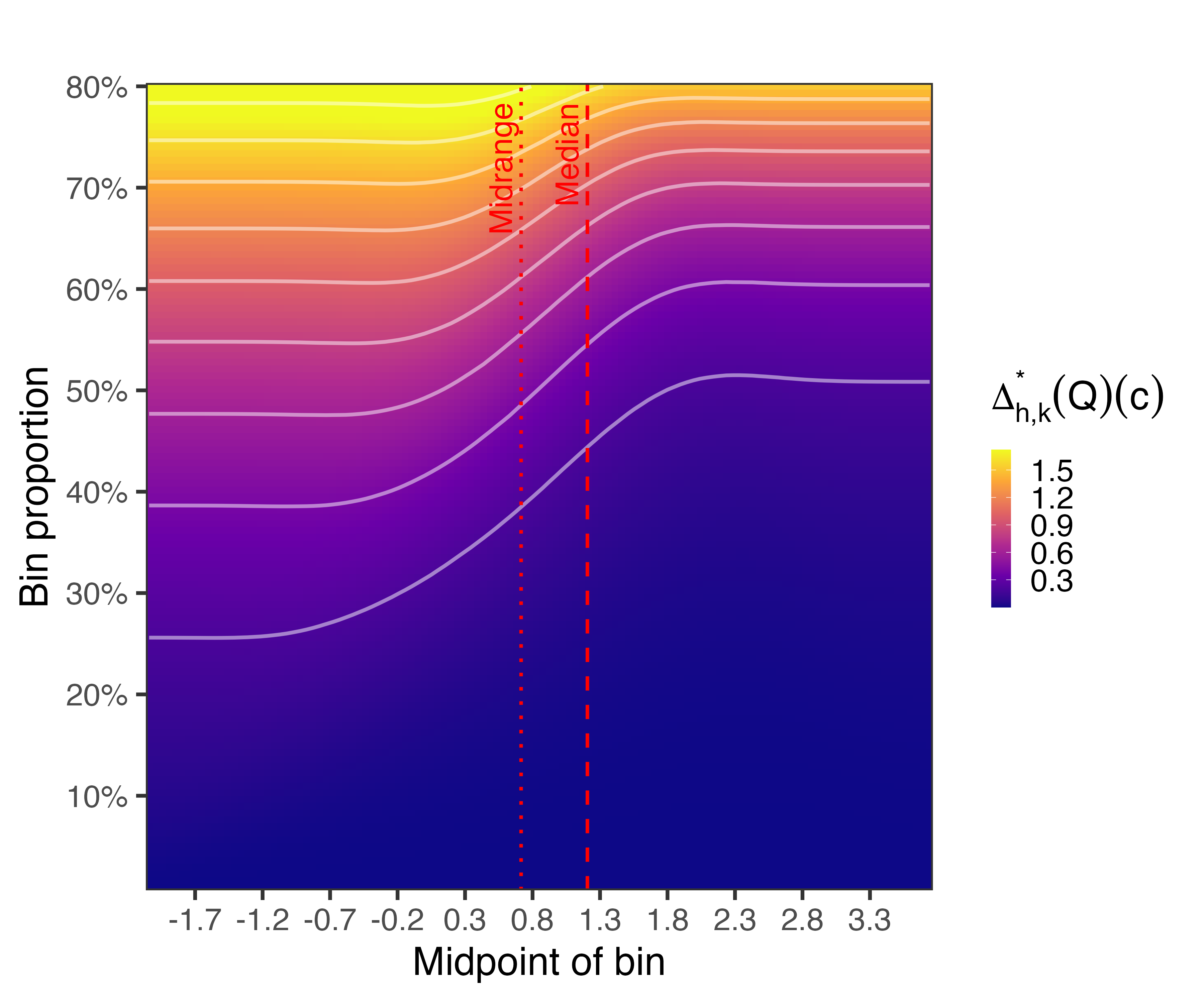}
    \caption{$\Delta^*_{h,k}(Q)(c)$.}
    \label{fig:bin_check_Delta_k}
\end{subfigure}
\caption{Diagnostic plots for mediator bins (covariate $C{=}0$). Panel (a) shows the empirical distribution of the mediator $M$ in a large simulated sample. Panel (b) shows the within-bin difference in conditional mediator means, $m_k(a_1,c)-m_k(a_0,c)$. Panel (c) shows the first-order coarsening-bias component, $\Delta^*_{h,k}(Q)(c)=\mu_m'(m_k(a_1,c),a_1,c)\{m_k(a_1,c)-m_k(a_0,c)\}g_k(a_0,c)$. Panels (b) and (c) are evaluated from the theoretical functions: for a bin with a given proportion on the $y$-axis, the color represents the corresponding value of the target quantity as the bin midpoint varies along the $x$-axis. The dotted line represents the midpoint of the range and the dashed line represents the median point.}
\label{fig:bin_check_diagnostics}
\end{figure*}

\textbf{Equal-width discretization.} The cut points are computed from the observed minimum and maximum (the observed range) of $M$ in each simulated dataset. Figure~\ref{fig:trend_n_width} shows that, for fixed $K$ except $K=2$, the bias of the coarsened estimator $\psi_h(\widehat Q)$ can increase with $n$. This behavior is driven by the unbounded support of the mediator. As the sample size grows, larger samples are more likely to contain extreme tail observations, so the empirical range of $M$ expands. Consequently, for fixed $K$, the equal-width bin length also increases. As shown in Table~\ref{tab:equal_width_range_prop}, the average empirical range increases from 8.11 at $n=500$ to 10.78 at $n=50{,}000$. Because $M$ is unimodally distributed, as illustrated in Figure~\ref{fig:M_hist_density}, this range expansion affects central and tail bins differently. Wider central bins can cover a larger high-density region of $M$ and therefore contain a larger proportion of observations, whereas tail bins will be more sparse and may shift farther into the tails. Table~\ref{tab:equal_width_range_prop} shows that the largest bin proportion among all bins tends to increase with $n$, especially for small $K$; this effect becomes weaker as $K$ increases. The case $K=2$ is an exception because there is only one interior cut point. Since the distribution of $M$ is only mildly skewed, with skewness $-0.29$, the split is relatively stable across $n$ and behaves more similarly to a coarse quantile-based discretization.

To clarify how wider empirical bins affect the coarsening error, we examine the corresponding population-level DGP quantities. As discussed in the main text, the first-order coarsening term in \eqref{eq:first_order_scaling} is driven by within-bin differences in conditional mediator means. In our DGP, the conditional mean of $M$ given $A=a_1$ and $C=c$ is uniformly larger than the conditional mean of $M$ given $A=a_0$ and $C=c$ across the relevant values of $c$. Therefore, the qualitative pattern is similar across covariate strata, and we use $C=0$ as an illustrative example. Figure~\ref{fig:bin_check_mk_difference} shows that, when the midpoint of a bin is fixed, a bin that contains a larger proportion of the mediator distribution, equivalently a wider bin under this construction, tends to exhibit a larger within-bin difference in conditional mediator means. This increase is more pronounced when the bin midpoint is close to the median. For example, when $K=3$, the midpoint of the central bin is located near the midpoint of the empirical range. As the central-bin proportion increases from 60.0\% to 74.5\%, the contrast $m_k(a_1,c)-m_k(a_0,c)$ also increases. By contrast, the tail bins contain less probability mass and move farther into the tails, with their midpoints shifting toward the extremes. As a result, their contributions to the within-bin mean difference are downweighted. Since the curves in Figure~\ref{fig:bin_check_mk_difference} are steeper in the central region, changes in the central bin dominate the overall error across bins. Moreover, the leading bias is not determined by $m_k(a_1,c)-m_k(a_0,c)$ alone; it also depends on the derivative of the outcome regression and the bin probability. We therefore examine $\Delta^*_{h,k}(Q)(c) =  \mu_m'\{m_k(a_1,c),a_1,c\} \{m_k(a_1,c)-m_k(a_0,c)\}g_k(a_0,c)$,  where $\Delta^*_{h}(Q)(c) = \sum_{k=1}^K \Delta^*_{h,k}(Q)(c)$ and $\Delta^*_{h}(Q)(c) \approx \Delta_{h}(Q)(c)$. The diagnostic in Figure~\ref{fig:bin_check_Delta_k} confirms that the first-order coarsening component is mainly influenced by the central bins. It also shows that, as $K$ increases, changes in bin proportions become smaller, and the corresponding changes in $\Delta^*_{h}(Q)(c)$ are attenuated. For example, when $K=10$, $\Delta^*_{h}(Q)(c)$ changes only modestly as the central-bin proportion increases from 22.4\% to 28.3\%. This behavior of the population-level DGP quantities explains the increasing bias trend of the coarsened estimator $\psi_h(\widehat Q)$ shown in Figure~\ref{fig:trend_n_width}.

Figure~\ref{fig:trend_K_width} further shows that the variance of $\psi_h(\widehat Q)$ under equal-width discretization can be nonmonotone in $K$. This behavior can be understood by decomposing the variability into two components: the usual sampling variability, as in equal-frequency discretization, and the additional variability induced by the random partition. Under empirical equal-width discretization, the partition is itself random because small changes in the sample range can move the interior cut points. These changes alter both the bin probabilities and the within-bin contrasts that enter the coarsened target. For larger $K$, the bins become more local and the sensitivity of the coarsened functional to any single boundary is reduced. In contrast, the debiased estimator $\widetilde \psi_h(\widehat Q)$ is substantially less affected by this random-partition behavior because it removes the leading first-order coarsening component within each bin. As $K$ increases, the variance pattern of $\psi_h(\widehat Q)$ becomes closer to the pattern observed under equal-frequency, where finer partitions reduce smoothing and tend to increase variance more gradually.

\textbf{Predefined-width discretization.} The cut points are fixed across simulated datasets. For each $K$, we partition the interval $[-1.5,3.3]$ into $K$ subintervals of equal length and set the outer boundary cut points to the observed minimum ($\ell$) and maximum values ($u$), so that observations outside the fixed interior interval are included in the first and last bins. For example, when $K = 3$, the bins are $[\ell, 0.1]$, $(0.1, 1.7]$, and $(1.7, u]$. Thus, for a given $K$, the same cut points are used in every Monte Carlo replication.

Because the interior bin widths do not depend on the empirical range, the bias of $\psi_h(\widehat Q)$ remains stable across $n$ for fixed $K$, as shown in Figure~\ref{fig:trend_n_fixed}. This pattern is similar to that observed under equal-frequency discretization. However, although the bias decreases as $K$ increases, it converges to a nonzero value. This occurs because increasing $K$ refines the partition only within the fixed interval $[-1.5,3.3]$, while the two tail bins always contain the regions $\leq -1.5$ and $\geq 3.3$. Once the central-region discretization bias is reduced, the remaining approximation error may be dominated by these tail bins. This explains why residual tail-induced bias can persist under predefined-width discretization even for large $K$.

\textbf{Choice of discretization scheme and discretization levels.}
These results highlight complementary practical considerations for choosing a discretization scheme. Equal-frequency discretization balances empirical bin masses and is often stable in finite samples, since the cut points converge to population quantiles. Empirical equal-width discretization is closely aligned with the bounded-support statement in the main paper, but it can be sensitive to sample extremes when the mediator has unbounded support, because the empirical range may expand with the sample size. Predefined-width discretization is useful when categories are fixed by scientific or clinical convention, as in settings where thresholds are externally defined (e.g. BMI) and should not vary across studies. However, this approach lacks the flexibility to adapt the choice of $K$ to a specific dataset. Moreover, if the mediator has non-negligible probability mass within coarse fixed intervals, such bins may induce residual coarsening bias. In such settings, the proposed debiased estimator $\widetilde{\psi}_h(\widehat Q)$ is particularly useful, since it reduces sensitivity to the chosen discretization scheme and substantially reduces bias for small $K$, while preserving the computational simplicity of discretization.

In practice, in addition to the choice of discretization scheme, the choice of the number of discretization levels $K$ is also an important consideration. From a theoretical perspective, increasing $K$ reduces approximation bias, but it may increase estimation variance and computational burden, particularly when some bins contain few observations. An implication of our results is that the proposed debiased estimator substantially reduces bias relative to the naive coarsened estimator even for small $K$. When scientifically meaningful categories are available, such as standard clinical thresholds, we recommend using the proposed debiased estimator with those bins. When no natural discretization is available, a practical strategy is to start with a small $K$ and gradually increase it to assess stability. If the estimates remain stable for larger values of $K$, one may select the smallest stable $K$ for simplicity.

\begin{table*}[t]
\centering
\caption{Summary of interior break points under equal-frequency discretization for $K = 2$, $K = 3$, and $K = 5$ across Monte Carlo replications. Values are reported as mean $\pm$ standard deviation. The boundary cut points are fixed at $-\infty$ (0\%) and $\infty$ (100\%).}
\label{tab:break_points_freq}
\footnotesize
\begin{tabular*}{\textwidth}{@{\extracolsep{\fill}}lccccccc@{}}
\toprule
& \multicolumn{1}{c}{$K = 2$} 
& \multicolumn{2}{c}{$K = 3$} 
& \multicolumn{4}{c}{$K = 5$} \\
\cmidrule(lr){2-2} \cmidrule(lr){3-4} \cmidrule(lr){5-8}
$n$ 
& 50.0\%
& 33.3\% & 66.7\%
& 20.0\% & 40.0\% & 60.0\% & 80.0\% \\
\midrule
500 
& 1.21 $\pm$ 0.08
& 0.53 $\pm$ 0.10 
& 1.81 $\pm$ 0.08 
& -0.16 $\pm$ 0.11 
& 0.82 $\pm$ 0.09 
& 1.57 $\pm$ 0.08 
& 2.33 $\pm$ 0.08 \\

5{,}000 
& 1.21 $\pm$ 0.03
& 0.53 $\pm$ 0.03 
& 1.81 $\pm$ 0.03 
& -0.16 $\pm$ 0.03 
& 0.82 $\pm$ 0.03 
& 1.57 $\pm$ 0.03 
& 2.34 $\pm$ 0.02 \\

50{,}000 
& 1.21 $\pm$ 0.01
& 0.53 $\pm$ 0.01 
& 1.81 $\pm$ 0.01 
& -0.17 $\pm$ 0.01 
& 0.82 $\pm$ 0.01 
& 1.57 $\pm$ 0.01 
& 2.34 $\pm$ 0.01 \\
\bottomrule
\end{tabular*}
\end{table*}

\begin{table*}[t]
\centering
\caption{Summary of the empirical range, minimum, maximum, and the largest bin proportions under equal-width discretization across Monte Carlo replications. Values are reported as mean $\pm$ standard deviation.}
\label{tab:equal_width_range_prop}
\footnotesize
\begin{tabular*}{\textwidth}{@{\extracolsep{\fill}}lcccccccc@{}}
\toprule
& & & & \multicolumn{5}{c}{Maximum proportion in bins (\%)} \\
\cmidrule(lr){5-9}
$n$ & Range & Min & Max & $K=2$ & $K=3$ & $K=5$ & $K=10$ & $K=20$ \\
\midrule
500      
& 8.11 $\pm$ 0.57  
& -3.30 $\pm$ 0.44
& 4.81 $\pm$ 0.39
& 61.6 $\pm$ 6.4
& 60.0 $\pm$ 4.9
& 39.6 $\pm$ 3.4
& 22.4 $\pm$ 2.0
& 12.3 $\pm$ 1.2 \\

5{,}000  
& 9.56 $\pm$ 0.51  
& -4.08 $\pm$ 0.38
& 5.48 $\pm$ 0.33
& 62.7 $\pm$ 5.6
& 68.3 $\pm$ 3.7
& 44.0 $\pm$ 3.0
& 25.4 $\pm$ 1.4
& 13.3 $\pm$ 0.8 \\

50{,}000 
& 10.78 $\pm$ 0.43 
& -4.73 $\pm$ 0.32
& 6.05 $\pm$ 0.28
& 63.6 $\pm$ 4.7
& 74.5 $\pm$ 3.0
& 48.4 $\pm$ 2.8
& 28.3 $\pm$ 1.2
& 14.8 $\pm$ 0.6 \\
\bottomrule
\end{tabular*}
\end{table*}

\end{document}